\documentclass[review, preprint, 12pt]{elsarticle}

\usepackage{amsmath}
\usepackage{lineno}
\usepackage{url}
\usepackage{float}
\usepackage{longtable}

\begin{document}

\begin{frontmatter}

%%Title
\title{A Phase-Field Discrete Element Method to study chemo-mechanical coupling in granular materials\tnoteref{t1}}
\tnotetext[t1]{It is important to notice that color need to be used for any figures in print.}

%%Authors
\author[1,2]{Alexandre Sac-\hspace {1pt}-Morane\corref{cor1}}
\ead{alexandre.sac-morane@uclouvain.be}

\author[1]{Manolis Veveakis}
\ead{manolis.veveakis@duke.edu}

\author[2]{Hadrien Rattez}
\ead{hadrien.rattez@uclouvain.be}

\cortext[cor1]{Corresponding author}

\affiliation[1]{organization={Multiphysics Geomechanics Lab, Duke University}, addressline={Hudson Hall Annex, Room No. 053A}, city={Durham}, postcode={27708}, country={NC, USA}}

\affiliation[2]{organization={Institute of Mechanics, Materials and Civil Engineering, UCLouvain}, addressline={Place du Levant 1}, city={Louvain-la-Neuve}, postcode={1348}, country={Belgium}}

%%Abstract
\begin{abstract}
This paper presents an extension of the discrete element method using a phase-field formulation to incorporate grain shape and its evolution. The introduction of a phase variable enables an effective representation of grain geometry and facilitates the application of physical laws, such as chemo-mechanical couplings, for modeling shape changes. These physical laws are solved numerically using the finite element method coupled in a staggered scheme to the discrete element model. The efficacy of the proposed Phase-Field Discrete Element Model (PFDEM) is demonstrated through its ability to accurately capture the real grain shape in a material subjected to dissolution only and compute the stress evolution. It is then applied to model the phenomenon of pressure solution involving dissolution and precipitation in granular materials at the microscale and enables to reproduce the creep response observed experimentally. This framework contributes to the enhanced understanding and simulation of complex behaviors in granular materials and sedimentary rocks for many geological processes like diagenesis or earthquake nucleation.
\end{abstract}

%%Research highlights
\begin{highlights}
\item Development of an extension of the discrete element method using a phase-field formulation to incorporate grain shape and its evolution.
\item Highlight the influence on the grain shape during an oedometer test under acid injection.
\item Model the phenomenon of pressure solution.
\end{highlights}

%%Keywords

\end{frontmatter}

%%=======================================================%%

\section{Introduction}

The intricate interplay between chemical and mechanical processes in soil and rocks has emerged as a key factor to consider for many engineering applications like underground storage or geothermal energy \cite{McCartney2016} or to understand geological processes like diagenesis or earthquake nucleation \cite{Rattez2020}. In particular, underground storage strategies envisioned to store large amounts of $CO_2$, captured from large point sources like power generation facilities, or $H_2$, produced during overproduction periods of renewable energies like solar or wind, involve the injection of a fluid into a reservoir rock. This fluid reacts with the surrounding rock inducing dissolution and/or precipitation that affects rocks' permeability \cite{Lesueur2020}, but also their mechanical behavior \cite{KUMAR2023}. To ensure the long-term success and safety of such storage solutions, the role of chemo-mechanical couplings needs to be understood and modeled as they can influence settlement at the surface \cite{Brzesowsky2014}, the stability of the caprock \cite{Rohmer2016, Rohmer2016b} or induced seismicity \cite{RATTEZ2021}. 

Chemo-mechanical couplings are also fundamental to many geological processes. For example, understanding of diagenetic transformations and in particular the phenomenon of pressure solution \cite{Bjorlykke1989} rely on a comprehensive grasp of these couplings. During the diagenetic processes sediments are transformed into sedimentary rocks. These processes influence reservoir quality in hydrocarbon exploration and the formation of economically valuable mineral deposits. Moreover, Pressure solution \cite{Bos2000}, has also a pivotal role in earthquake nucleation and recurrence. It involves three chemo-mechanical processes at the micro-scale \cite{Rutter1976}: dissolution due to stress concentration at grain contacts, diffusive transport of dissolved mass from the contact to the pore space, and precipitation of the solute on the less stressed surface of the grains. These processes lead to a time-dependent compaction of the rock by changing its microstructure, pore structure and composition\cite{Rutter1976}. It also induces a modification of the strength of a wide range of geological materials. As such, it has been suggested as a mechanism for long-term evolution of fault strength during interseismic periods  \cite{Sleep1992}, but also to control faults' frictional behavior at low speed and thus influences earthquake nucleation\cite{Bos2000}.

Chemo-mechanical couplings have already been investigated with finite-element models at the meso/macro scale \cite{Tang2023} but this kind of formulation strongly depends the chosen constitutive equations, which are not well constrained as they depend on the specific microstructure and chemical reactions considered. To better understand the driving processes simulations on the microstructure should be considered. To do so, the discrete model (DEM) \cite{Cundall1980} is used in the present study as it enables the micromechanical analysis of granular materials and sedimentary rocks by reproducing the interactions among individual grains and grains' rearrangement \cite{OSullivan2011}. Traditional DEM formulations assume grains as disks in two dimensions (2D) or spheres in three dimensions (3D). The dissolution/precipitation is then taken into account as a homogeneous decrease/increase of the grain diameter \cite{Cha2019, Alam2022} or with an addition/subtraction of a dissolved/precipitated layer thickness at the contacts \cite{vanDenEnde2018}. The extrapolation of the results obtained from those approaches to geological applications is limited as they only consider disks or spheres as grains, but real particles often exhibit highly irregular shapes. These complex geometries of the grains significantly influence the macroscopic mechanical behavior of granular materials \cite{Guevel2022, Binaree2019}. Therefore, accurate models should aim to capture this complexity. Various approaches have been developed within the framework of DEM to account for irregular grain shapes. These include incorporating rolling resistance calibrated with grain geometry \cite{Mollon2020, Rorato2021}, utilizing grain clusters \cite{Garcia2009}, using ellipsoids \cite{Rothenburg1991}, employing superquadric particles \cite{Podlozhnyuk2018}, and even considering polyhedral shapes \cite{Cundall1988, Nezami2004, AlonsoMarroquin2009}. More recently, a level-set discrete element model was developed to capture the complex shapes of grains and reproduce experimental results \cite{VanDerMeer2015, Kawamoto2018}. However, none of those approaches consider a grain shape evolution due to chemo-mechanical couplings. 
%Thus, the aim of the present study is to develop an extension of the discrete element model (which verifies the micro-scale, the irregular shape criteria and the localized precipitation/dissolution criteria) to track the solute generation and diffusion. 

Considering the granular material as a phase, the phase-field theory (PF) \cite{Landau1936, Cahn1958, Allen1979, Wheeler1992} provides a suitable framework for modeling the local addition or reduction of material quantity using physics-based laws. Dissolution/precipitation at the contact interfaces is controlled by introducing mechanical and chemical energy into the Allen-Cahn formulation of the phase variables, while the mass conservation and the solute diffusion are addressed through a coupled diffusion formulation of the solute concentration \cite{Guevel2020}. 

Previous studies have already explored the coupling between finite element modeling and discrete element methods to study deformable grains \cite{Munjiza2004} or grain growth and sintering \cite{Shinagawa2014}.
In this study, we present an extension of the discrete element method to simulate irregular particle shapes in granular materials and their heterogeneous evolution using the phase-field variable as a geometric descriptor. We apply this method to two cases: (i) an oedometric test with partial dissolution and (ii) the pressure solution phenomenon involving two or multiple grains. The first case highlights the influence of grain shape and dissolution on the mechanical behavior, while the second case investigates the effects of heterogeneous dissolution and precipitation on the rate of material compaction.

%%=======================================================%%

\section{Methodology}

The objective of this paper is to present a novel coupling between a Phase-Field (PF) model and a Discrete Element Model (DEM). 
In this section, the algorithms and constitutive equations for discrete element models and the phase field are presented separately. The framework of the model is then presented to give an overview of the algorithm and how each model involved in the coupling scheme is used. 

\subsection{DEM}

\emph{Time integration}

The kinematics of grains are described by their center $\mathbf{C}$, vertices $p$, and mass $m$. Various integration methods can be used \cite{Samiei2013}, but for computational efficiency, an explicit method is chosen. The Symplectic Euler method, presented in Equation \ref{Time Integration Equations}, is employed. This algorithm approximates the velocity using a forward difference scheme and the position using a backward difference scheme. It is selected for its suitability and accuracy in the simulation, with time discretization playing a significant role in the simulation outcome.

\begin{eqnarray}
a^{t-1} &=& \frac{\sum F^{t-1}}{m} \nonumber \\
v^{t} &=& v^{t-1} + a^{t-1} \times dt \nonumber \\
p^{t} &=& p^{t-1} + v^{t} \times dt
\label{Time Integration Equations}
\end{eqnarray}

Here, $F$ represents the force applied to the grain, $m$ is the mass of the grain, $\mathbf{p}$ denotes the coordinates of the grain vertices, and $dt$ is the time step.

The rotation of the grains is also considered, especially in shearing problems. The rotation integration follows the same scheme, as shown in Equations \ref{Time Integration Equations Rotation}. A rigid body rotation, as described in Equation \ref{Rigib body rotation}, is used to compute the coordinates of the grain vertices.

\begin{eqnarray}
\omega^{t} &=& \frac{\sum M^{t}}{I} \nonumber \\
\theta^{t} &=& \theta^{t-1} + \omega^{t} \times dt 
\label{Time Integration Equations Rotation}\\
\mathbf{p} &=& R \cdot (\mathbf{p} - \mathbf{C}) + \mathbf{C}
\label{Rigib body rotation}
\end{eqnarray}

In Equations \ref{Time Integration Equations Rotation}, $M$ represents the moment applied to the grain, and $I$ is the grain's moment of inertia.
In Equation \ref{Rigib body rotation}, $\mathbf{p}$ denotes the coordinates of the grain vertex, $R$ is the rotational matrix computed from the rotation increment $\theta^t-\theta^{t-1}$, and $\mathbf{C}$ represents the coordinates of the grain center.

\vskip\baselineskip

\emph{Contact detection}

As the particles are polygonal, an algorithm is needed to detect contact between grains $i$ and $j$. First, the nearest vertex of grain $j$ to grain $i$ needs to be determined. This is achieved by applying a virtual displacement, as described in Equation \ref{Virtual Displacement}, to the vertices of grain $j$.

\begin{equation}
\mathbf{p}^v = \mathbf{p} + \max(R_i,R_j) \times \frac{\mathbf{C}_j-\mathbf{C}_i}{\lVert \mathbf{C}_j-\mathbf{C}_i \rVert}
\label{Virtual Displacement}
\end{equation}

Here, $\mathbf{p}^v$ represents the virtual coordinates of the vertex, $\mathbf{p}$ denotes the coordinates of the vertex, $R_{i,j}$ is the maximum radius of the grain $i,j$ (grains have different radii depending on the direction), and $\mathbf{C}_{i,j}$ denotes the coordinates of the center.
This virtual translation assumes that the grains are not in contact (even if they were before this operation). After the virtual displacement, the distances between vertices of grains $i$ and $j$ are computed. The vertices used for contact detection are determined by considering the nearest ones after the virtual displacement.

Next, the algorithm of the common plane, described in \ref{CP Algorithm Appendix}, is applied to compute the contact normal vector $\overrightarrow{n}$ \cite{Cundall1988, Nezami2004, Podlozhnyuk2018}. Finally, the distance $d^{i,j}$ between each grain and the common plane needs to be calculated. As shown in Figure 1 of Nezami et al. \cite{Nezami2004}, space is divided into two parts. The normal distance between grains is defined as $\delta_n = d^j-d^i$ (with particle $i$ mainly in the negative half-space and particle $j$ mainly in the positive half-space). It is important to note that particles are in contact only if the overlap $\delta_n$ is negative, as depicted in Figure 2 of Nezami et al. \cite{Nezami2004}.

\vskip\baselineskip

\emph{Interaction between grains}

\begin{figure}[ht]
\centering
\includegraphics[width=0.8\linewidth]{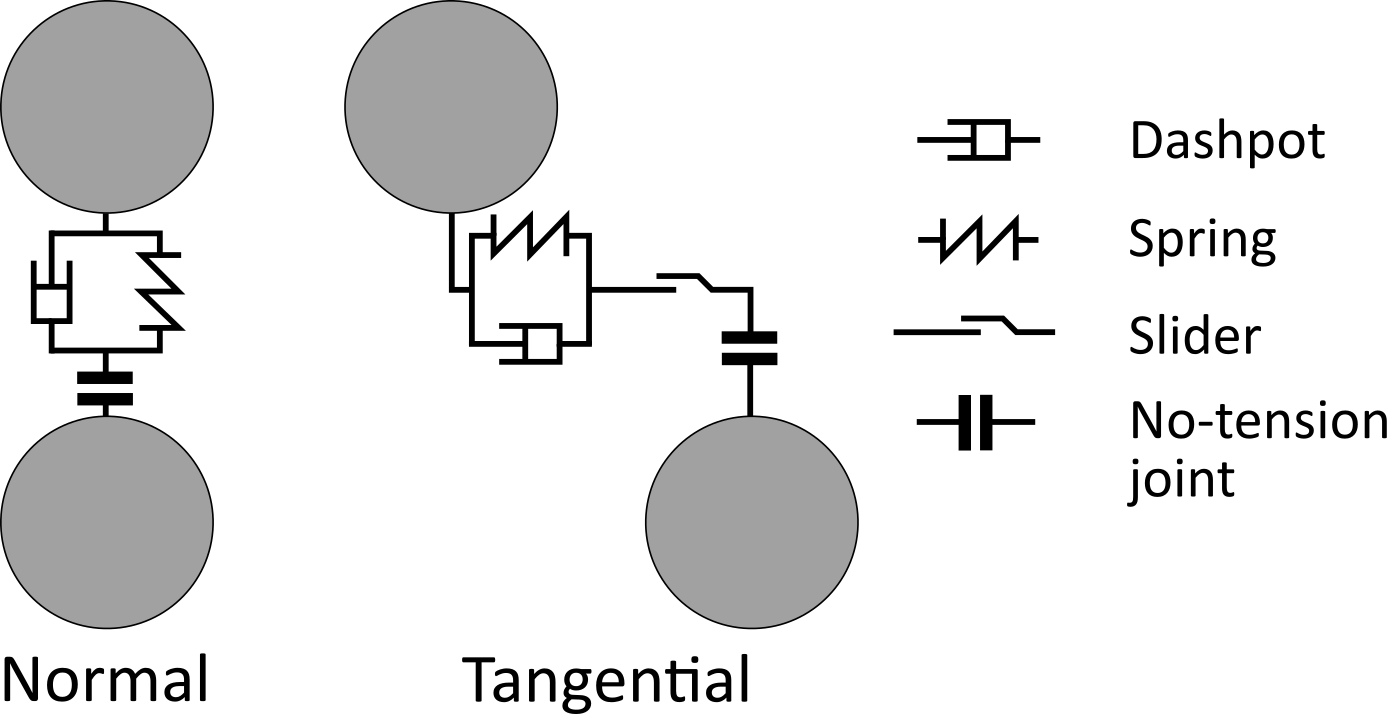}
\caption{Model for the interaction between two particles.}
\label{Model Interaction Figure}
\end{figure}

The interaction between two grains is illustrated in Figure \ref{Model Interaction Figure} and described by the following equations. The normal response consists of a nonlinear spring and a dashpot in parallel, along with a no-tension joint. Similarly, the tangential response includes a nonlinear spring and a dashpot in parallel, a slider, and a no-tension joint.

The normal spring in the contact model is defined by Equation \ref{From Material to Stiffness Equations}. It is worth noting that some equivalent parameters, such as the Young's modulus $Y_{eq}$ and the radius $R_{eq}$, are used and depend on the grain properties.

\begin{eqnarray}
k &=& \frac{4}{3}Y_{eq}\sqrt{R_{eq}} \nonumber \\
\text{where } \frac{1}{Y_{eq}} &=& \frac{1-\nu_i^2}{Y_i}+ \frac{1-\nu_j^2}{Y_j} \nonumber \\
\text{and } \frac{1}{R_{eq}} &=& \frac{1}{R_i} + \frac{1}{R_j}
\label{From Material to Stiffness Equations}
\end{eqnarray}

Here, $k$ represents the stiffness of the normal spring, $Y_{i,j}$ is the Young's modulus of the grain $i$ or $j$, $\nu_{i,j}$ is the Poisson's ratio of the grain $i$ or $j$, and $R_{i,j}$ is the mean grain radius of the grain $i$ or $j$.

To introduce energy dissipation and reach a equilibrium, a linear damping term is added. The damping coefficient $\eta$ is defined in Equation \ref{From restitution coefficient to damping} based on a restitution coefficient ($e=1$ for a perfect elastic contact, $e=0$ for complete energy dissipation).

\begin{eqnarray}
\eta &=& 2 \times \gamma \sqrt{m_{eq} \times k } \nonumber \\
\text{where } \gamma &=& -\frac{\ln(e)}{\sqrt{\pi^2+\ln(e)^2}}\nonumber \\
\text{and } m_{eq} &=& \frac{m_i \times m_j}{m_i + m_j}
\label{From restitution coefficient to damping}
\end{eqnarray}

Here, $\eta$ denotes the damping coefficient, $m_{i,j}$ is the mass of the grain $i$ or $j$, $k$ is the stiffness of the normal spring, and $e$ is the restitution coefficient.

Next, the Hertz theory is applied in Equation \ref{Normal DEM Law} to obtain the normal force \cite{Hertz1882, Johnson1985, DiRenzo2004}. It is important to note that this mechanical response is nonlinear with respect to the overlap. The stiffness increases as the overlap becomes larger. The damping term is also included.

\begin{eqnarray}
F_n = Fs_n + Fd_n = -k \times \delta_n^{3/2} - \eta \times (\overrightarrow{v_i}-\overrightarrow{v_j}) \cdot \overrightarrow{n}
\label{Normal DEM Law}
\end{eqnarray}

Here, $\delta_n$ is the distance between grains, $\overrightarrow{v_{i,j}}$ is the speed of the grain $i$ or $j$, and $\overrightarrow{n}$ is the normal vector of the contact.

The tangential spring in the contact model is defined by Equation \ref{From Material to Stiffness Equations 2}. Similar to the normal spring, it uses equivalent parameters (shear modulus $G_{eq}$ and radius $R_{eq}$) that depend on the grain properties. Additionally, this parameter is obtained using the Mindlin and Deresiewicz theory \cite{Johnson1985, DiRenzo2004, Mindlin1953}. The tangential stiffness depends on the normal overlap $\delta_n$, with a stiffer spring for larger overlaps. Moreover, it is nonlinear, depending on the value of the tangential overlap $\delta_t$, with a softer spring for larger tangential overlaps.

\begin{eqnarray}
kt &=& kt0 \sqrt{1 -\frac{2 \times kt0 \times \delta_t}{3 \times \mu \times Fs_n}} \nonumber \\
\text{where } kt0 &=& 8 \times G_{eq} \sqrt{R_{eq} \times \delta_n} \nonumber \\
\text{and } \frac{1}{G_{eq}} &=& \frac{1-\nu_i}{G_i}+ \frac{1-\nu_j}{G_j} \nonumber \\
\text{and } G_i &=& \frac{Y_i}{2(1+\nu_i)}
\label{From Material to Stiffness Equations 2}
\end{eqnarray}

Here, $kt$ represents the stiffness of the tangential spring, $\mu$ is the friction coefficient, and $G_{i,j}$ is the shear modulus of the grain $i$ or $j$.

Similar to the normal contact, a linear damping term is added for energy dissipation, and the damping coefficient is defined in Equation \ref{tangential damping equation}.

\begin{eqnarray}
\eta_t &=& 2 \times \gamma \sqrt{m_{eq} \times kt}
\label{tangential damping equation}
\end{eqnarray}

To obtain the tangential force, an incremental relation defined in Equation \ref{Tangential DEM Law} is used. It is important to consider rotation, as grain rotation is the primary phenomenon in shearing conditions. The Coulomb criterion is applied, allowing sliding between grains.

\begin{eqnarray}
\Delta \delta_t &=& \left((\overrightarrow{v_i}-\overrightarrow{v_j}) \cdot \overrightarrow{t} + R_i \times \omega_i + R_j \times \omega_j\right) \times dt \nonumber \\
Fs_t^t &=& Fs_t^{t-1} - kt \times \Delta \delta_t \nonumber \\
Fs_t &\leq& \mu Fs_n
\label{Tangential DEM Law}
\end{eqnarray}

Here, $\Delta \delta_t$ denotes the increment in tangential overlap, $\overrightarrow{t}$ is the tangential vector of the contact, $Fs_t$ is the tangential force, and $\mu$ is the friction coefficient. It is important to notice that the orientation of the tangential vector can evolve with time. An update of the incremental tangential force $Fs_t$, defined in Equation \ref{Update Tangential Force}, is applied.

\begin{equation}
    Fs_t^{t-1} = Fs_t^{\prime,t-1}\,\overrightarrow{t}_{old}.\overrightarrow{t}_{new}
    \label{Update Tangential Force}
\end{equation}
Here, $Fs_t^{t-1}$ is the tangential force updated, $Fs_t^{\prime,t-1}$ is the tangential force not updated, $\overrightarrow{t}_{old}$ is the tangential vector of the contact at the previous iteration, and $\overrightarrow{t}_{new}$  is the tangential vector of the contact at the current iteration.

The total tangential force is the sum of the spring term and the damping term, as explained in Equation \ref{total tangential force}.

\begin{equation}
F_t = Fs_t + Fd_t = -kt \times \delta_t^{3/2} - \eta_t\times(\overrightarrow{v_i}-\overrightarrow{v_j}).\overrightarrow{t}
\label{total tangential force}
\end{equation}

Note that the equations for the Discrete Element Method outlined in this section and used in the present paper are frequently applied in the literature for geomaterials \cite{OSullivan2011}. Multiple adaptations and expansions of this model exist, tailored to the specific applications targeted or couplings considered \cite{Feng2023} and could be easily added to the present framework.

\vskip\baselineskip

\subsection{Phase-Field}

\begin{figure}[ht]
    \centering
      \includegraphics[width=0.7\linewidth]{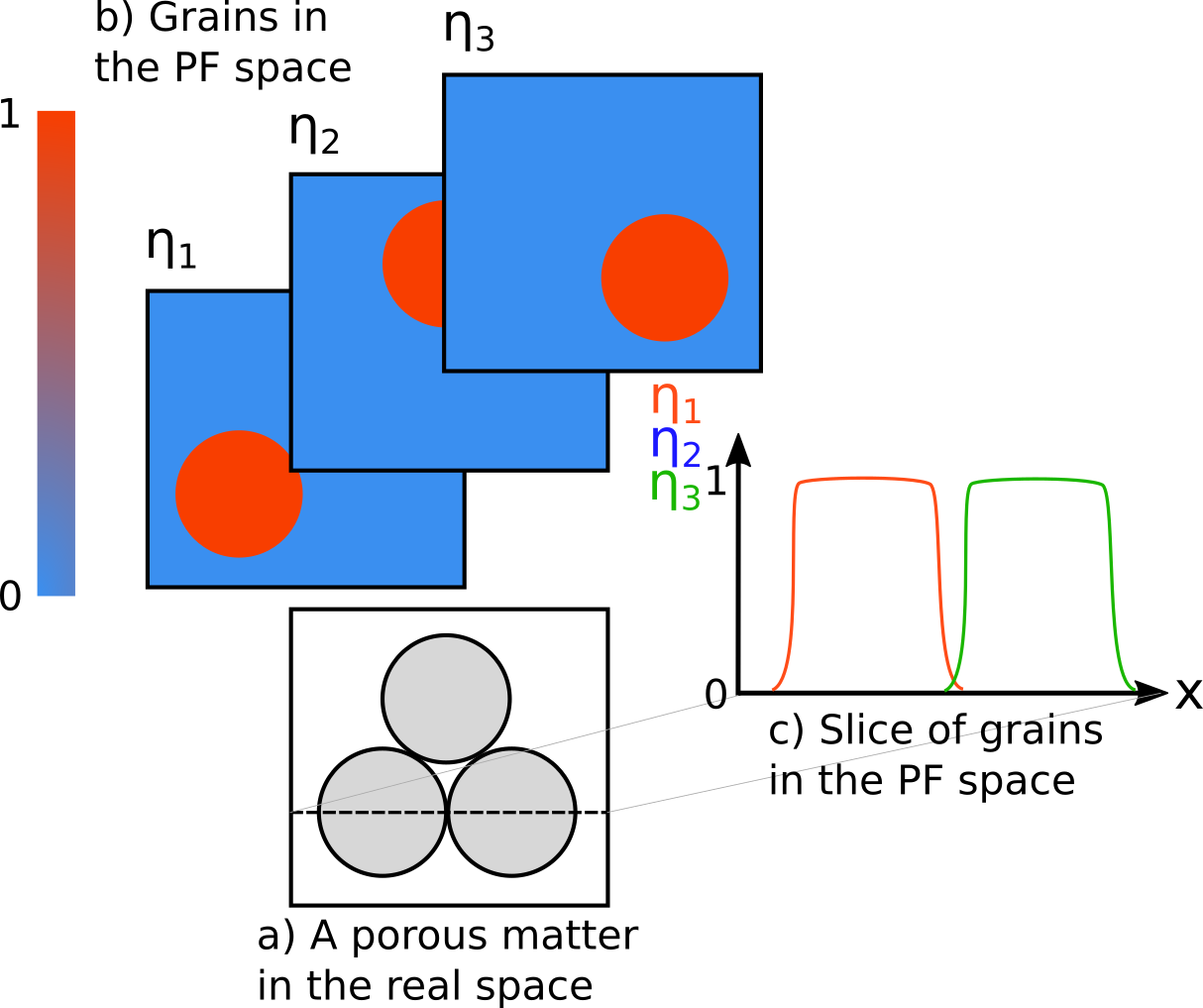}
    \caption{The phase-field method is applied to granular matter: a) a granular matter is composed of grains and by pore, b) one phase variable $\eta_i$ is generated by grain, c) the phase variable is equal to $1$ if it is inside the grain associated and equal to $0$ elsewhere.}
    \label{PF for Geomaterial}
\end{figure}

A granular matter is composed of grains and pores. As illustrated in Figure \ref{PF for Geomaterial}, this porous matter is represented here in the phase-field space by a combination of phase variables (one phase per grain). This phase variable is equal to $1$ if it is inside the grain associated and equal to $0$ elsewhere. 

The phase-field method, specifically the Allen-Cahn equation described in Equation \ref{AC formulation}, is employed to describe the dissolution/precipitation of the material \cite{Allen1979}. In this approach, nonconserved order parameters $\eta_j$ are utilized to represent the phase transformation. 

\begin{equation}
\frac{\partial \eta_j}{\partial t} = -L_j\left(\frac{\partial \left(f_{loc}+E_d\right)}{\partial \eta_j} - \kappa_j \nabla^2 \eta_j \right)
\label{AC formulation}
\end{equation}

\begin{figure}[ht]
    \centering
    \includegraphics[width=0.5\linewidth]{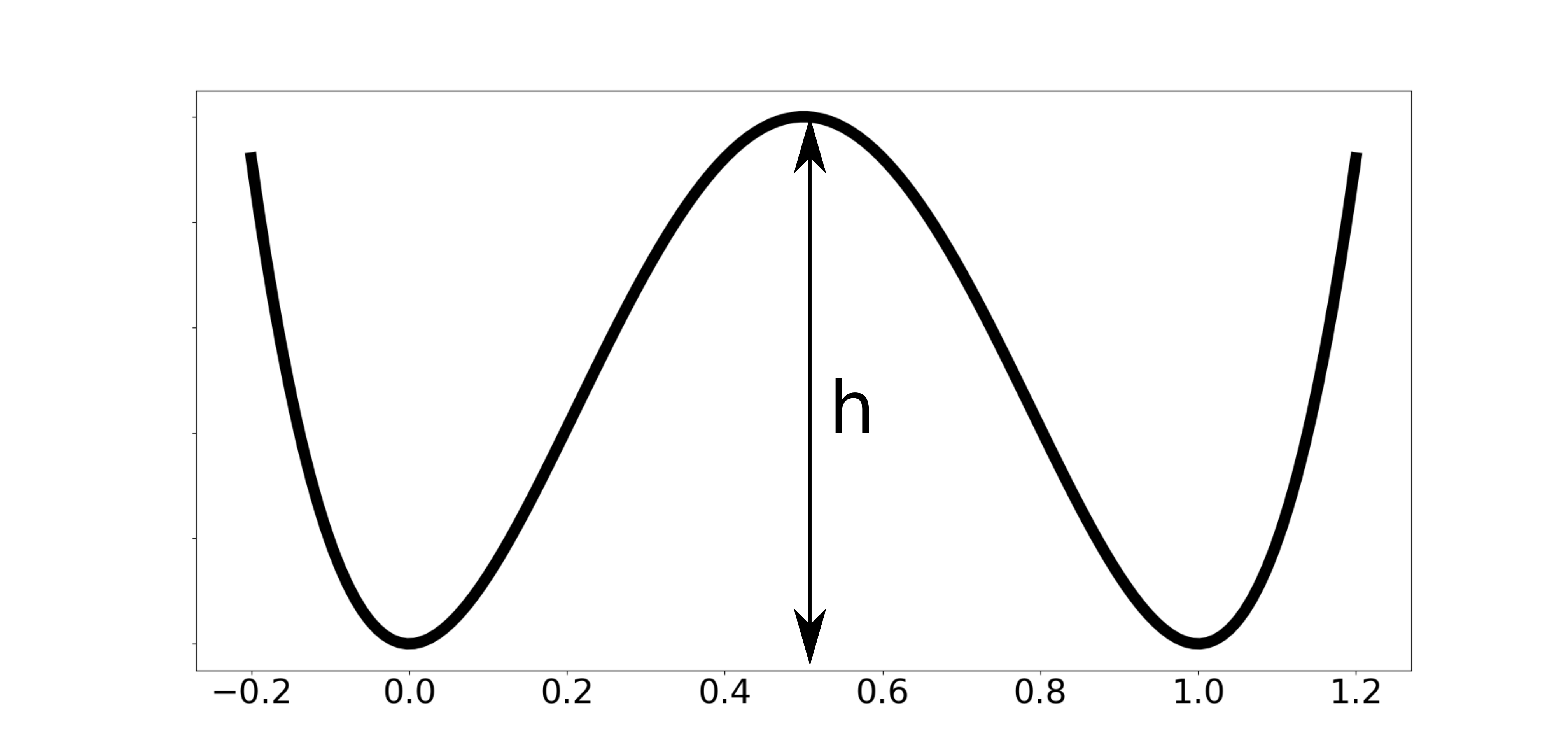}
    \caption{A double-well function as a local free energy density $f_{loc}$.}
    \label{Free Energy Figure}
\end{figure}

In this equation, the terms $L_j$ and $\kappa_j$ correspond to the order parameter mobility and the gradient energy coefficient, respectively. The order parameter mobility affects the overall dissolution/precipitation speed, while the gradient energy coefficient influences the interface width and reaction. The term $E_d$ encompasses additional sources of energy introduced into the system, such as mechanical, chemical, or thermal loading. The local free energy density $f_{loc}$ is specified in Equation \ref{Free energy} and is visualized in Figure \ref{Free Energy Figure}.

\begin{equation}
f_{loc} = 16 \times h \times \left(\sum_j \eta_j^2(1-\eta_j)^2\right)
\label{Free energy}
\end{equation}

The local free energy density $f_{loc}$ employs a double-well function with a barrier height $h$. Notice that minima of this potential energy are located at $\eta_j=0$ and $\eta_j=1$. At the equilibrium and without external destabilization, the phase variables stay at $\eta_j=0$ or $\eta_j=1$ (no dissolution/precipitation).
Figure \ref{Free Energy Figure} illustrates the double-well function as a representation of the local free energy density $f_{loc}$.

\begin{figure}[ht]
    \centering
    \includegraphics[width=0.45\linewidth]{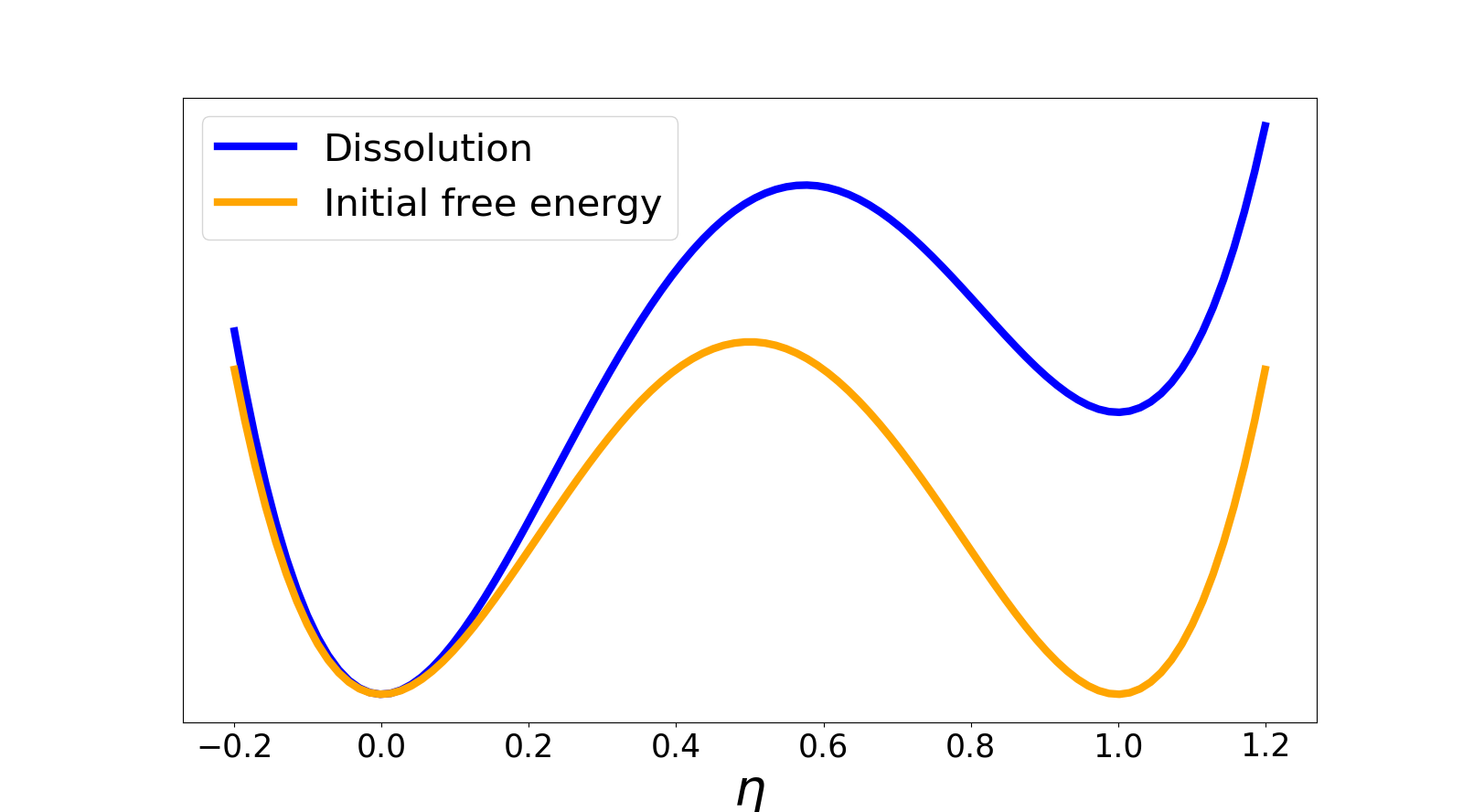}~
    \includegraphics[width=0.45\linewidth]{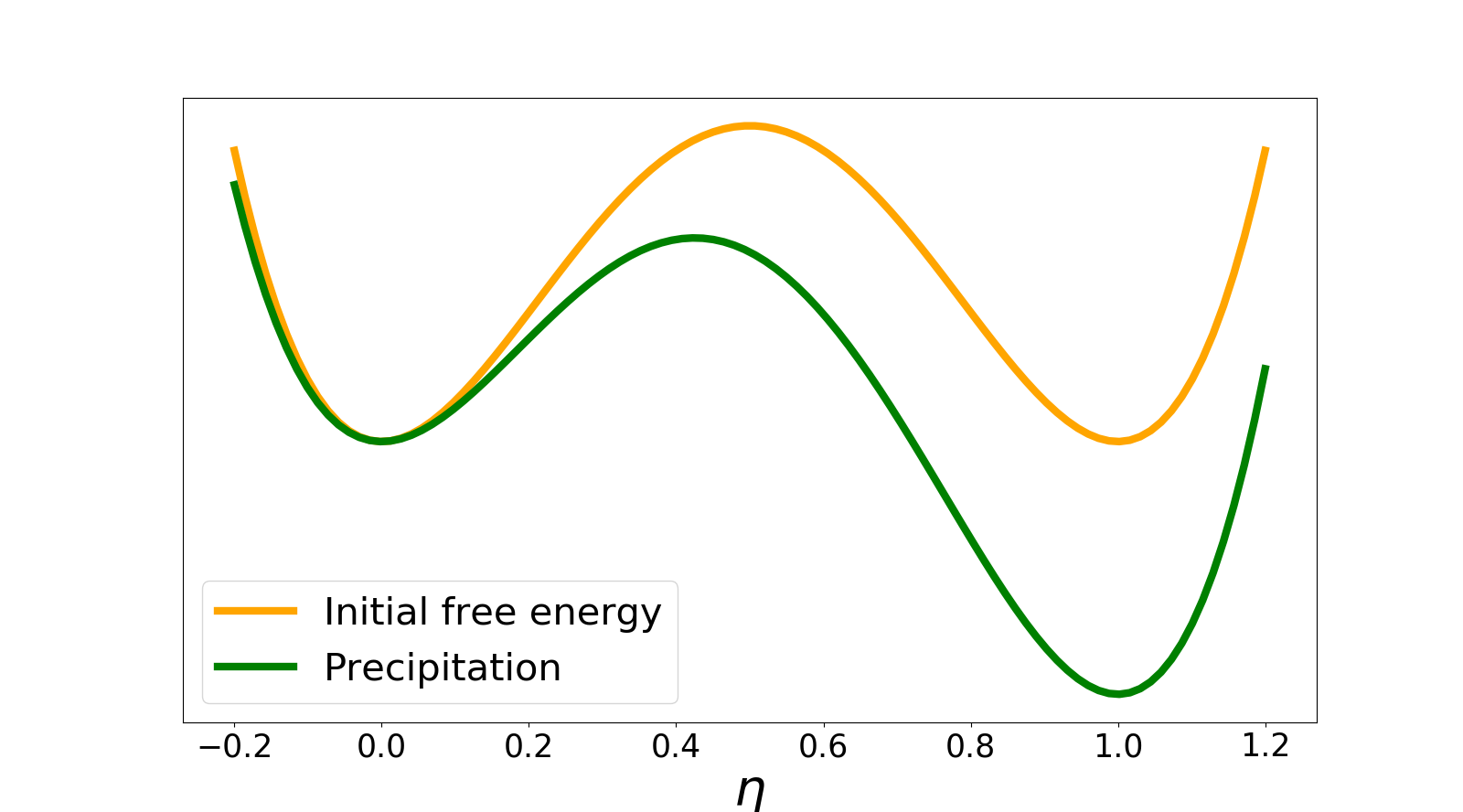}
    Dissolution destabilization \hspace{0.1\linewidth} Precipitation destabilization
    \caption{The double-well function is destabilized to favor dissolution/precipitation.}
    \label{Destabilized Double Well}
\end{figure}

To obtain localized dissolution/precipitation, the idea is to add external source term $E_d$ to tilt the initial free energy to favor dissolution/precipitation, as depicted in Figure \ref{Destabilized Double Well}. 
When the dissolution phenomenon dominates, the double-well function is tilted towards $\eta_j = 0$, causing the material to aim for dissolution. Conversely, when the precipitation phenomenon dominates, the double-well function is tilted towards $\eta_j = 1$, indicating the material's tendency to precipitate.

To model solute concentration in the pore fluid, a new variable, denoted as $c$, is introduced in the phase-field formulation. The Equation \ref{AC formulation} is solved along with a new equation for the variable $c$, as given in Equation \ref{Solute formulation}.

\begin{equation}
	\frac{\partial c}{\partial t} = -\sum\limits_j\left(\frac{\partial \eta_j}{\partial t}\right) + \kappa_c \nabla^2 c    
    \label{Solute formulation}
\end{equation}
The Equation \ref{Solute formulation} represents the conservation of solute mass and is solved using the variations of phase-field variables $\eta_j$ as a source term. Here, $\kappa_c$ represents the gradient energy coefficient that controls the diffusion rate. Notably, the gradient energy coefficient $\kappa_c$ is assumed to be 0 within the grains, so that the solute concentration $c$ is nonzero outside the grains and diffuse from the contact area to the pore surface. 
The algorithm's implementation for building the heterogeneous gradient energy coefficient $\kappa_c$ is described in \ref{Build kappa c}.
The terms $\frac{\partial c}{\partial t} = - \sum\limits_j \frac{\partial \eta_j}{\partial t}$ in Equation \ref{Solute formulation} ensure the conservation of mass. When a grain ($\eta_j$) dissolves, solute concentration ($c$) is generated, and vice versa, when $\eta_j$ is generated through precipitation, some solute disappears.

%%=======================================================%%

\vskip\baselineskip

\subsection{PF-DEM Couplings}

\begin{figure}[ht]
\centering
\includegraphics[width=0.6\linewidth]{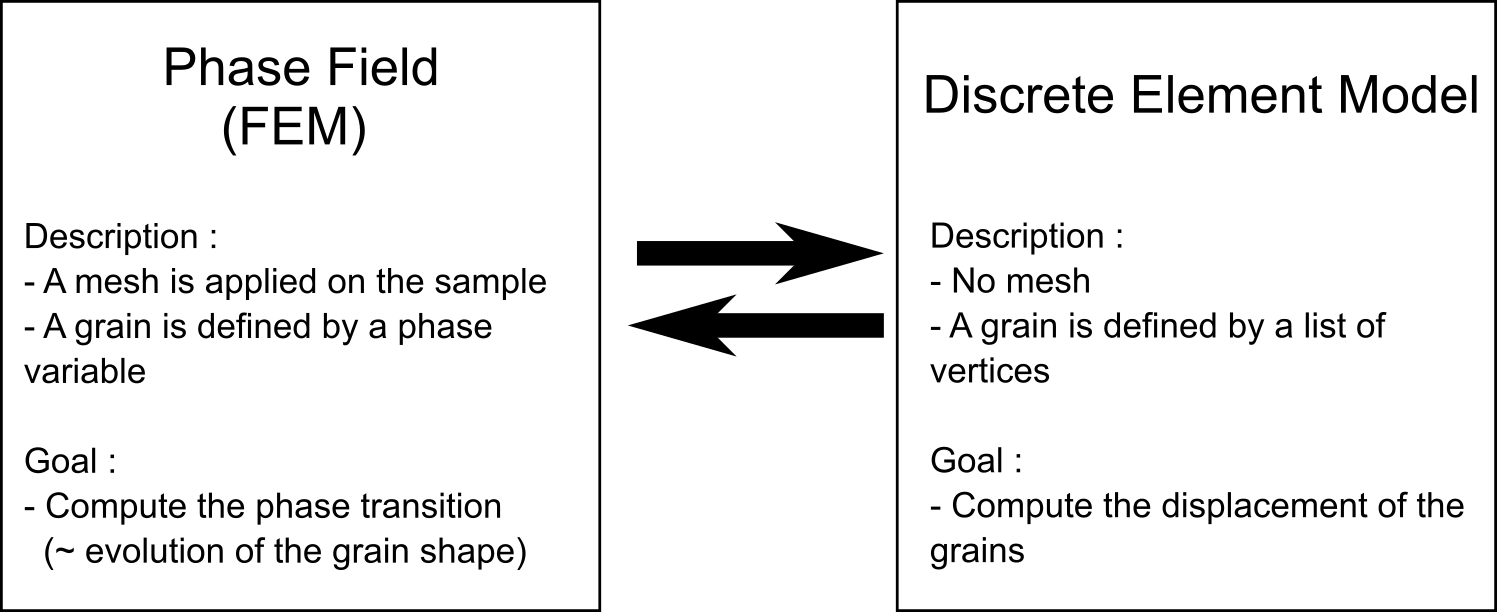}
\caption{Data exchange between the phase-field simulation and discrete element modelization.}
\label{Data Exchange Figure}
\end{figure}

\begin{figure}[ht]
\centering
\includegraphics[width=0.9\linewidth]{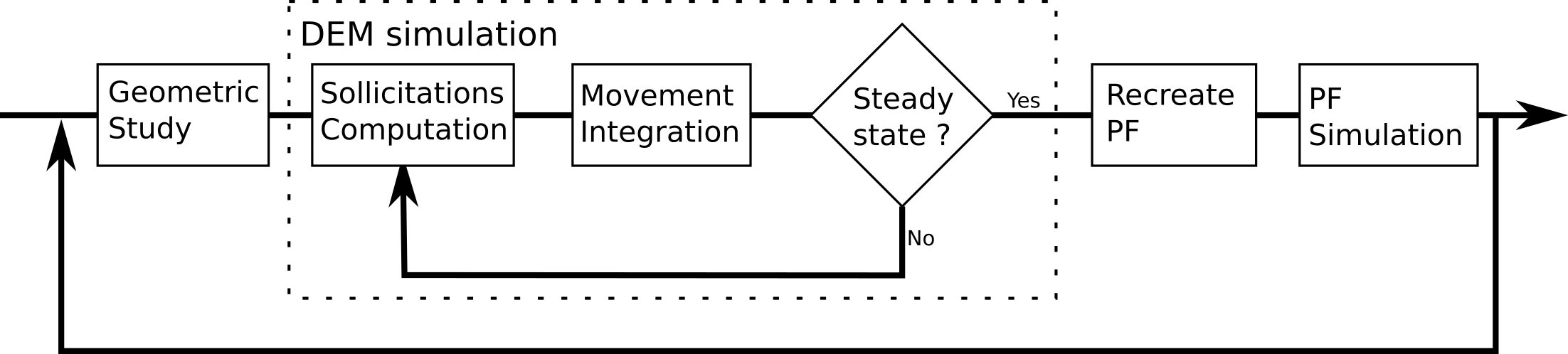}
\caption{Global scheme of the simulation. Two time scales are used: one for the discrete element modelization and one for the phase-field modelization.}
\label{Global Scheme Figure}
\end{figure}

The data exchange and global scheme of the PF-DEM couplings are illustrated in Figures \ref{Data Exchange Figure} and \ref{Global Scheme Figure}, respectively.

In the phase-field simulation, the sample geometry is discretized into a mesh, and the grains are represented by phase variable maps. A grain detection algorithm is applied to determine the new grain shape and solute configuration based on the phase-field outputs. On the other hand, in the discrete element modelization, grains are represented as a collection of vertices. The objective of this model is to compute the new positions of the grains and the mechanical energy at each contact.

It is important to note that although PF and DEM operate on different time scales, they are still connected. The mechanical equilibrium (DEM) is assumed to be instantaneous, while grain dissolution or precipitation occurs over longer time periods, typically several hours to days.

\begin{figure}[ht]
\centering
\includegraphics[width=0.3\linewidth]{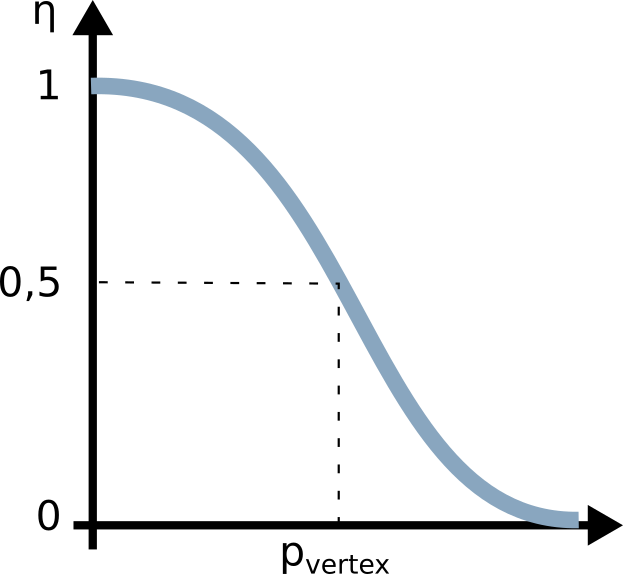}
\caption{The phase field is divided into layers, and the grain boundary is assumed to occur at $\eta_j=0.5$.}
\label{PFtoDEM Figure}
\end{figure}

After a phase-field step, the system does not immediately reach a mechanical equilibrium due to ongoing dissolution or precipitation. Therefore, the new grain boundaries need to be detected. Figure \ref{PFtoDEM Figure} illustrates the division of the phase-field into multiple layers (along the x- and y-axes), assuming the grain boundary occurs at $\eta_j=0.5$. A polygonal particle is obtained based on the vertex positions. To reduce computational cost, only a certain number of vertices are used. The discretization of the grain boundary is crucial, and the quality of overlap estimation depends strongly on this parameter, as explained in \ref{Find n border}. The geometric properties of the grain, such as its center, surface area, and inertia, can be computed from the list of vertices using a Monte Carlo method described in \ref{Monte Carlo method Appendix}.

Once the grain boundaries are known, a series of DEM iterations takes place until the system reaches an equilibrium. In order to dissipate mechanical energy, some damping is introduced, as explained earlier in this section. The convergence criteria for detecting the equilibrium may vary depending on the problem. However, a maximum number of iterations is set to avoid an infinite loop. Once the equilibrium is reached, new phase maps are constructed.

In the simulations presented in Section \ref{Section Irregular Shape}, a Deconstruction-Rebuild algorithm is employed, assuming that the interface width ($w$) remains constant throughout the study. This parameter is proportional to the square root of the ratio between the gradient energy coefficient and the double-well's height, both of which are constant. Therefore, $w\propto \sqrt{\kappa/h}$ \cite{Guevel2020}. Given the knowledge of the center and vertices of a grain, a distribution of the radius following the angle can be computed (the choice of the angle origin is arbitrary; for this study, the +x-axis is considered). Subsequently, the new phase variable $\eta_j$ is calculated using a cosine profile described by Equation \ref{Cosine Profile} for all nodes of the mesh.

\begin{equation}
\eta_j = \left\lbrace
\begin{array}{ll}
1 &\text{for } r \leq R-\frac{w}{2} \\
0.5\times \left(1+\cos\left(\pi\frac{r-R+\frac{w}{2}}{w}\right)\right) &\text{for } R-\frac{w}{2} \leq r \leq R+\frac{w}{2} \\
0 &\text{for } r \geq R+\frac{w}{2}
\end{array}
\right.
\label{Cosine Profile}
\end{equation}

Here, $r$ represents the distance between the node and the center, $R$ is the radius in the considered direction, and $w$ denotes the interface width.

\begin{figure}[ht]
\centering
\includegraphics[width = 0.8\linewidth]{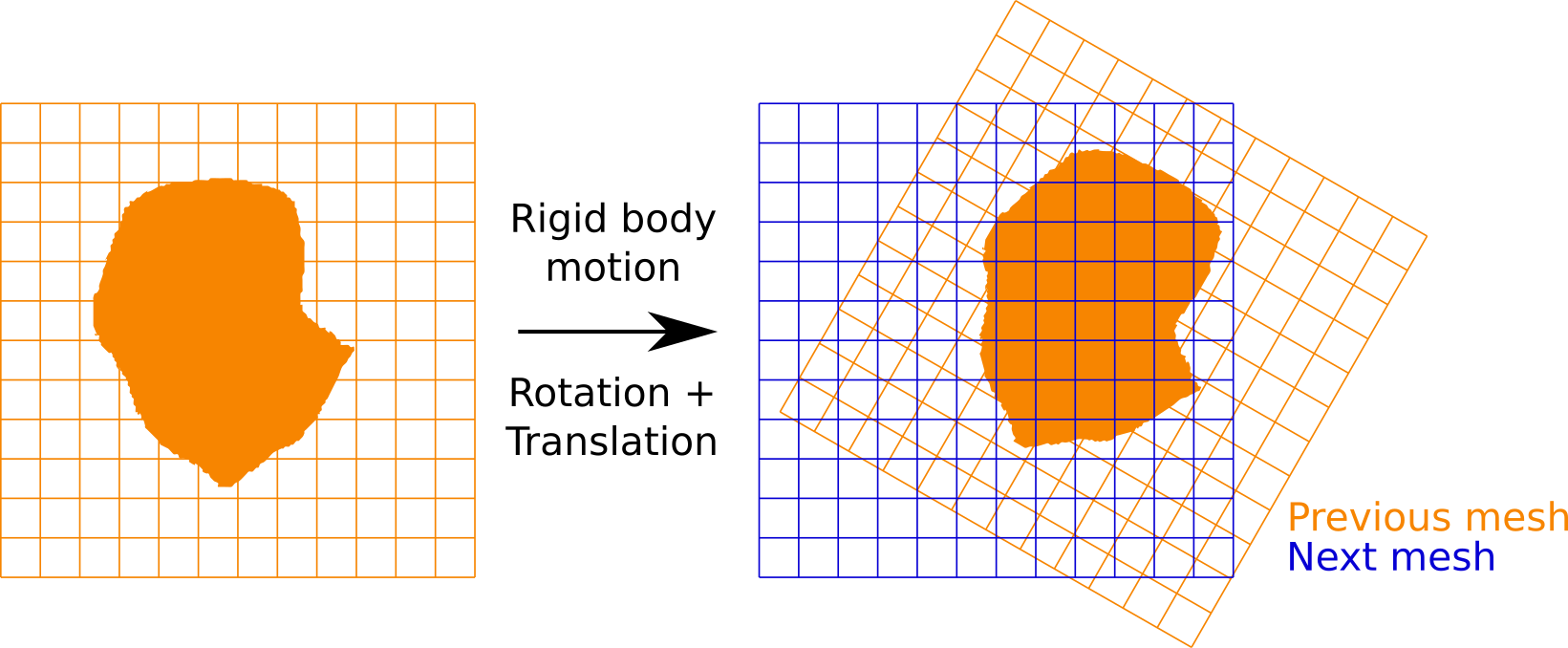}
\caption{Scheme of the interpolation algorithm. The next mesh is computed from the previous mesh by applying an inverse rigid body motion.}
\label{Interpolation Scheme}
\end{figure}

In the simulations presented in Section \ref{Section Heterogeneous dissolution/precipitation}, a different algorithm called Interpolation is used. This algorithm considers that the interface width ($w$) is not constant during the study, particularly increasing in the contact zone during pressure solution modeling. The new phase map is obtained through interpolation from the previous one. As illustrated in Figure \ref{Interpolation Scheme}, at each iteration, the next mesh is computed based on the previous mesh by applying an inverse rigid body motion. Since the two meshes often do not perfectly match, bilinear interpolation is performed to estimate the values of the next mesh.

\section{Application to Irregular Shapes and homogeneous dissolution }
\label{Section Irregular Shape}

\begin{figure}[ht]
    \centering
    \includegraphics[width=0.3\linewidth]{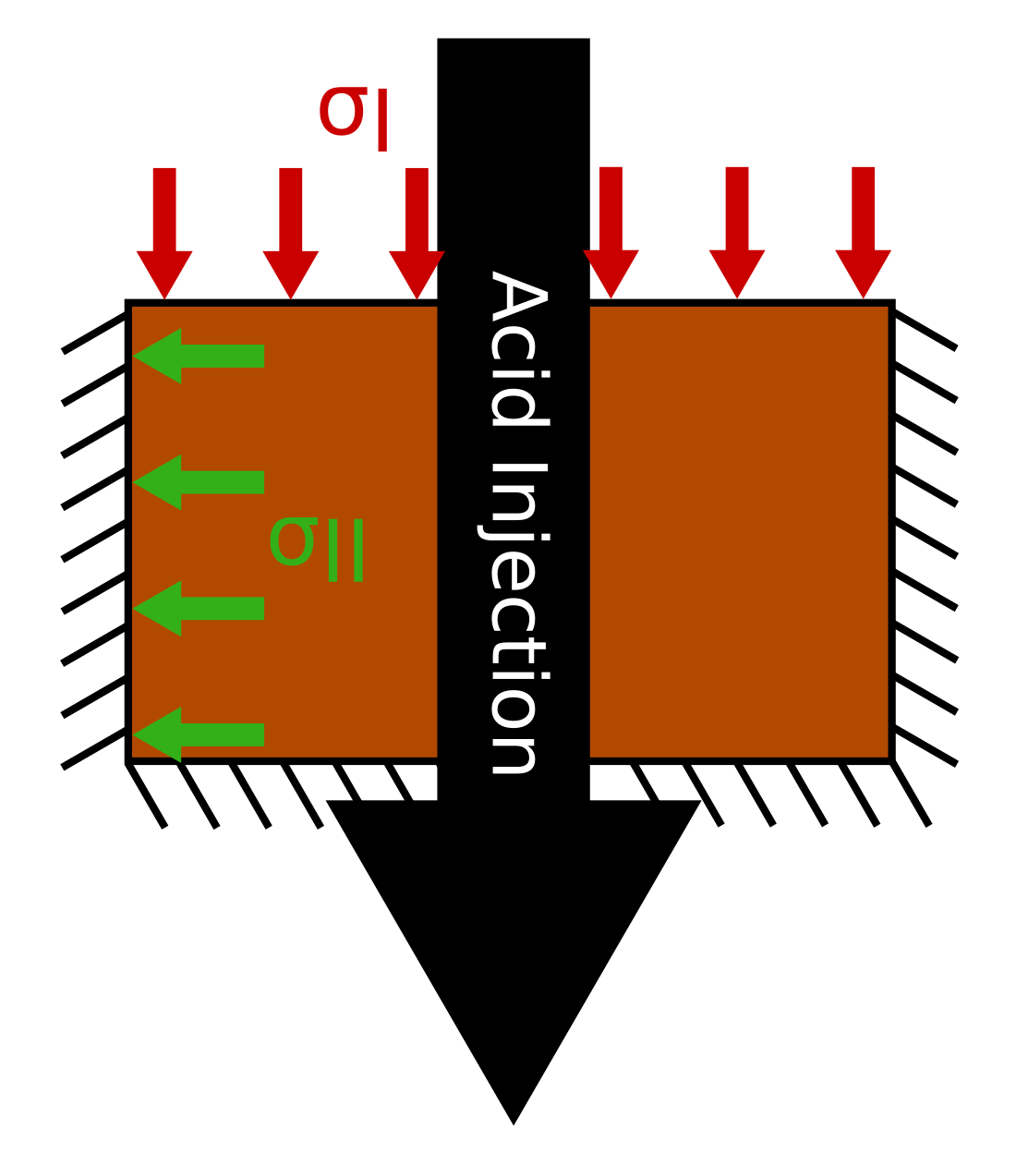}
    \caption{Scheme of the sample under oedometric condition and acid injection.}
    \label{Scheme Oedometric Condition Acid Injection}
\end{figure}

This section aims to demonstrate the capability of the PFDEM to handle irregular grain shapes. The simulations conducted here are intended to be compared with previous experiments and numerical simulations conducted by Shin et Santamarina and Cha et Santamarina \cite{Santamarina2009, Santamarina2014}. Figure \ref{Scheme Oedometric Condition Acid Injection} illustrates the setup where a sample is subjected to a constant stress in oedometric conditions while acid is injected. The granular material consists of two types of grains: dissolvable and undissolvable.

The distribution between dissolvable and undissolvable grains is determined by the ratio of the total surface area of dissolvable grains to the total surface area of all grains. As the acid is injected, it tends to reduce the size of the dissolvable grains. It has been observed that the coefficient $k_0=\sigma_{II}/\sigma_I$ evolves with the dissolution process due to internal reorganization around the dissolvable grains.

The simulation campaign parameters are provided in Table \ref{Parameters Acid Oedometer}. The mean behavior is obtained by extrapolating results from three simulations for each configuration.

\begin{longtable}{| p{8.5cm} | l | p{2.5cm} |}
        \hline
        Parameters&Unit&Value\\
        \hline
        Number of grains&-&300\\
        Particle Size Distribution for undissolvable grains (radius and percentage of the number of undissolvable grains)&$\mu m\,( \%)$&420 (17), 385 (33), 315 (33), 280 (17)\\
        Percentage of grain dissolvable \hspace{0.4\linewidth}($=S_{dissolvable}/(S_{dissolvable}+S_{undissolvable})$) &\%& 5 or 15\\
        Particle Size Distribution for disks as dissolvable grains (radius and percentage of the number of dissolvable grains)&$\mu m\,( \%)$&360 (17), 330 (33), 270 (33), 240 (17)\\
        Particle Size Distribution for squares as dissolvable grains (dimension and percentage of the number of dissolvable grains)&$\mu m\,( \%)$&500 (17), 460 (33), 380 (33), 440 (17)\\
        \hline
        Young modulus& GPa&70\\
        Poisson's ratio&-&0.3\\
        Density&$kg/m^3$&2500\\
        Friction grain-grain&-&0.5\\
        Friction grain-wall&-&0\\
        Restitution coefficient&-&0.2\\
        \hline
        Mobility for phase-field&-&1\\
        Gradient coefficient for phase-field&-&3\\
        \hline
        Time step for discrete element model&sec&$\Delta t_{crit}/8$\\
        \hline
        Linear force applied on the upper wall&N/m&$4\times Y\,R_{mean}$/2000\\
        Percentage of the mean dimension dissolved between each iteration&\%&0.5\\
        \hline
    \caption{Parameters used for the simulation campaign, concerning the acid injection inside a sample under oedometric condition.}
    \label{Parameters Acid Oedometer}
\end{longtable}

\begin{figure}[ht]
    \centering
    \includegraphics[width=0.4\linewidth]{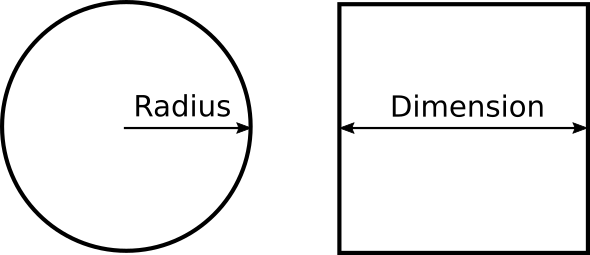}
    \caption{Definition of the characteristic length for a disk and for a square.}
    \label{Definition Grain Geometry}
\end{figure}

In the simulations presented in this study, the undissolvable grains are always assumed to be disks, while the dissolvable grains can be either disks or squares (although other shapes are possible, the simulation is restricted to these two). The main objective of this section is to discuss the influence of grain shape on overall behavior. The latter choice is only possible in the present framework and represents more accurately the experimental setup as the undissolvable grains are glass beads and the dissolvable ones are salt grains with a cubic shape. The size of the grains is defined as shown in Figure \ref{Definition Grain Geometry}, where a disk is characterized by its radius and a square is determined by the length of one side.

The choice of grain shape affects certain material properties, such as surface density. For disk particles, the surface density is given by $\sigma_m = \frac{4}{3} \rho R_{\text{mean}}$, while for square particles, it is $\sigma_m = \rho L_{\text{mean}}$, where $R_{\text{mean}}$ is the mean radius, $L_{\text{mean}}$ is the mean dimension, and $\rho$ is the density.

The definition of the particle size distribution depends on the specific case and is presented in Table \ref{Parameters Acid Oedometer}. The total number of grains has been chosen to minimize computational cost while respecting the assumption of a representative element volume, as determined by a preliminary study presented in \ref{Look for REV}.

To establish the initial configuration, perfect disks are assumed (using the circumscribed circle for non-disk grains). These disks are randomly generated without overlap within the sample and then loaded. Once an initial equilibrium is reached, the disks are discretized, and their shapes are maintained. Another loading step is performed to reach a second equilibrium. A sample geometry with a height-to-diameter ratio of 0.6 is chosen to minimize boundary effects at the top and bottom \cite{Santamarina2009}, and the sample generation process ensures this geometry is achieved.

The time step is computed based on the critical time step defined for a sphere using the Hertzian contact model, as shown in Equation \ref{Critical Time Step} \cite{Sheng2004}:

\begin{equation}
\Delta t_{\text{crit}} = \frac{\pi L_{\text{min}}}{0.1631\nu + 0.88}\sqrt{\frac{\rho}{G}}
\label{Critical Time Step}
\end{equation}

Here, $L_{\text{min}}$ represents the minimal characteristic length for the grains (half of the dimension for squares), $\nu$ is the Poisson's ratio, $\rho$ is the density, $G=\frac{Y}{2(1+\nu)}$ is the shear modulus, and $Y$ is the Young's modulus.

A linear vertical force is applied to the upper wall to confine the sample. The intensity of this force is proportional to the normal stiffness, considering an isotropic 3D material model with $k_n = 4 \times Y \times R_{\text{mean}}$ \cite{Cundall2004}, where $Y$ is the Young's modulus and $R_{\text{mean}}$ is the mean radius of the undissolvable grains.

In this case, the external source term $E_d$ in the phase-field formulation (as described in Equation \ref{AC formulation}) models the acid. It is assumed that the solute quantity generated by dissolution is instantaneously evacuated (the diffusive-conservative Equation \ref{Solute formulation} is not solved). The formulation of this source term is explained in Equation \ref{Ed formulation acid}. It should be noted that this term is applied only to dissolvable grains, as there is no interaction between grains during the phase-field simulation. Therefore, it is possible to extract dissolvable grains individually for separate simulations, and undissolvable grains are not included in the phase-field simulation as they remain unchanged.

\begin{equation}
E_d= e_{\text{diss}} \times \left(\sum_j \eta_j^2(3-2\eta_j)\right)
\label{Ed formulation acid}
\end{equation}

Here, $e_{\text{diss}}$ represents the energy of dissolution, and $\eta_j$ is a phase variable that models a dissolvable grain.

The amount of material dissolved during each phase-field simulation step depends on the grain size, the energy of dissolution $e_{\text{diss}}$, and the duration of the simulation. It is important to ensure proper discretization between DEM steps. In this regard, the DEM step models mechanical equilibrium, while the PF step models dissolution as the source of perturbation. Mechanical equilibrium is assumed to be instantaneous, and the amount of material dissolved between DEM simulations must be sufficiently small. A simulation campaign presented in \ref{Look for RTS} was conducted to verify this criterion.

\begin{figure}[ht]
    \centering
    \includegraphics[width= 0.9\linewidth]{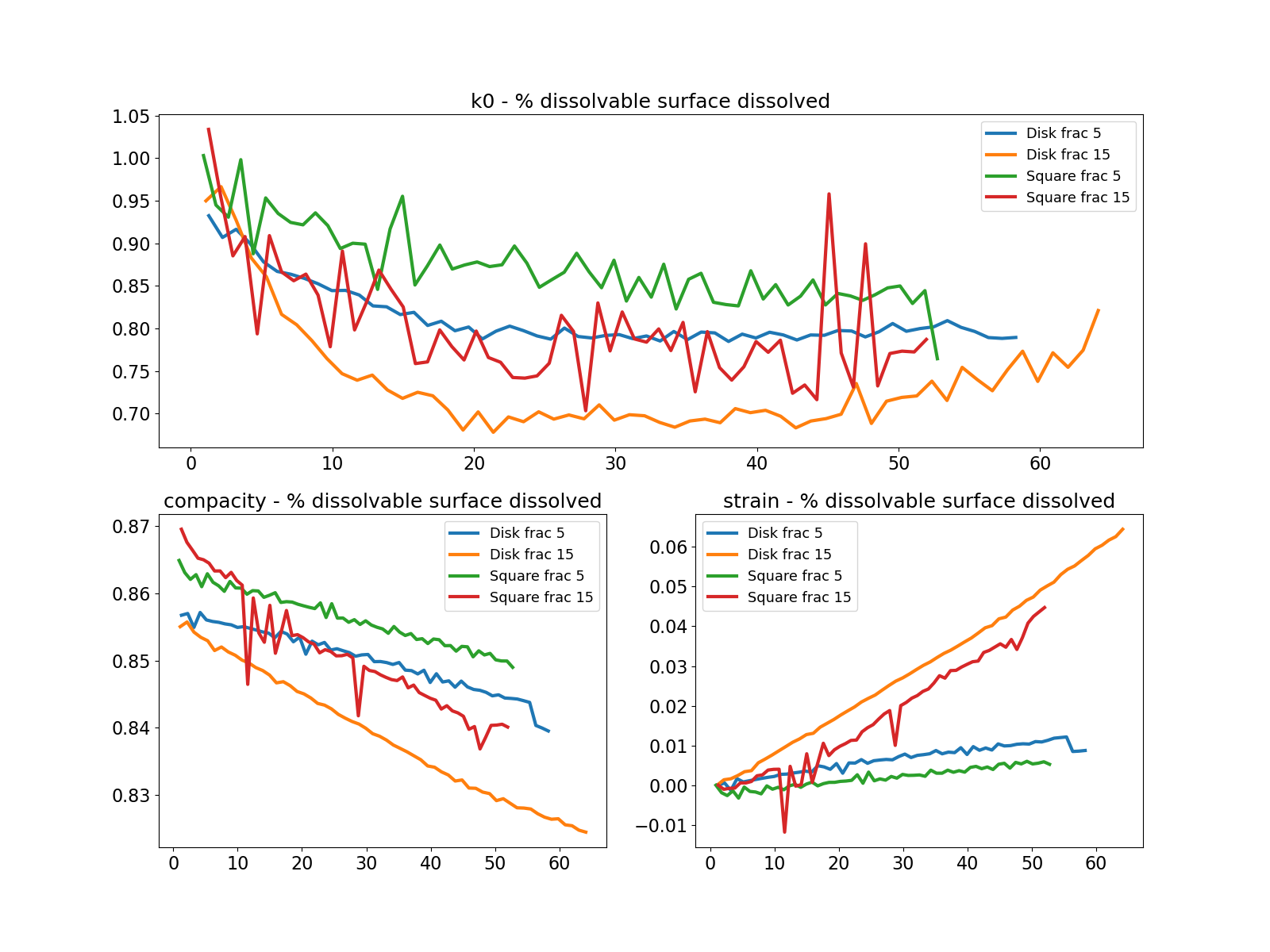}
    \caption{Mean evolution of $k0$ (top), of the compacity (left) and the vertical strain (right) with the dissolution of the dissolvable material, considering dissolvable grains as disks or as squares. Two ratios of the dissolvable surface over the total surface are studied.}
    \label{k0 Campaign Acid Oedometer}
\end{figure}

\begin{figure}[ht]
    \centering
    \includegraphics[width=0.9\linewidth]{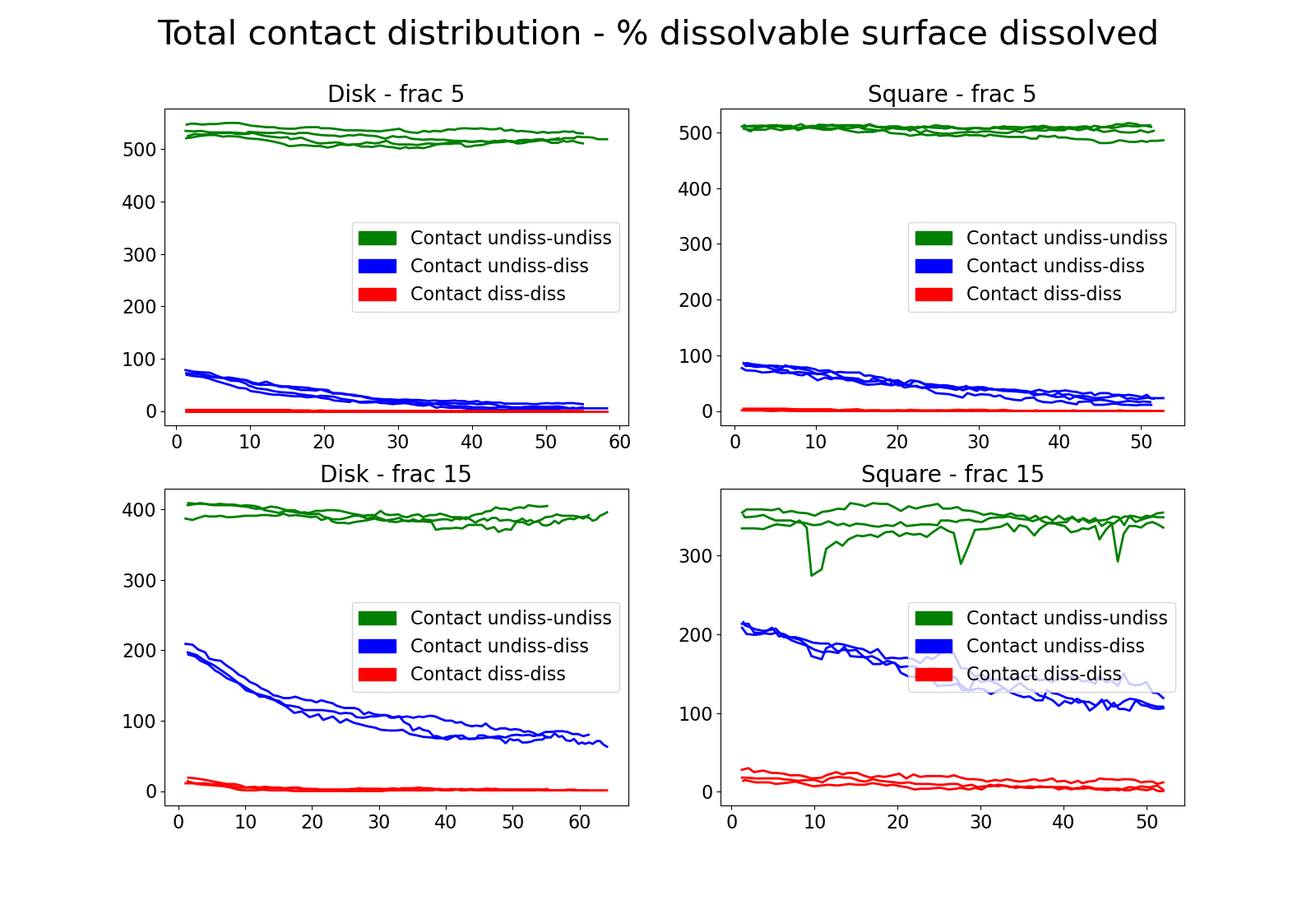}
    \caption{Evolutions of contact distribution with the dissolution of the dissolvable material, considering dissolvable grains as disks or as squares. Two ratios of the dissolvable surface over the total surface are studied.}
    \label{Contact Distribution Campaign Acid Oedometer}
\end{figure}

The results of the simulation campaign are shown in Figures \ref{k0 Campaign Acid Oedometer} and \ref{Contact Distribution Campaign Acid Oedometer}. Additionally, four example movies are linked to this article, demonstrating the evolution of grain configuration and chain forces for both disk-shaped and square-shaped dissolvable grains with a dissolvable surface ratio of 0.15 relative to the total surface.

Figure \ref{k0 Campaign Acid Oedometer} illustrates the mean evolution of $k_0$ (top), compacity (left), and vertical strain (right) with the dissolution of dissolvable material, considering dissolvable grains as disks or squares. Two different ratios of dissolvable surface to total surface are studied. As highlighted by Shin et Santamarina and Cha et Santamarina \cite{Santamarina2009, Santamarina2014} and observed in Figure \ref{k0 Campaign Acid Oedometer}, the dissolution of grains leads to an evolution (increase or decrease) of the coefficient $k_0 = \frac{\sigma_{II}}{\sigma_I}$. In this paper, a stronger reduction in $k_0$ is observed when the percentage of dissolvable grains is larger. This reduction occurs due to the reorganization of grains around the dissolved grains, as shown in Figure \ref{Contact Distribution Campaign Acid Oedometer}. The number of undissolvable-dissolvable contacts decreases with material dissolution, while the number of undissolvable-undissolvable contacts remains relatively constant.

It should be noted that the simulations by Shin et Santamarina and Cha et Santamarina constrained particle rotation to account for grain angularity and interlocking, while in our simulations, particle rotation remains free. Figure 3 from Cha et Santamarina \cite{Santamarina2014} highlights that hindering the grains rotation tends to decrease the coefficient $k_0$. Hindering particle rotation is often used as a proxy for grains angularity \cite{SacMorane2023}, so their simulations results tend to show that more angular grains would have a larger decrease of $k_0$, which is the opposite that what we observe on \ref{k0 Campaign Acid Oedometer}. This highlights the importance of considering explicitly the grain shape to obtain an accurate response of the system. A parametric study would be relevant to better understand the influence of confinement, particle size distribution, rolling model, and initial compressibility to better understand in which conditions a rolling model can be used as a proxy for grains angularity when chemo-mechanical couplings are involved.

Furthermore, simulations with square dissolvable grains exhibit more noise, even though the grain discretization was increased to 80 vertices for these simulations. This increased noise makes it harder to reach a equilibrium due to the applied moments on square particles, arising from the arm moment and the force at the contacts. With disk particles, the forces at the contacts are more often directed towards the center of the grains, resulting in smaller applied moments. Figure \ref{Contact Distribution Campaign Acid Oedometer} also reveals that the number of undissolvable-undissolvable contacts is smaller for square grains compared to disk grains. While this type of contact does not introduce noise (as it involves two disk grains), the contact undissolvable-dissolvable becomes more significant as it includes a square-shaped grain. The difference in contact numbers arises from the fact that there are more dissolvable grains in the case of squares than disks. Consequently, the percentage of dissolvable grains is defined as the ratio of dissolvable surface area ($S_{\text{dissolvable}}$) to the sum of dissolvable and undissolvable surface areas ($S_{\text{dissolvable}} + S_{\text{undissolvable}}$), and the square-shaped grain has a smaller surface area than the disk-shaped grain. For a percentage of 15\%, a simulation with 300 grains contains 58 grains (in the case of disks) or 83 grains (in the case of squares).

Figure \ref{Contact Distribution Campaign Acid Oedometer} demonstrates that the number of undissolvable-dissolvable contacts evolves differently depending on the shape of the dissolvable grains, especially when the dissolvable surface ratio is 0.15. The number of contacts for disks decreases in two steps: a sharp slope (around 5 contacts lost per percentage) from 0\% to 15\% dissolvable surface dissolved, followed by a slight slope (around 1 contact lost per percentage) from 15\% to 60\% dissolvable surface dissolved. On the other hand, the number of contacts for squares decreases in a single step, with a slight slope (around 1-2 contacts lost per percentage). It is worth mentioning that the initial configuration algorithm assumes perfect disks. Once the first equilibrium is reached, the shapes of the particles are generated, and the circumscribed disks are replaced by disks or squares. In the case of squares, this operation suddenly reduces the number of undissolvable-dissolvable contacts. When dissolution begins in the case of disks, many contacts with small overlaps are deleted in the initial iterations. In contrast, these contacts with small overlaps are already deleted during the initial configuration generation for squares.

Figure \ref{k0 Campaign Acid Oedometer} illustrates that the evolution of compacity $\left(= \frac{S_{\text{grains}}}{S_{\text{box}}}\right)$ and vertical strain seems to depend solely on the percentage of dissolvable grains. The slope is sharper when the percentage of dissolvable grains is larger. For a given percentage, regardless of the shape (disk or square), the same amount of material is dissolved at each iteration.

\vskip\baselineskip

In conclusion, this section emphasizes the ability of PFDEM to handle irregular grains. The shape of the grains has a significant influence on granular behavior, and considering the true shape instead of perfect disks/spheres is important.

%%=======================================================%%

\section{Application to Irregular Shapes, heterogeneous dissolution and precipitation}
\label{Section Heterogeneous dissolution/precipitation}

In the context of PFDEM simulations, the heterogeneous dissolution/precipitation process can be modeled. This section aims to demonstrate the capability of PFDEM to simulate such processes. In particular, we consider the example of the pressure solution phenomenon that involves both processes at the microscale and compare the simulation results with experimental findings reported by De Meer et al. \cite{DeMeer1997}.

The pressure solution phenomenon arises from the premise that in densely packed granular materials intergranular forces are substantial and this mechanism is dependent on temperature and mineral composition. This process involves the dissolution of material in high-stress areas and precipitation in low-stress areas, facilitated by diffusive mass transfer through the pore fluid \cite{Spiers1990, Gratier1993, Urai2008}.

\begin{figure}[ht]
    \centering
    \includegraphics[width=0.7\linewidth]{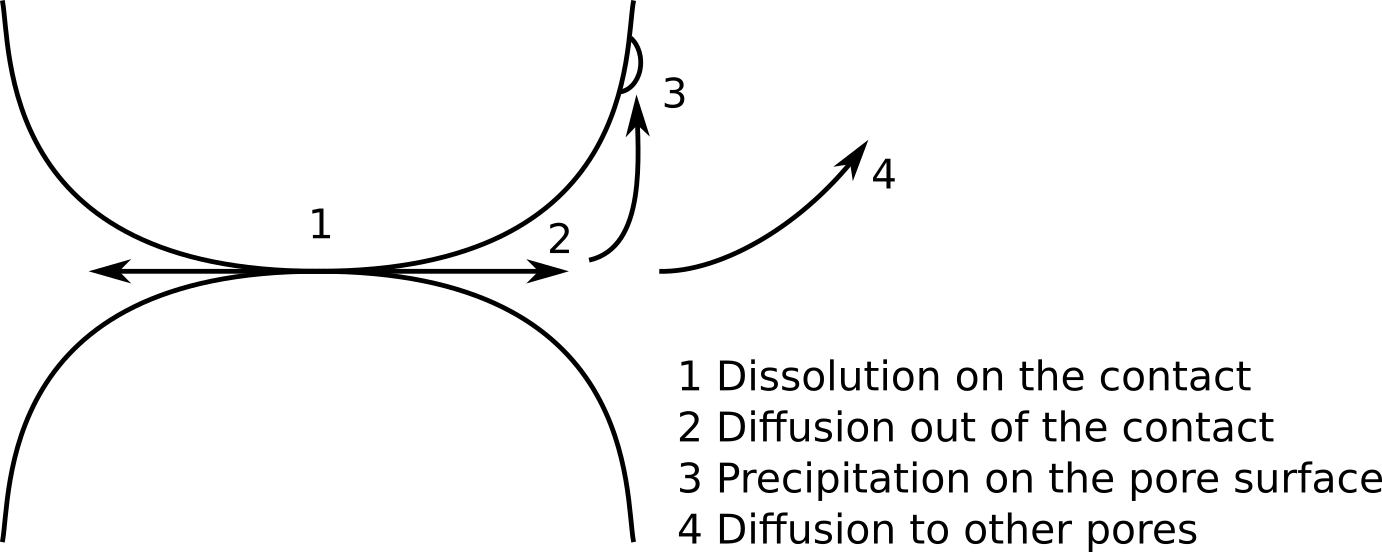}
    \caption{Scheme of pressure solution phenomena.}
    \label{Scheme Pressure Solution}
\end{figure}

Figure \ref{Scheme Pressure Solution} provides a schematic representation of the pressure solution process at the boundaries between two grains. In this scheme, the compacted granular medium is subjected to stress, leading to the formation of high-stress and low-stress regions. Within the high-stress regions at the contact, dissolution takes place due to a modification of the chemical potential, resulting in the removal of material. This dissolved material is then transported through the pore fluid to the low-stress regions far from the contact, where precipitation occurs, leading to the formation of new material.

Here, the Equation \ref{Solute formulation} is solved to track the solute generation, and its diffusion, and to ensure the total mass is conserved.

In the phase-field formulation, Equation \ref{AC formulation}, the external source term $E_d$ (presented in Equation \ref{AC formulation}) models both the mechanical energy at the contact level ($E_{mec}$) and the chemical energy due to the solute ($E_{che}$). As explained in Figure \ref{Destabilized Double Well}, the mechanical energy induces dissolution, whereas the chemical energy induces precipitation. The formulation of this source term is given by Equation \ref{Ed formulation pressure solution}.

\begin{equation}
E_d = E_{mec}-E_{che} = \sum\limits_j\alpha\, \text{min}(\eta_i)\,e_{mec}\times\eta_j^2(3-2\eta_j) - \chi c\times\eta_j^2(3-2\eta_j)
\label{Ed formulation pressure solution}
\end{equation}
In Equation \ref{Ed formulation pressure solution}, $E_d$ is defined as the difference between the mechanical energy ($E_{mec}$) and the chemical energy ($E_{che}$). The mechanical energy term is computed based on the contact level from the DEM simulation and is applied to the contact area. The solute concentration $c$ is multiplied by a coefficient $\chi$ representing the chemical energy term. A function of interpolation $\eta^2_j(3-2\eta_j)$ is used to apply the source term $E_d$ only at the interface of the phase variables ($\eta_j \neq 1 \text{ or } 0$). 

The source term $E_{mec}$ is non-zero only if both phase variables $\eta_i$ involved in the contact are non-zero, approximately representing the contact area. This assumption generates a heterogeneous dissolution focused on the contact area. To ensure stability, it is recommended to work with a source term $E_{mec}$ lower than 0.2. The coefficient $\alpha$ is used to normalize the mechanical energy term. The term $e_{mec}$, defined in Equation \ref{Potential Mechanical Energy}, represents the potential mechanical energy at the contact level and is computed as a potential normalized according to the relation \ref{From Material to Stiffness Equations}.

\begin{equation}
    e_{mec} = \frac{\int Fs_n d\delta_n}{\sum\limits_{\Omega} \text{min}(\eta_i)}= \frac{\int k\delta_n^{3/2} d\delta_n}{\sum\limits_{\Omega} \text{min}(\eta_i)} = \frac{\frac{2}{5}k\delta_n^{5/2}}{\sum\limits_{\Omega} \text{min}(\eta_i)}
    \label{Potential Mechanical Energy}
\end{equation}

The quantity $\sum\limits_{\Omega} \text{min}(\eta_i)$ represents the sum over the contact area of the minimal value between phase variables involved $\eta_i$ at each node. This variable is equivalent to the contact surface and ensures that the amount of energy introduced in the system from the interaction between grains remains constant. For example, if only two grains are in contact, the introduced energy term $\sum\limits_{\Omega} E_{mec}$ is proportional to $\frac{2}{5}k\delta_n^{5/2}$. Thus, the term remains constant with time as $\sum\limits_{\Omega} E_{mec}=\sum\limits_{\Omega}\alpha\,\text{min}(\eta_i)\,e_{mec} = \sum\limits_{\Omega}\alpha\,\text{min}(\eta_i)\,\frac{2/5\,k\,\delta_n^{5/2}}{\sum\limits_{\Omega} \text{min}(\eta_i)} = \alpha\,\frac{2}{5}k\,\delta_n^{5/2}\frac{\sum\limits_{\Omega} \text{min}(\eta_i)}{\sum\limits_{\Omega} \text{min}(\eta_i)} = \alpha\,\frac{2}{5}k\delta_n^{5/2}$.

In the phase-field simulation, the time step (relative to shape evolutions) is adaptive to avoid numerical issues with the coupled problem. The deformation induced by the phase-field simulation and the grain displacements computed in the discrete element modelization are closely interconnected. It is important to ensure that the deformation remains in the small-strain range. For fluid-structure interaction problems, an Aitken method \cite{Aitken1938} is commonly used to adapt the deformation of the structure. In this problem, the mean absolute value per contact node of the total energy $E_d$ is estimated. By initializing the sum of the absolute total energy as $E_d$ and the number of contact nodes as $node\_contact$, the mesh is scanned. If a contact zone is detected (where $\eta_i$ and $\eta_j$ are both larger than 0.5), the absolute value of the total energy $E_d$ is added to the indicator $E_{d,abs}$, and the number of contact nodes is incremented. The mean absolute value per contact node of the total energy is then computed by dividing $E_{d,abs}$ by $node\_contact$. This parameter is related to the deformation known in the phase-field simulation. Based on the value of this parameter, the duration of the simulation, and consequently the grain deformation during the phase-field simulation, can be adjusted.

\vskip\baselineskip

A first simulation campaign was conducted using only two grains with the same radius, and the parameters used for this pressure solution simulation are listed in Table \ref{Parameters Pressure Solution}. Since the configuration is relatively simple, the discrete element method (DEM) step can be simplified by using an analytical solution. In this case, the steady overlap between the two grains can be computed from the applied external load, and only an overlap is applied to move the grains.

\begin{table}[ht]
    \centering
    \begin{tabular}{|l|l|l|}
        \hline
        Parameters&Unit&Value\\
        \hline
        Mobility for phase-field $\eta_j$&-&1\\
        Gradient coefficient for phase-field $\kappa_j$&-&0.01\\
        Gradient coefficient for solute $\kappa_c$&-&50\\
        Initial time step for phase-field simulation&$s$&0.2\\
        Duration of the phase-field simulation&$s$&6 $\times$ the time step\\
        \hline
        Overlap applied&$\mu m$&10\% of the initial radius\\
        Chemical energy coefficient $\chi$&-&0.5\\
        Mechanical energy coefficient $\alpha$&-&Initial $\sum\limits_{\Omega} \text{min}(\eta_i)$\\
        \hline
    \end{tabular}
    \caption{Parameters used for the two grains pressure solution simulation.}
    \label{Parameters Pressure Solution}
\end{table} 

\begin{figure}[ht]
\centering
\includegraphics[width=0.48\linewidth]{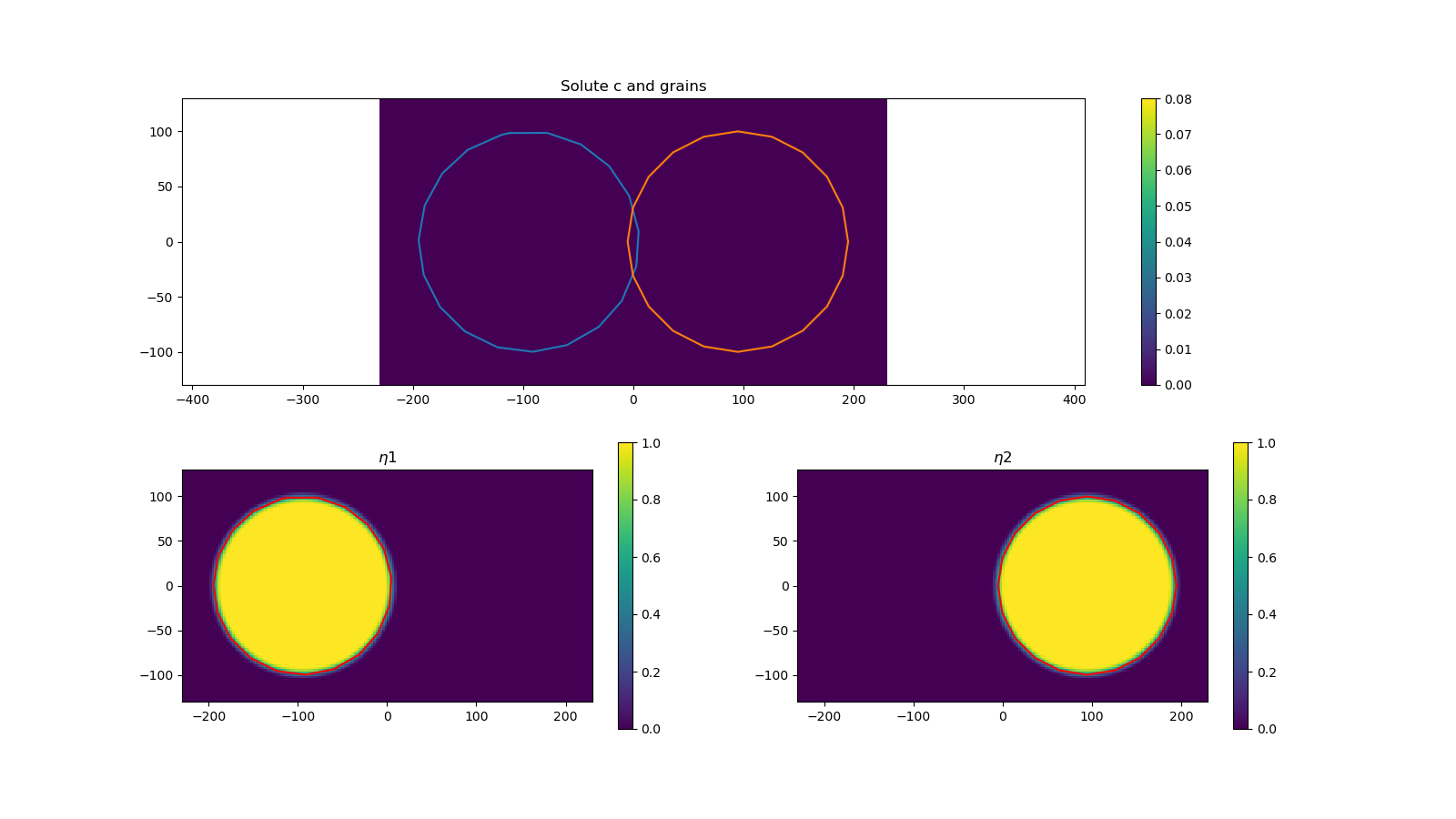}~
\includegraphics[width=0.48\linewidth]{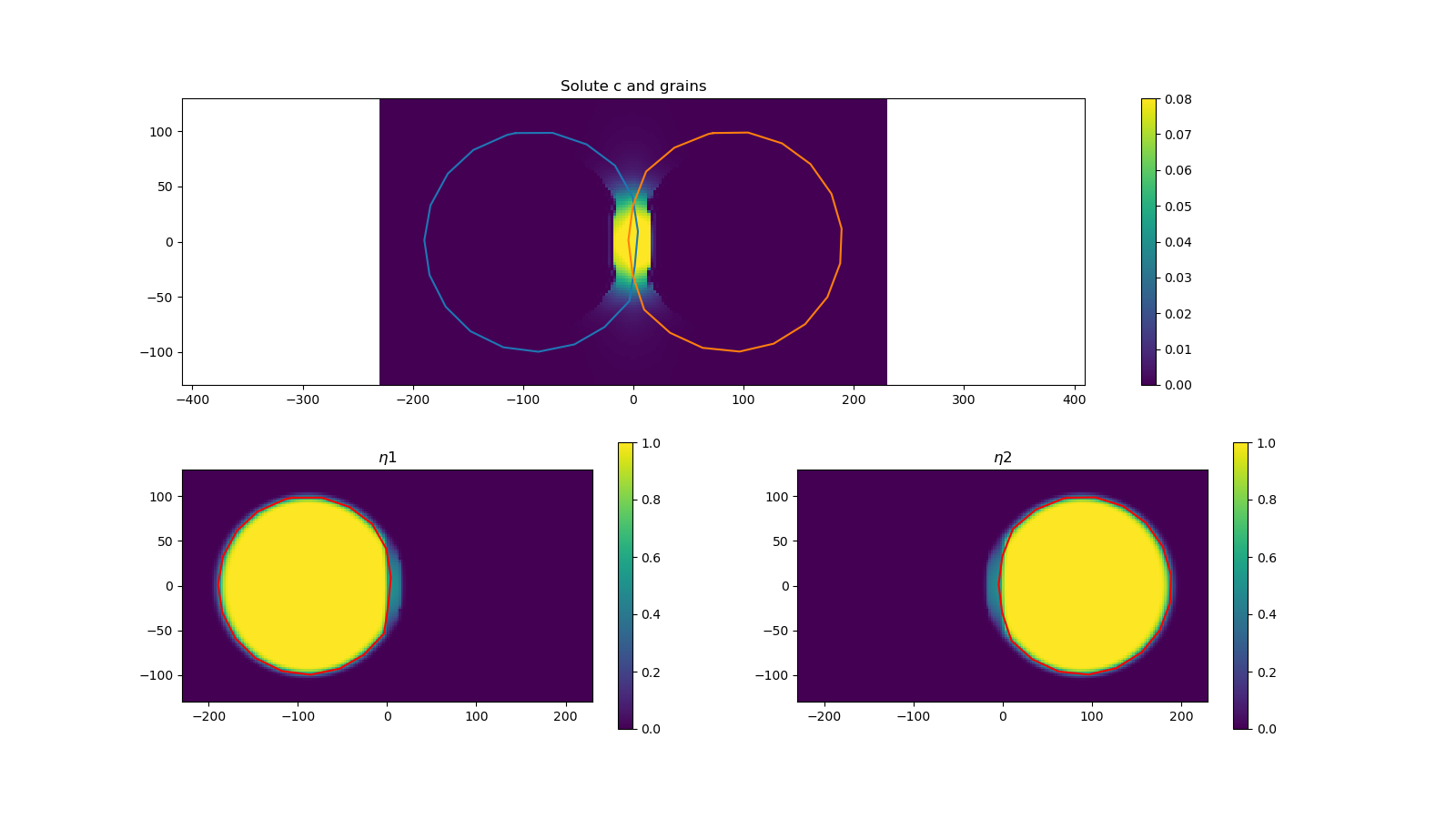}\\
t0 \hspace{0.45\linewidth} t1

\includegraphics[width=0.48\linewidth]{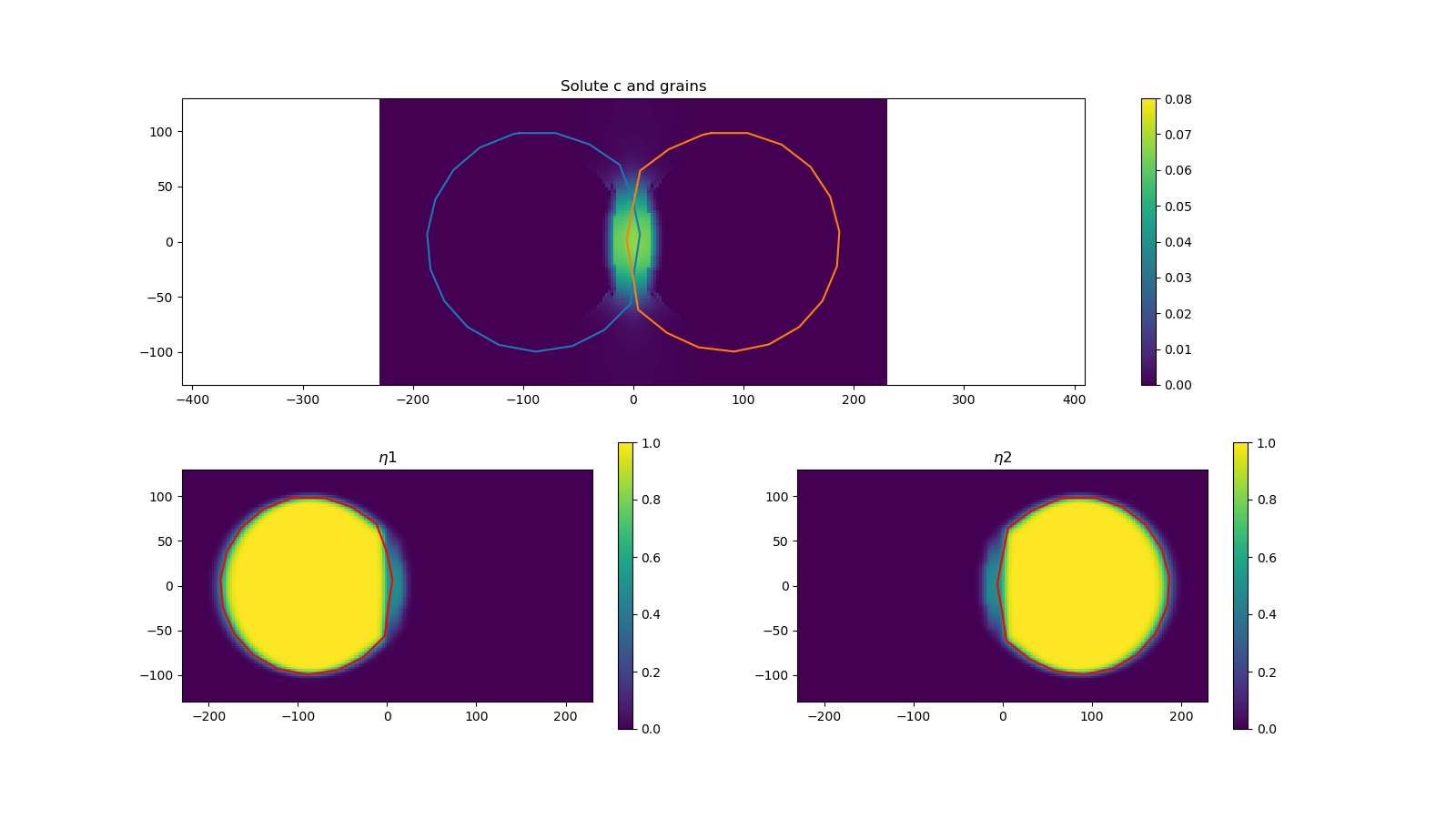}~
\includegraphics[width=0.48\linewidth]{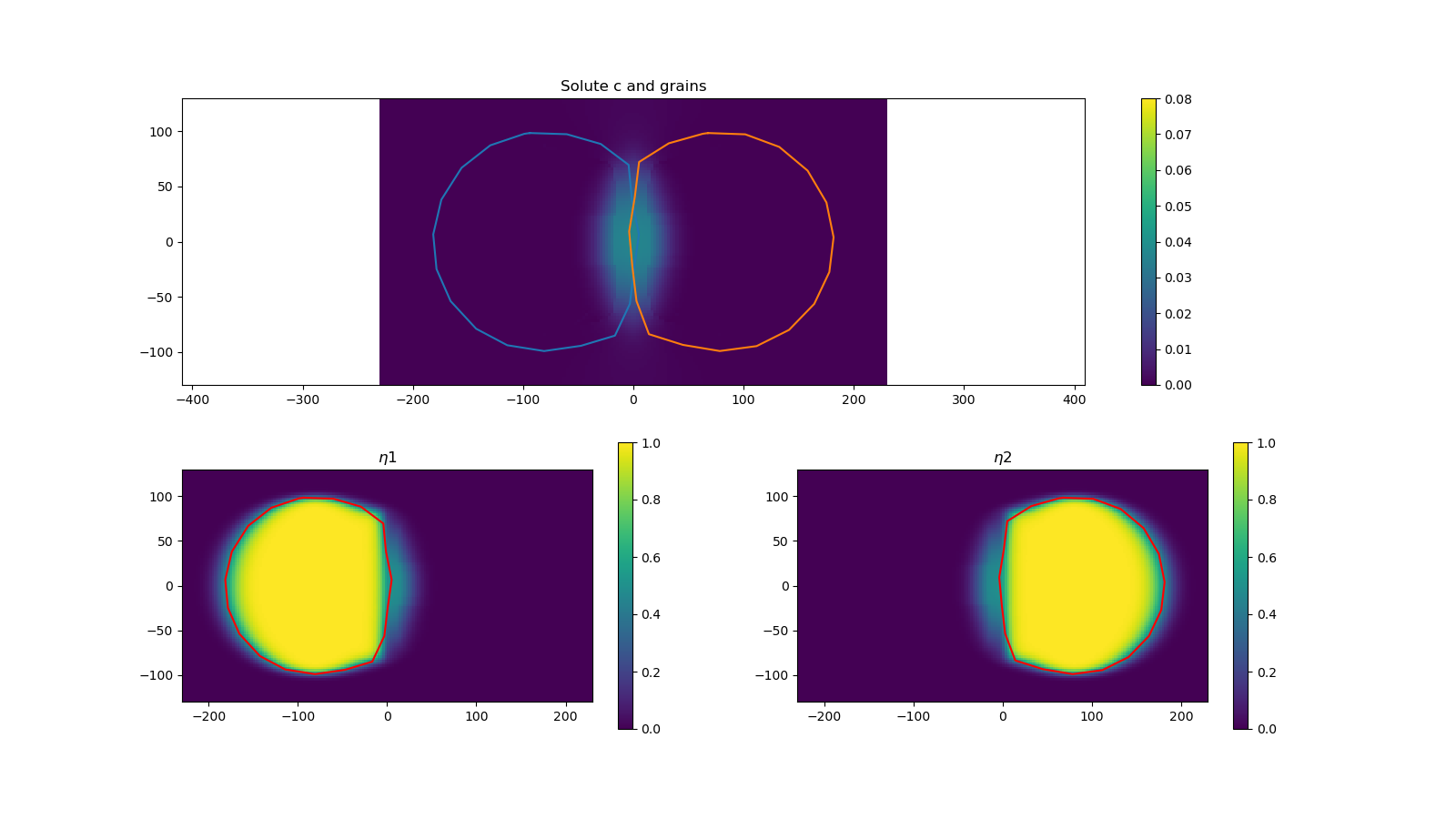}\\
t2 \hspace{0.45\linewidth} t3

\caption{Pressure solution phenomena between two grains. For each time, the upper plot represents the position of the grains and the solute concentration whereas the lower plots represent the phase-field variable of each grain.}
\label{Configuration PS 2G}
\end{figure}

The configuration of the two-grain system at different times is depicted in Figure \ref{Configuration PS 2G}, and a movie is linked to the article for a more comprehensive visualization. The initial configuration consists of two grains in contact with no solute in the sample. As the simulation progresses, dissolution occurs at the contact, resulting in a decrease in the phase-field variables and the generation of solute. The dissolution process continues, and the solute diffuses throughout the sample. The amount of solute generated decreases as the chemical energy at the contact balances the mechanical energy. Eventually, a steady-state is reached, where grains continue to dissolve at the contact, while an equal amount of solute precipitates far from the contact.

\begin{figure}[ht]
    \centering
    \includegraphics[width = 0.48\linewidth]{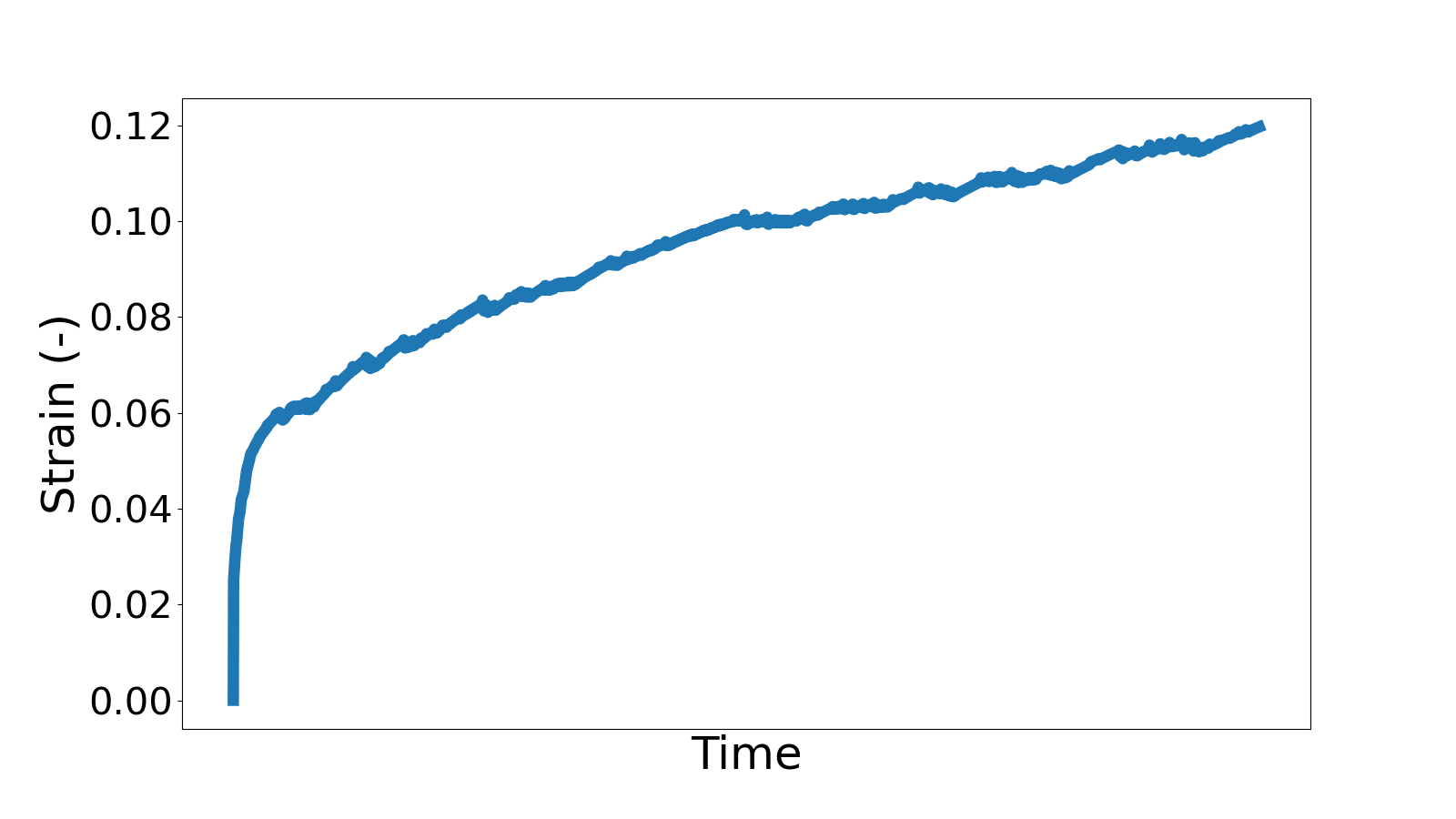}~
    \includegraphics[width = 0.48\linewidth]{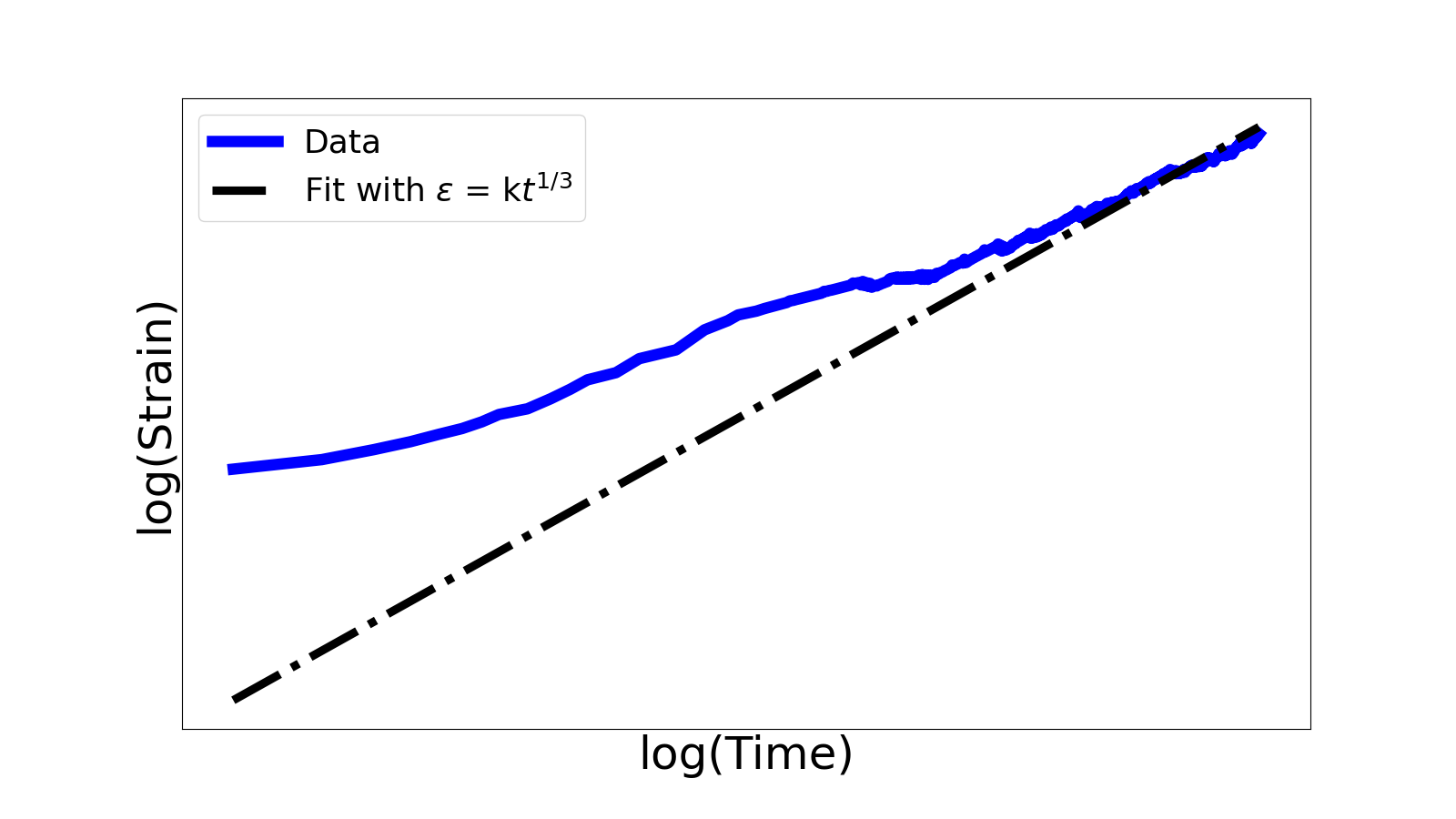}
    \caption{Evolution of the vertical strain with times (left) and fit with the Andrade creep law (right).}
    \label{Result 2G Pressure Solution Strain}
\end{figure}

\begin{figure}[ht]
    \centering
    \includegraphics[width = 0.7\linewidth]{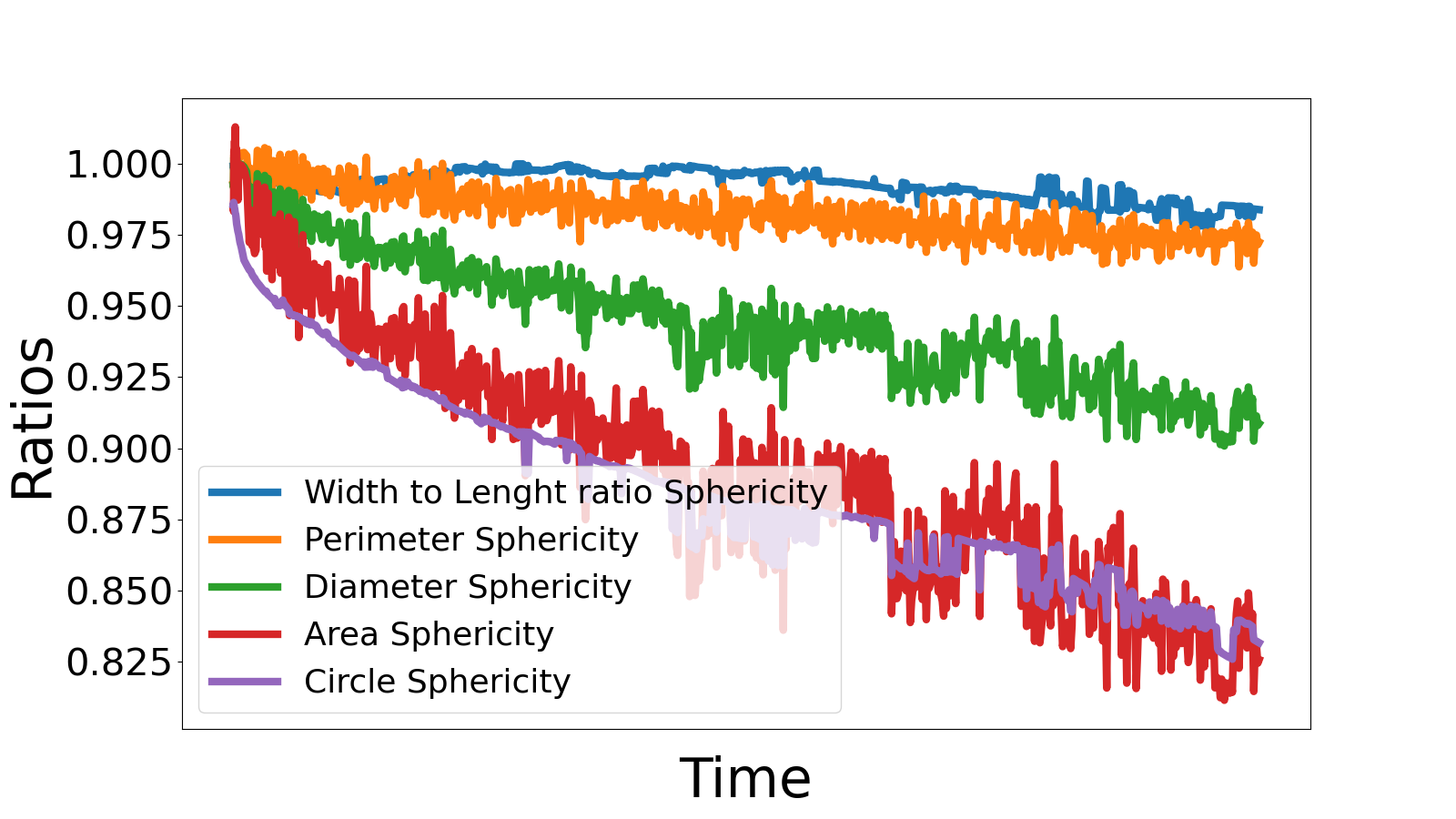}
    \caption{Evolution of different indices to appreciate the sphericity of the grain 1.}
    \label{Result 2G Pressure Solution Sphericity}
\end{figure}

\begin{figure}[ht]
    \centering
    \includegraphics[width = 0.7\linewidth]{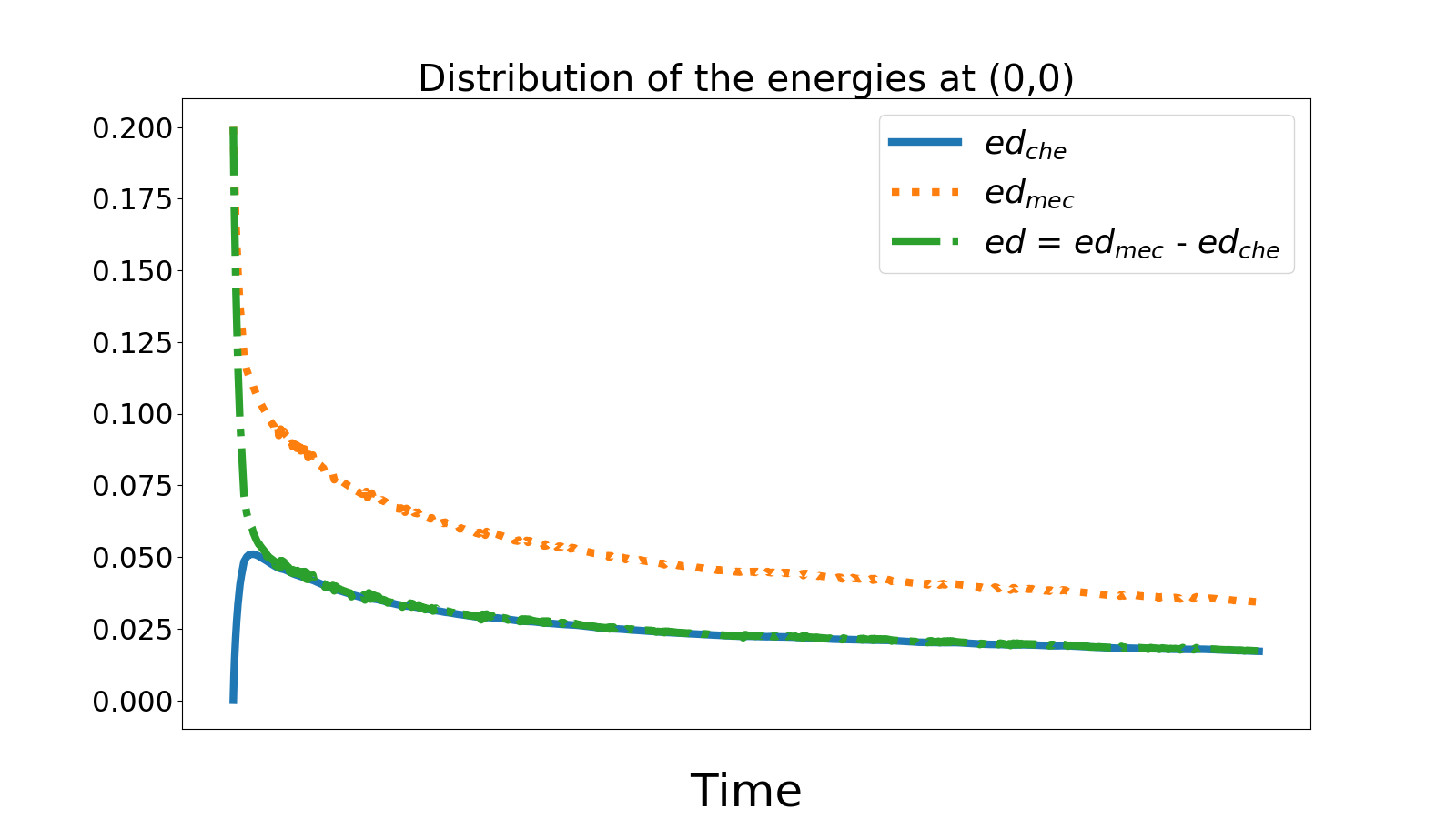}
    \caption{Evolution of the chemical, mechanical and total energy at the center point of the sample (always in the contact area).}
    \label{Result 2G Pressure Solution Energy}
\end{figure}

\begin{figure}[ht]
    \centering
    a)
    \includegraphics[width = 0.45\linewidth]{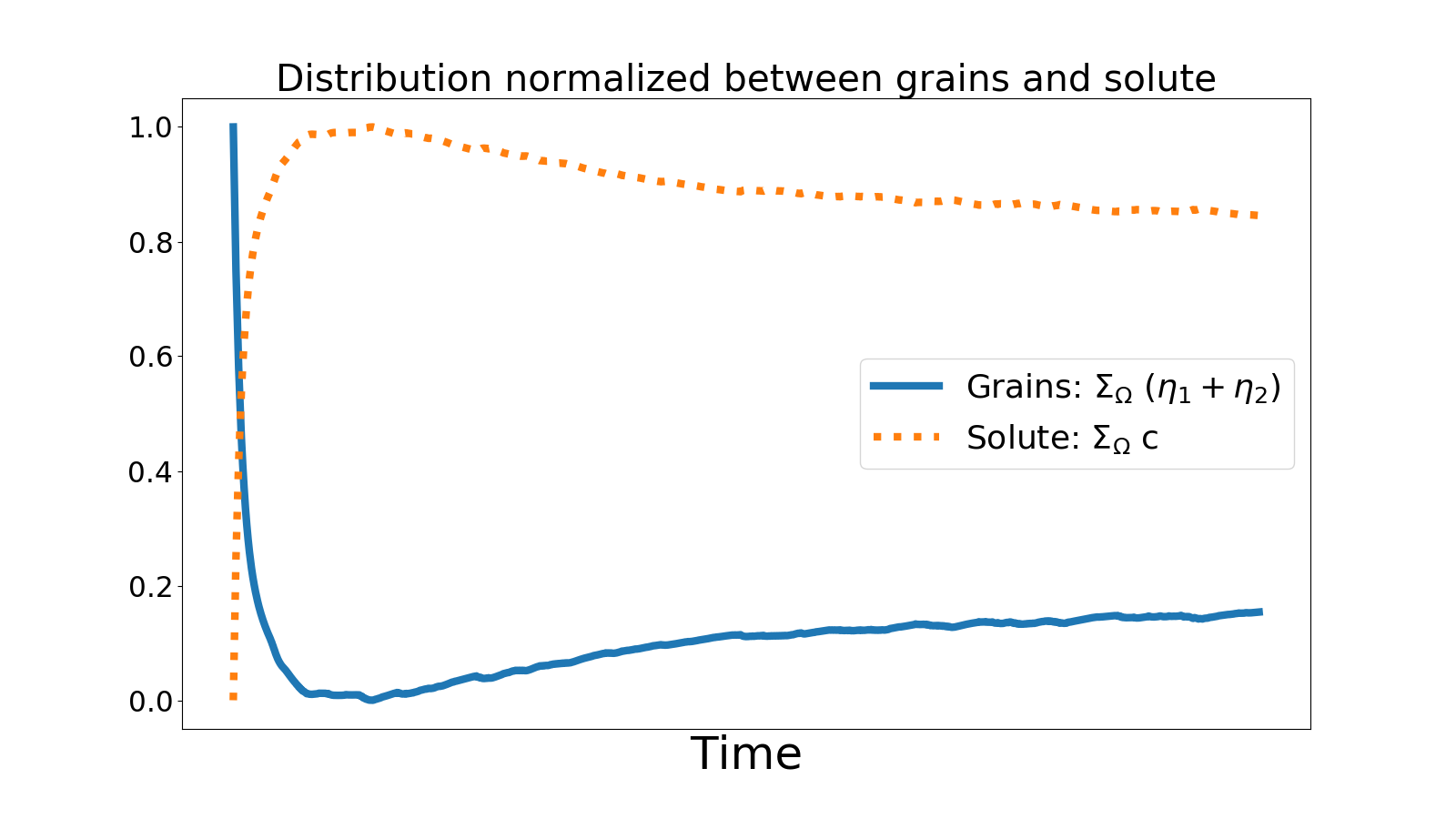}~
    b)
    \includegraphics[width = 0.45\linewidth]{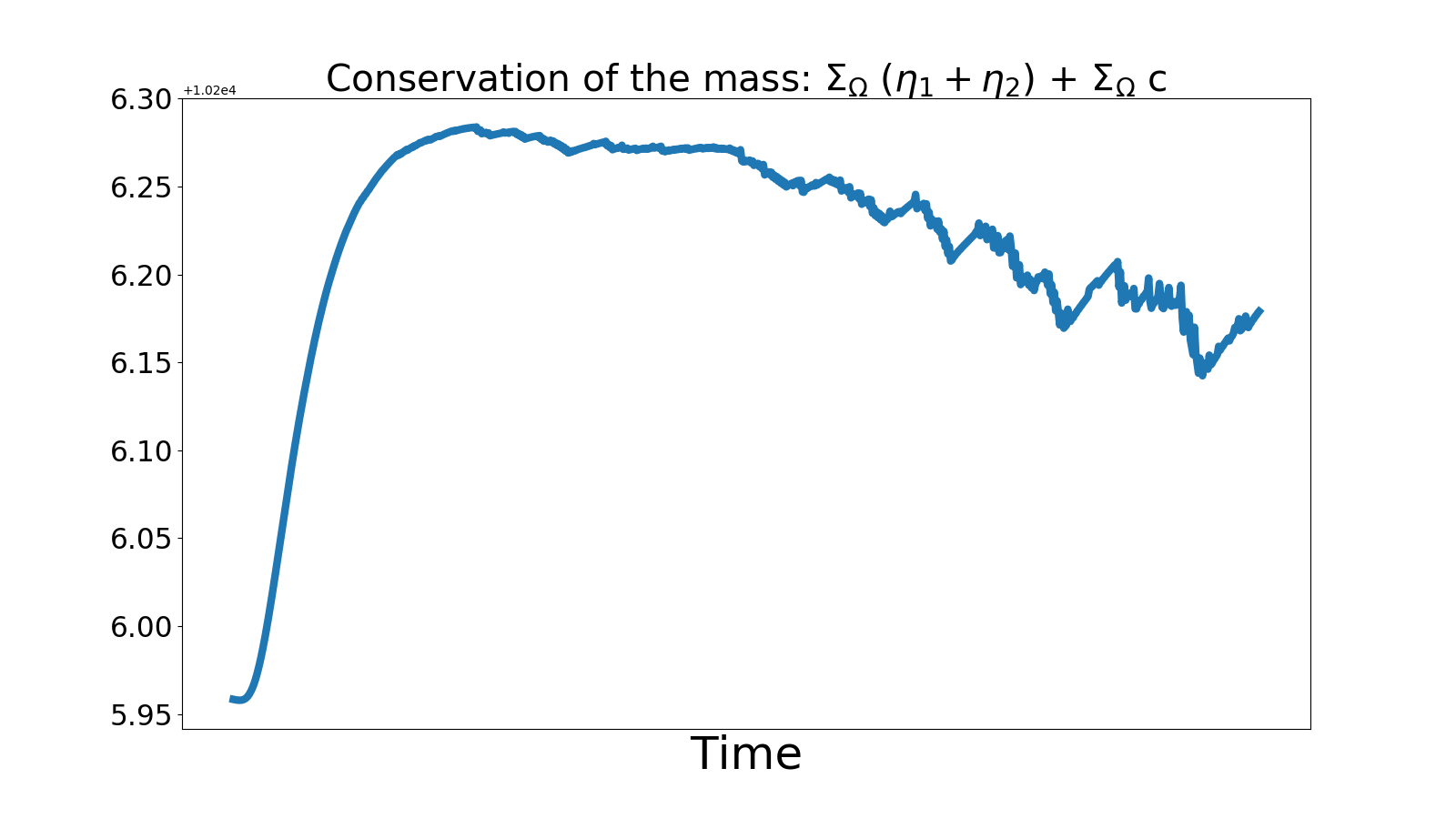}
    \caption{Evolution of the grains and solute quantities (first steps have been removed to facilitate the curve reading): a) A decrease in the grain quantity leads to an increase in the solute one. b) The conservation of the mass is verified.}
    \label{Result 2G Pressure Solution Quantity}
\end{figure}

Various curves presented in Figures \ref{Result 2G Pressure Solution Strain}, \ref{Result 2G Pressure Solution Sphericity}, \ref{Result 2G Pressure Solution Energy}, and \ref{Result 2G Pressure Solution Quantity} provide further insights into the simulation. Figure \ref{Result 2G Pressure Solution Strain} shows the evolution of vertical strain with time, highlighting a creep phenomenon. This behavior is consistent with previous research findings on the topic \cite{DeMeer1997, Guevel2020}. The slowdown in the strain evolution can be attributed to two scenarios: (i) the expanding contact area reduces the introduced mechanical energy, and (ii) the pore water becomes saturated with the solute concentration. This creep phenomenon is known as the Andrade creep response, and it has been observed in various geomaterials \cite{Dysthe2002}. The vertical strain evolves proportionally with the cubic root of time ($\epsilon = k\, t^{1/3}$), and the slope of 1/3 is verified for long-term times.

Figure \ref{Result 2G Pressure Solution Sphericity} illustrates the shape evolution of the grains using different sphericity indices. The shape of a 2D grain is characterized by five sphericities, as defined in Equation \ref{Sphericity Definition} \cite{Zheng2015}. The sphericities decrease over time, indicating that the shape of grain 1 deviates from a perfect circle as pressure solution occurs. It is worth noting the evolution of the width-to-length ratio sphericity, which initially decreases, then increases, and finally decreases again. This suggests that the particle shape tends to transition from a circle to a square and then to a rectangle.

\begin{align}
    \text{Area sphericity  }&\frac{A_s}{A_{cir}} \nonumber\\
    \text{Diameter sphericity  }&\frac{D_c}{D_{cir}} \nonumber\\
    \text{Circle ration sphericity  }&\frac{D_{ins}}{D_{cir}}& \nonumber\\
    \text{Perimeter sphericity  }& \frac{P_c}{P_{s}}\nonumber\\
    \text{Width to length ratio sphericity  }&\frac{d_2}{d_{1}}     
    \label{Sphericity Definition}
\end{align}

Here, $A_s$ is the surface of the grain, $A_{cir}$ is the surface of the minimum circumscribing circle, $D_c$ is the diameter of the perfect circle having the same area as the grain $\left(=2\sqrt{A_s/\pi}\right)$, $D_{cir}$ is the diameter of the minimum circumscribing circle, $D_{ins}$ is the diameter of the largest inscribing circle, $P_c$ is the perimeter of the perfect circle having the same area as the grain $\left(=2\sqrt{A_s\pi}\right)$, $P_s$ is the perimeter of the grain, and $d_1$ and $d_2$ are the length and the width of the grain.

The evolution of the chemical $ed_{che}$, mechanical $ed_{mec}$, and total energy $ed$ at the center point of the sample (0, 0) is depicted in Figure \ref{Result 2G Pressure Solution Energy}. This point is selected because it always remains in the contact area during the simulation. As explained in Equation \ref{Ed formulation pressure solution}, the chemical and mechanical energy terms are proportional to the solute concentration at the selected point and the contact surface, respectively. The simulation starts with significant dissolution of the grains, followed by solute diffusion and precipitation. These phenomena increase the size of the contact area, leading to a decrease in the mechanical energy term. By choosing appropriate parameters, the two limiting scenarios described by Lu et al. \cite{Lu2021} can be reproduced.
  
Figure \ref{Result 2G Pressure Solution Quantity} presents the material distribution between the grains and the solute. A normalized distribution is shown for a clearer plot. The normalization is presented in Equation \ref{Normalized Distribution}.

\begin{equation}
    v^n(t) = \frac{v(t)-v_{min}}{v_{max}-v_{min}}
    \label{Normalized Distribution}
\end{equation}
where $v^n(t)$ is the normalized value (grain or solute quantity in the entire domain) at the time $t$, $v(t)$ is the value at the time $t$, $v_{min}$ is the minimum reached of the value and $v_{max}$ is the maximum reached of the value.

The evolution of the grain quantity is computed by summing the phase-field variables over the sample, while the evolution of the solute quantity is computed by summing the solute concentration over the sample. Figure \ref{Result 2G Pressure Solution Quantity}b demonstrates the mass conservation, as the sum of the two variables remains constant (with a negligible difference due to mass loss during phase-field generations). In conjunction with Figure \ref{Result 2G Pressure Solution Energy}, it appears that a steady-state is reached. At the center point (in the contact area), material dissolution occurs (indicated by the positive sum of energy), but at the same time, the sum of the solute concentration over the sample remains constant. This indicates that the same quantity of solute generated at the contact is precipitated far from the contact, ensuring mass conservation.

\vskip\baselineskip

Once a standard configuration with only two grains has been studied, a multiple grains configuration is considered using the parameters presented in Table \ref{Parameters Pressure Solution MG}. In the multiple grains configuration, a new algorithm described in \ref{Move out solute grain} is applied to move the solute with the grains.

\begin{longtable}{|p{0.55\linewidth}|l|p{0.2\linewidth}|}
        \hline
        Parameters&Unit&Value\\
        \hline
        Number of grains&-&30\\
        Particle Size Distribution for grains (radius and percentage of the number of grains)&$\mu m\,( \%)$&420 (17), 385 (33), 315 (33), 280 (17)\\
        \hline
        Young modulus& GPa&70\\
        Poisson's ratio&-&0.3\\
        Density&$kg/m^3$&2500\\
        Friction grain-grain&-&0.5\\
        Friction grain-wall&-&0\\
        Restitution coefficient&-&0.2\\
        \hline
        Time step for discrete element model&sec&$\Delta t_{crit}/8$\\
        \hline
        Mobility for phase-field&-&1\\
        Gradient coefficient for phase-field $\kappa_j$&-&0.01\\
        Gradient coefficient for solute $\kappa_c$&-&50\\
        Initial time step for phase-field simulation&$s$&0.01\\
        Duration of the phase-field simulation&$s$&8 $\times$ the time step\\
        \hline
        Linear force applied on the upper wall&N/m&$4\times Y\,R_{mean}$/1000\\
        Chemical energy coefficient $\chi$&-&0.5\\
        Mechanical energy coefficient $\alpha$&-&Calibrated $\sum\limits_{\Omega} min(\eta_i)$\\
        \hline
    \caption{Parameters used for multiple grains pressure solution simulation.}
    \label{Parameters Pressure Solution MG}
\end{longtable}

To define the initial configuration, perfect disks are assumed. They are randomly generated without overlap between them in the sample and then loaded. Once a equilibrium is reached, the disks are discretized. The sample generation aims to achieve a ratio between the sample diameter ($D$) and the sample height ($H$) equal to 0.6.

The coefficient $\alpha$ is computed from a calibration simulation. Two grains with a radius equal to the mean radius of the particle size distribution are generated. Then, a vertical confinement force is applied to the two grains to obtain an overlap. Finally, the coefficient $\alpha$ is equal to $0.06\times \sum\limits_{\Omega} \min(\eta_i)$, where $\sum\limits_{\Omega} \min(\eta_i)$ is the sum over all the calibration sample of the minimal value of $\eta_i$ at each node.

\begin{figure}[ht]
    \centering
    \includegraphics[width=0.48\linewidth]{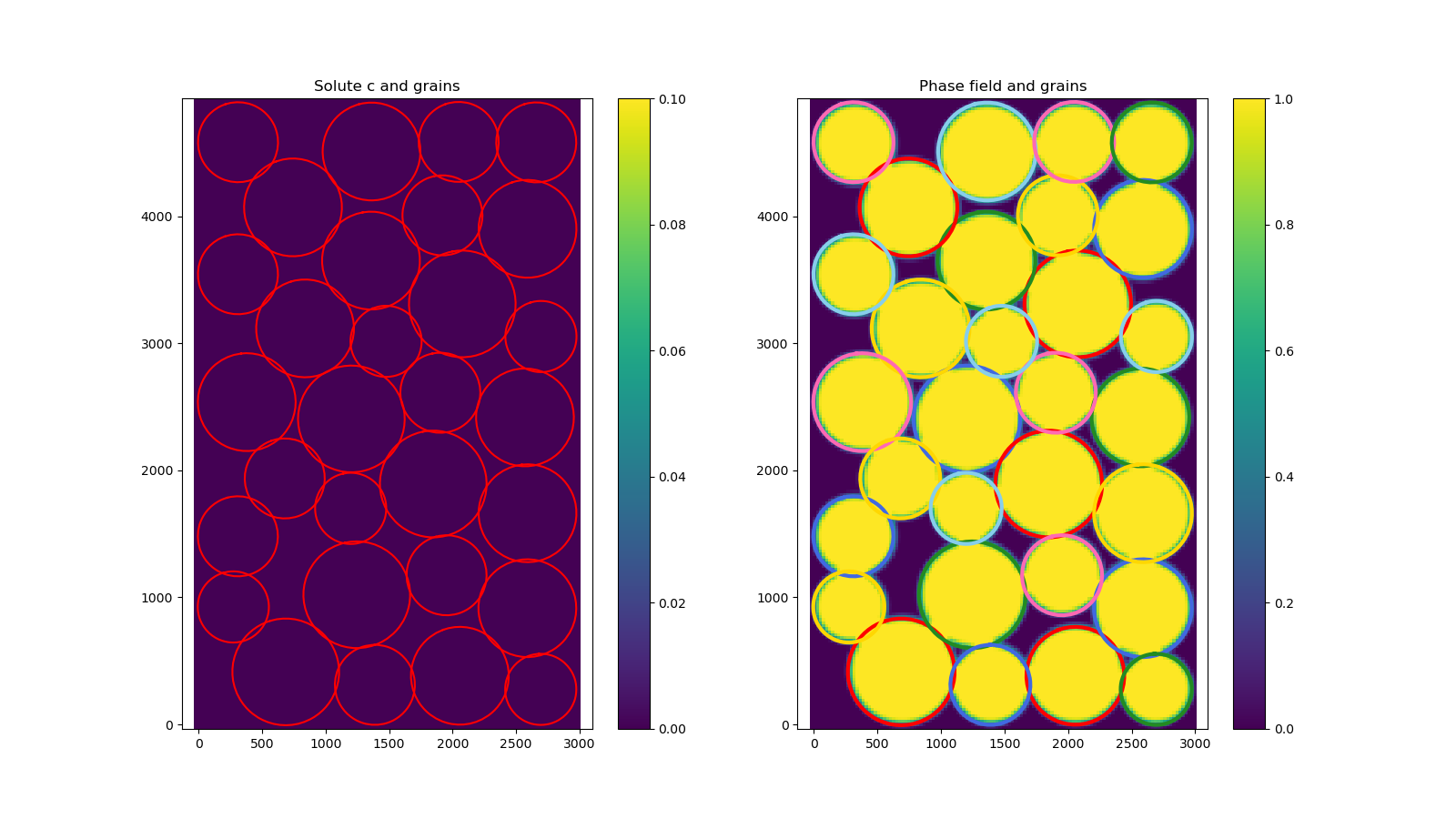}~
    \includegraphics[width=0.48\linewidth]{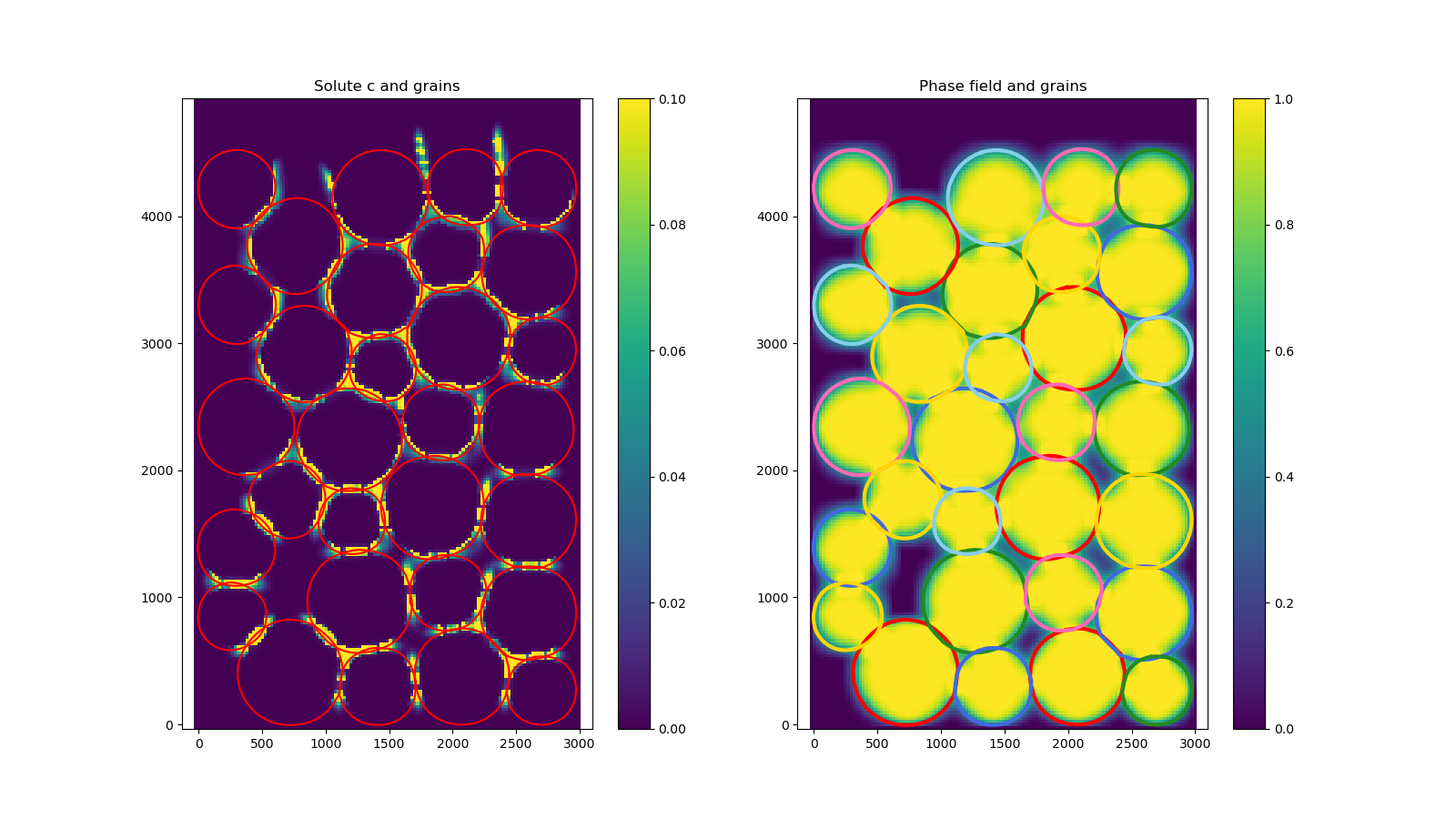}\\
    t0 \hspace{0.45\linewidth} t1
    \caption{Pressure solution phenomena between multiple grains. For each time, the left plot represents the position of the grains and the solute concentration whereas the right plots represent the sum of the phase-field variables of each grain.}
    \label{Configuration PS Multiple Grain}
\end{figure}

Figure \ref{Configuration PS Multiple Grain} shows some pictures of the configuration, and a movie is linked to the article for a more comprehensive visualization. Initially, there is no solute in the sample. After several iterations, some solute is generated and localized at the contact level. Grain reorganization occurs as shape deformation takes place. Strong concentrations of solute appear, and solute leakage can be observed at the top of the sample. In this configuration, the interaction with the top wall is not considered, and nothing forces the generated solute to follow the movement of the grains.

\begin{figure}[ht]
    \centering
    \includegraphics[width=0.5\linewidth]{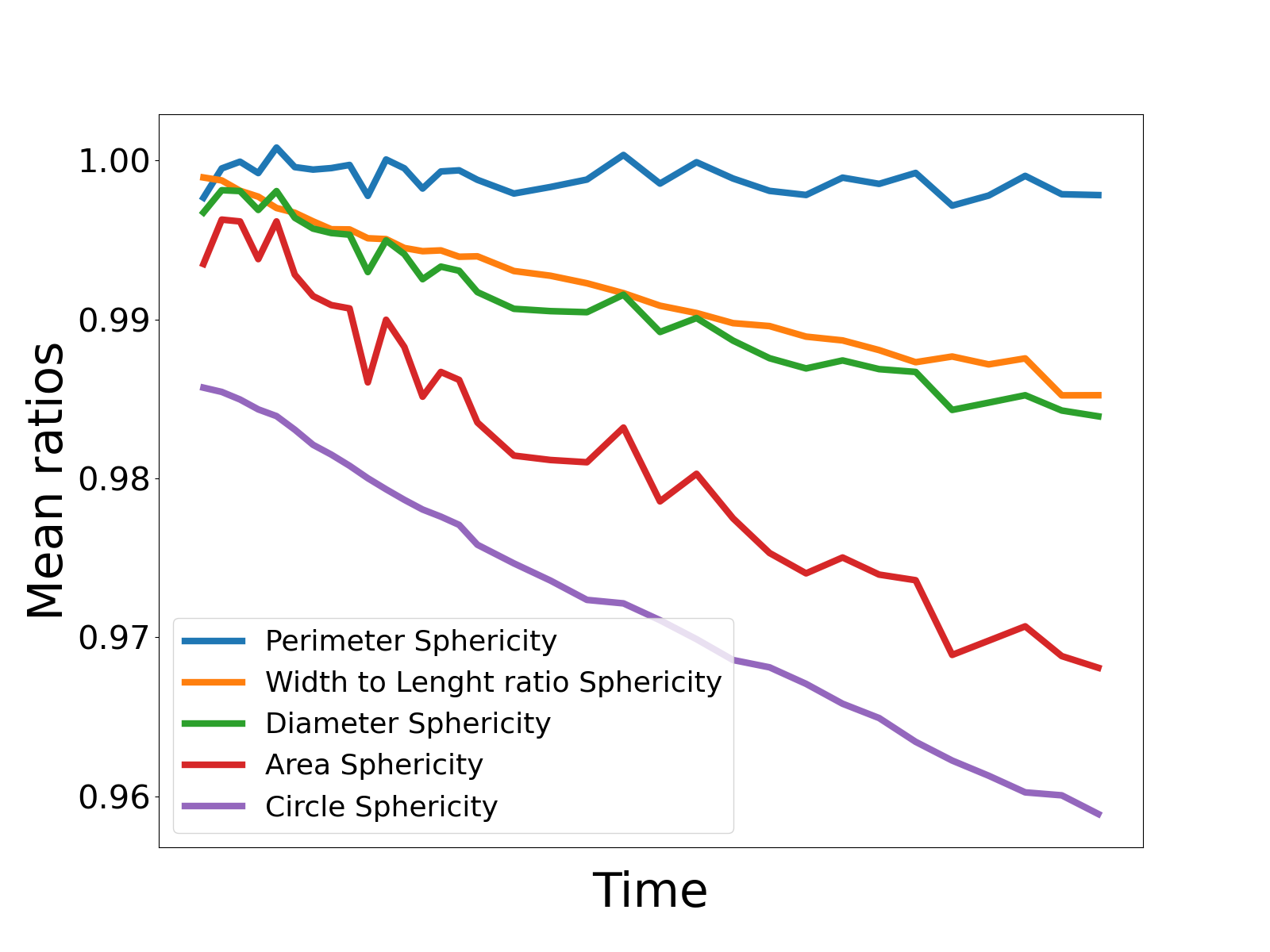}
    \caption{Evolution of different mean indices to appreciate the sphericity of the grains.}
    \label{Sphericity MG PS}
\end{figure}

The evolution of the shape of the grains is illustrated in Figure \ref{Sphericity MG PS} by the evolution of the mean sphericity indices. In the case of multiple grains, the mean value over the grains is considered. It can be observed that the grains become less and less similar to a sphere over time.

\begin{figure}[ht]
    \centering
    \includegraphics[width = 0.48\linewidth]{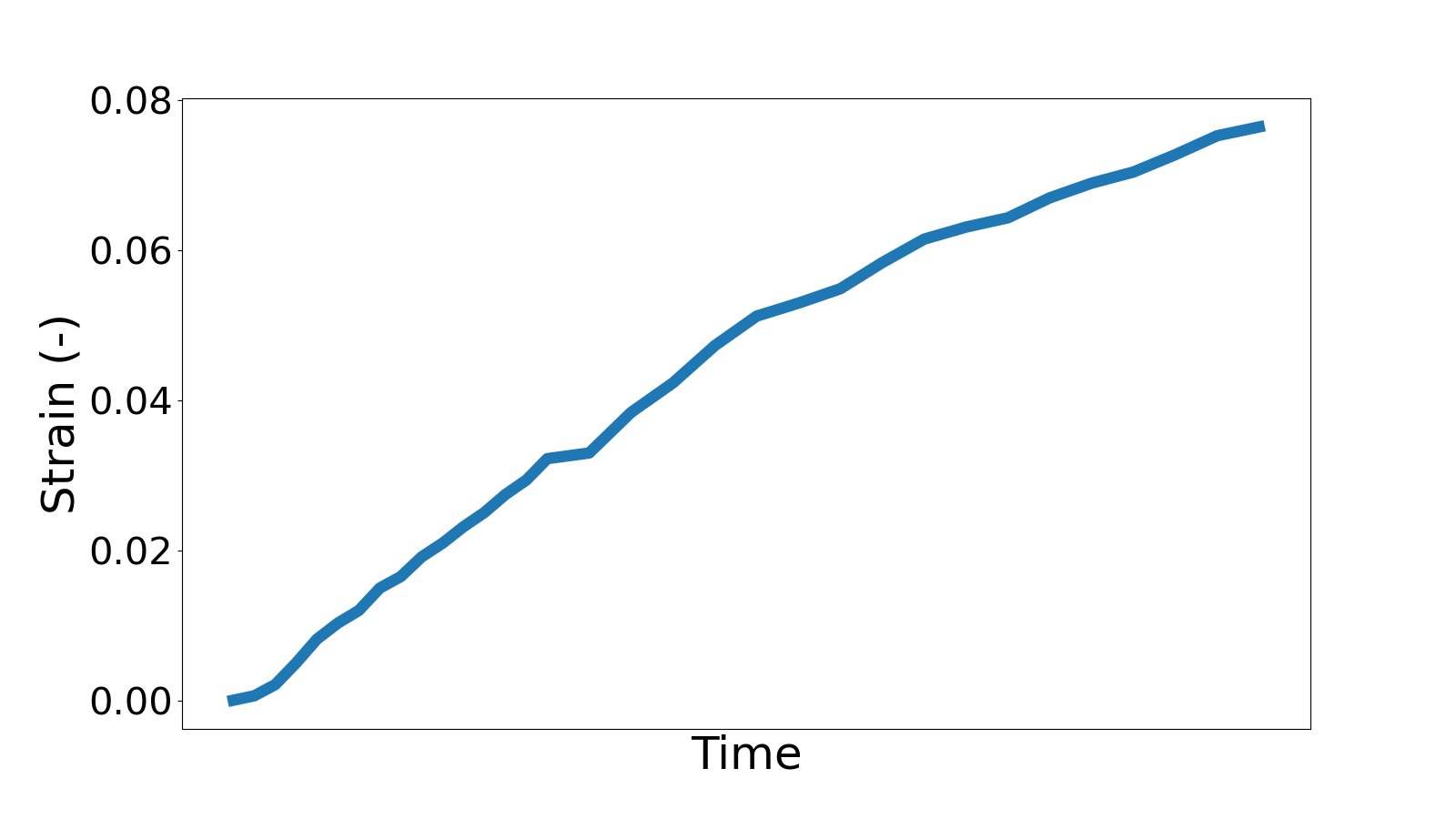}~
    \includegraphics[width = 0.48\linewidth]{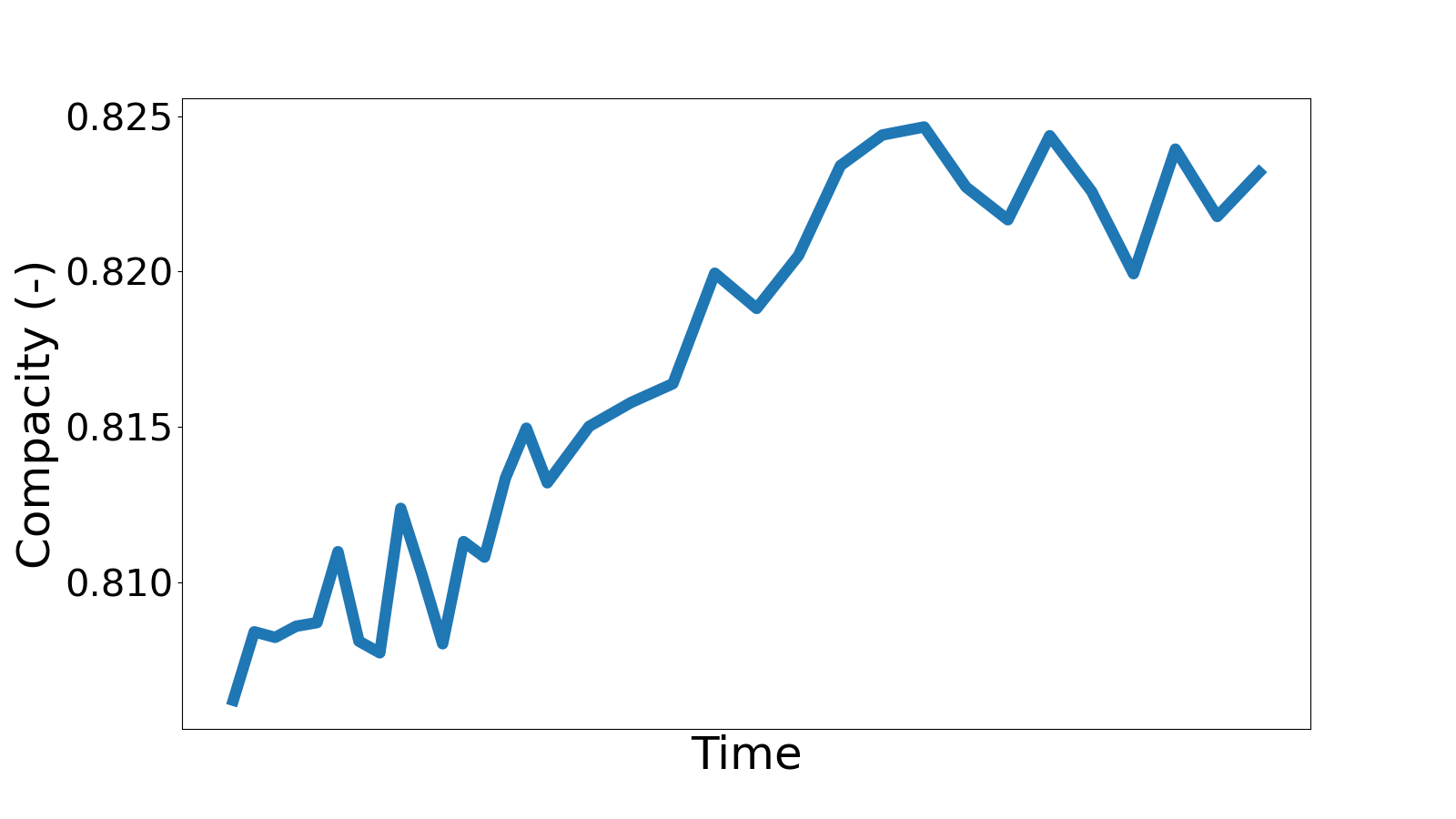}
    \caption{Evolution of the vertical strain and the sample compacity.}
    \label{Vertical strain and Compacity}
\end{figure}

Figure \ref{Vertical strain and Compacity} shows the sample's vertical strain and compacity. The evolution of compacity is important to comment on, as it is noisy and does not evolve linearly with the vertical strain. It is computed with the ratio of the surface grains over the sample surface. In this configuration, the confinement force is large, and overlaps are quite large as well. To obtain a corrected compacity index, it becomes important to reduce the contact surface and not count the material surfaces twice.

\begin{figure}[ht]
    \centering
    \includegraphics[width = 0.5\linewidth]{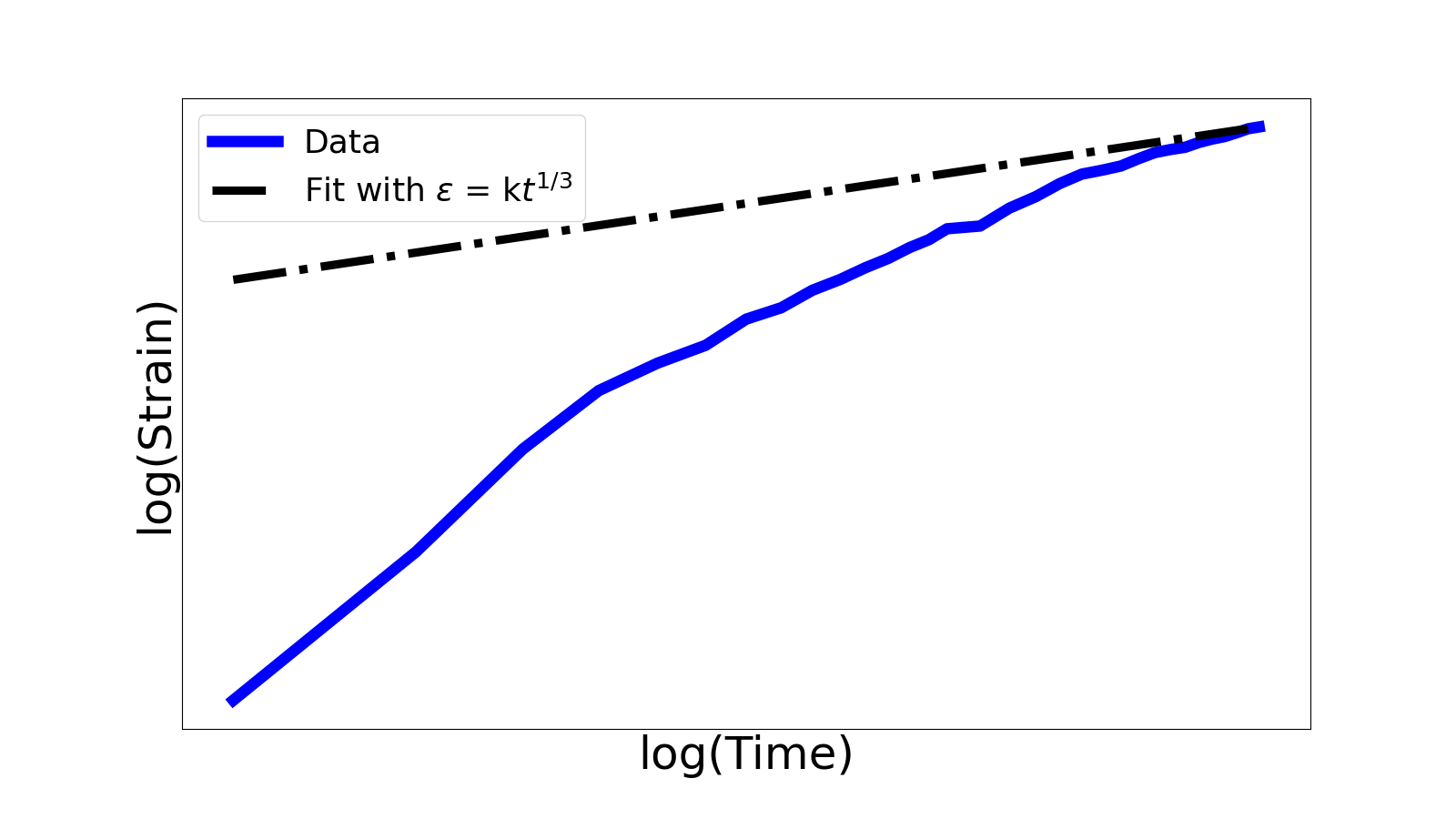}~
    \caption{Comparison between the numerical results and the Andrade creep law for the evolution of strain with time.}
    \label{Andrade Fit}
\end{figure}

A fit with an Andrade creep response has been attempted in Figure \ref{Andrade Fit}. The results of the simulation seems to be aiming to reach the tendency of the Andrade fit.

\begin{figure}[ht]
    \centering
    a)
    \includegraphics[width=0.45\linewidth]{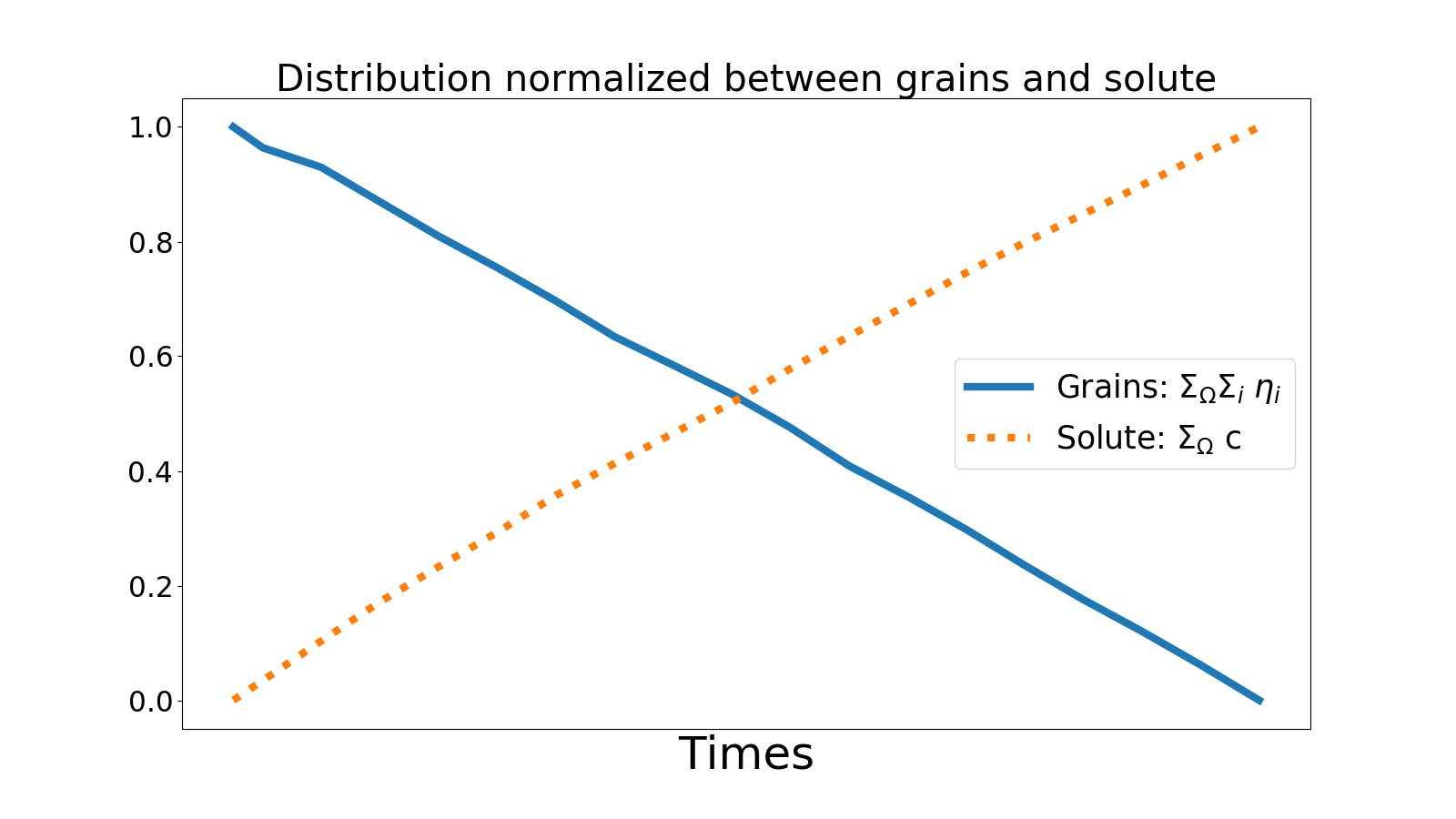}
    b)
    \includegraphics[width=0.45\linewidth]{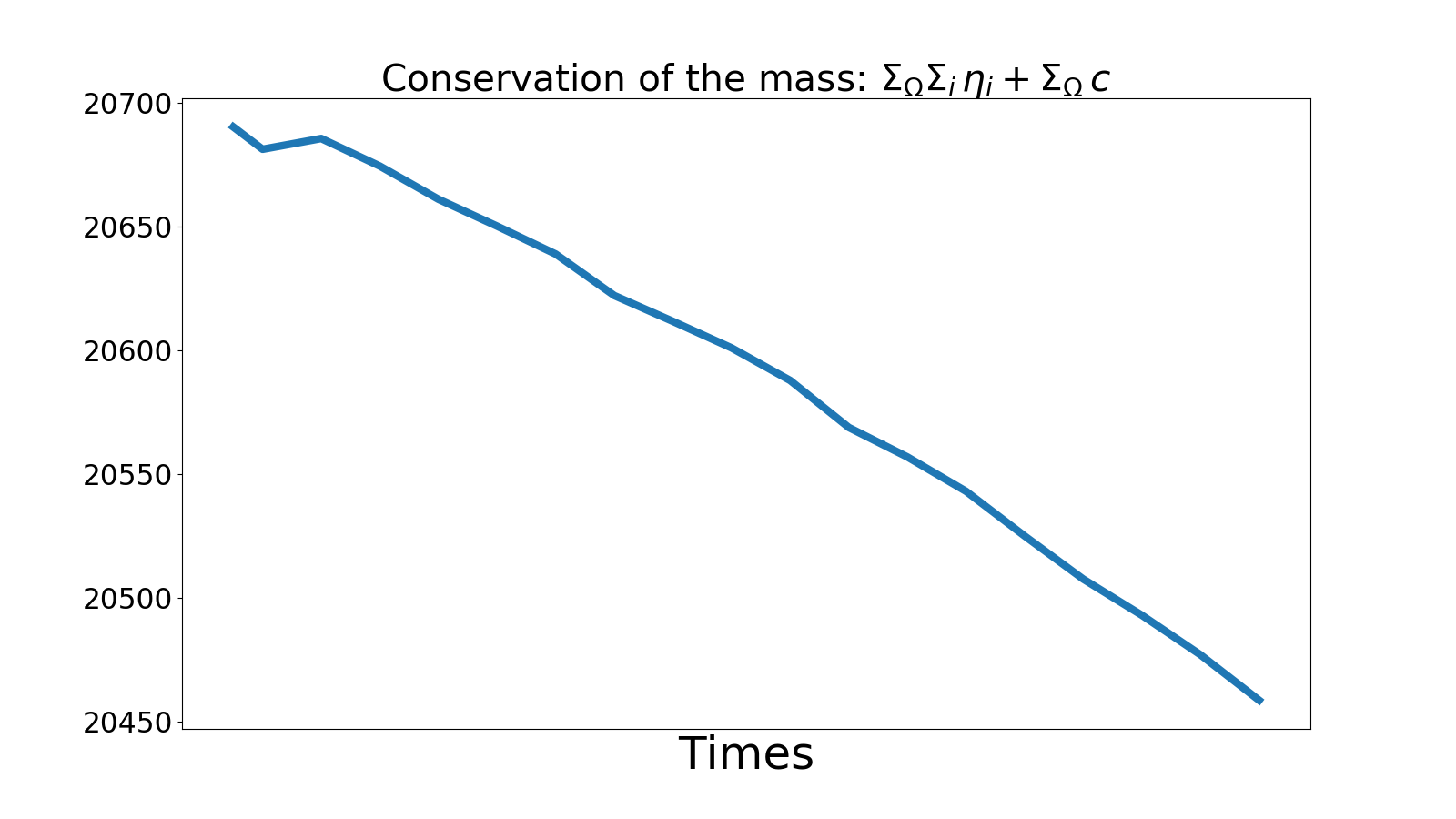}
    \caption{Evolution of the grains and solute quantities. a small loss of mass (because of the grain displacement and detection algorithms) is detected.}
    \label{Result MG Pressure Solution Quantity}
\end{figure}

The normalized distribution (presented Equation \ref{Normalized Distribution}) between grains and solute is illustrated in Figure \ref{Result MG Pressure Solution Quantity}a. The mass conservation is studied in Figure \ref{Result MG Pressure Solution Quantity}b. It is observed that a small loss of mass occurs, approximately 1\% of the initial mass. This loss is due to the grain displacement and detection algorithms, as these algorithms use a window around the grain to limit the computational cost and cut everything outside this window.

\vskip\baselineskip

The goal of this section was to highlight the PFDEM's capability to handle heterogeneous dissolution/precipitation processes. The observed grain shape evolution during the pressure solution phenomena is not uniform, as it varies from the contact point (high-stress area) to the pore surface (low-stress area). To capture this behavior, a solution concentration and a chemo-mechanical formulation have been developed and incorporated into the phase-field equations. The results demonstrate that the sphericities of the grains decrease over time, indicating that the grains become less disk-like. Additionally, the sample tends to an Andrade creep response, which is commonly observed in geomaterials during pressure solution. These findings illustrate the PFDEM's ability to accurately simulate and analyze the complex behavior of pressure solution phenomena, taking into account heterogeneous dissolution/precipitation effects.

%%=======================================================%%

\section{Conclusions}

This work presents an extension of the classical discrete element model by coupling it to the phase-field theory, enabling the capture of real grain shape and its evolution based on physical laws. The study was conducted in two steps. Firstly, the effect of grain shape was investigated through a sample subjected to constant stress in oedometric conditions with partial dissolution. The results highlighted the importance of considering the actual grain shape instead of relying solely on modeling with rolling resistance as the increase of grain angularity showed opposite evolution for the lateral pressure coefficient evolution $k_0$. Secondly, the pressure solution phenomena was examined, considering a heterogeneous dissolution/precipitation simulation. A comparison with experimental data was performed, demonstrating that the simulated data tends to the Andrade creep response, a characteristic behavior observed in granular materials undergoing pressure solution.

The proposed formulation opens up avenues for further exploration into the influence of various parameters on the creep behavior and the effect of grains characteristics. Parameters such as solute diffusivity, chemical energy factor, and mechanical energy factor can now be systematically studied to gain insights into their impact on the overall behavior. This coupling of discrete element modeling with phase-field theory provides a useful framework for investigating the complex interplay between grain characteristics and the mechanical response of geomaterials undergoing dissolution/precipitation processes.

\section{Software}

The PF model is solved using the MOOSE software \cite{Moose2020}, while the DEM is solved using a custom-developed Python software.
The scripts and algorithms used in this study are available on GitHub at the following links:

\begin{itemize}
\item \url{https://github.com/AlexSacMorane/PFDEM_AC}
\item \url{https://github.com/AlexSacMorane/PFDEM_ACS}
\item \url{https://github.com/AlexSacMorane/PFDEM_ACS_MultiGrains}
\end{itemize}

%%=======================================================%%

\section{Acknowledgements}

This research has been partially funded by the Fonds Spécial de Recherche (FSR) and by the Wallonia-Bruxelles Federation.

%%=======================================================%%

\appendix

\section{Find the grain discretization}
\label{Find n border}

The discretization of the grain (= the number of vertices) is the main issue for the quality of the result. Hence, if the discretization is coarse, the shape is not well described and some errors can occur for contact computing. The goal of this appendix is to study the influence of this parameter on the results quality. 

\begin{figure}[ht]
    \centering
    \includegraphics[width = 0.45\linewidth]{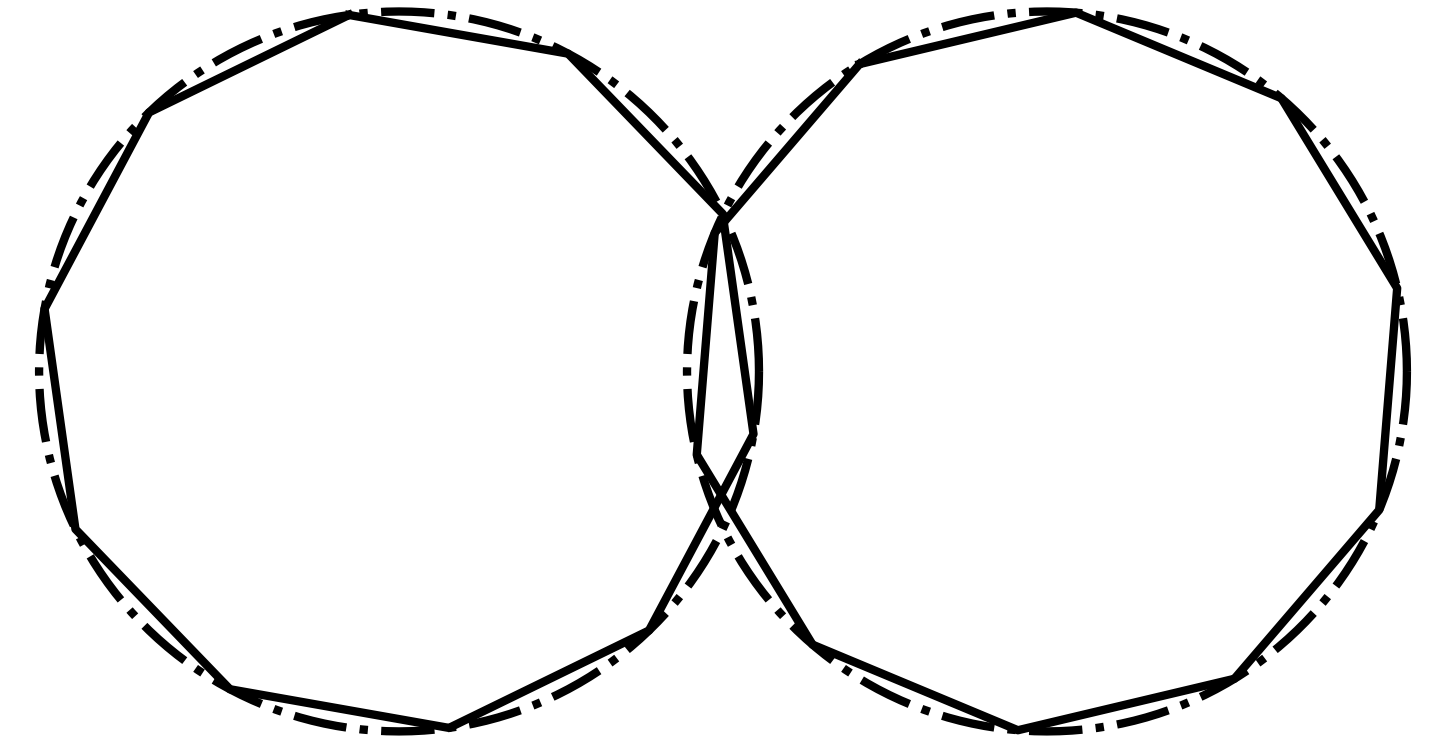} ~
    \includegraphics[width = 0.45\linewidth]{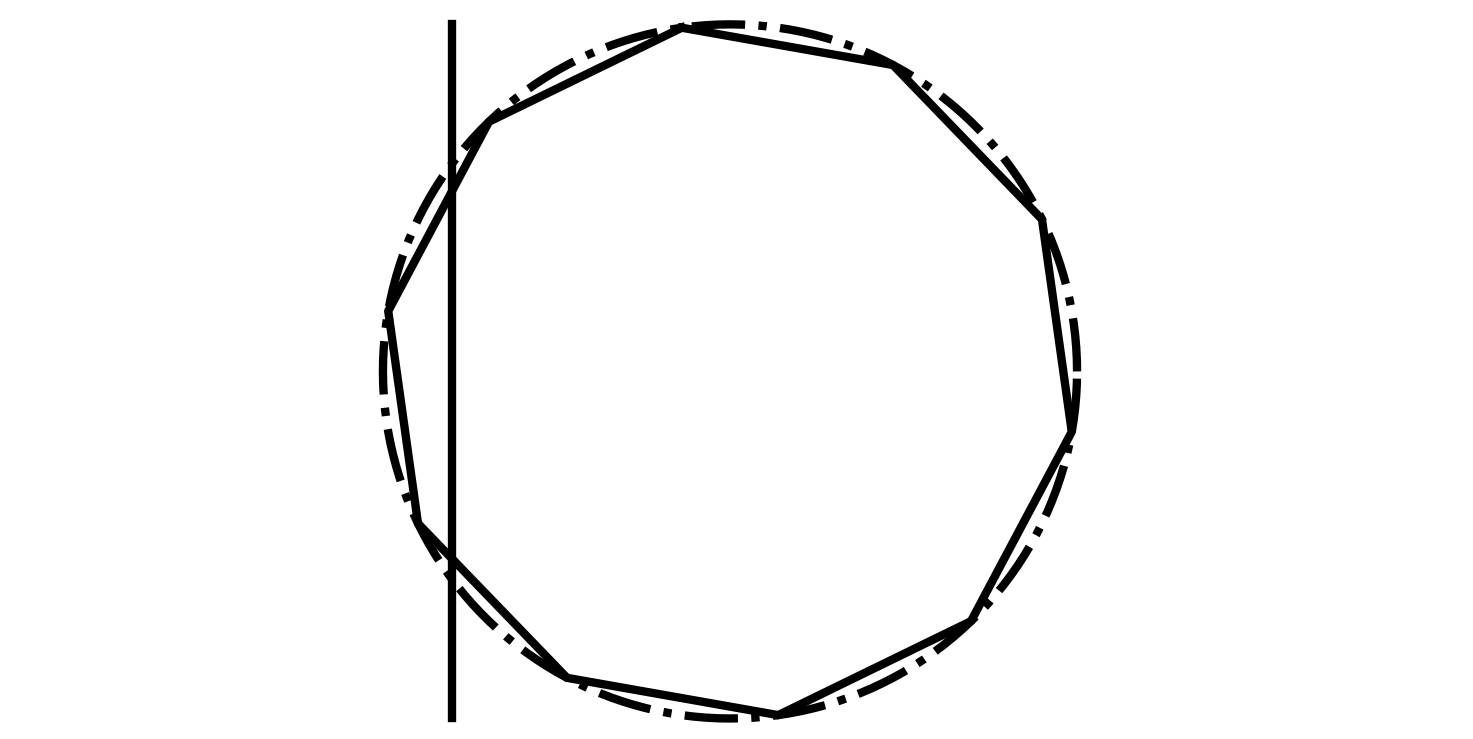}
    \caption{Two cases are considered: a contact grain-grain and a contact grain-wall. The dashed lines represent perfectly circular grains.}
    \label{2 simple cases n border}
\end{figure}

First, two simple cases illustrated in figure \ref{2 simple cases n border} are considered: two grains in contact and one grain in contact with a wall. Those cases represent the elementary phenomena in a granular sample.

\begin{figure}[ht]
    \centering
    \includegraphics[width = 0.8\linewidth]{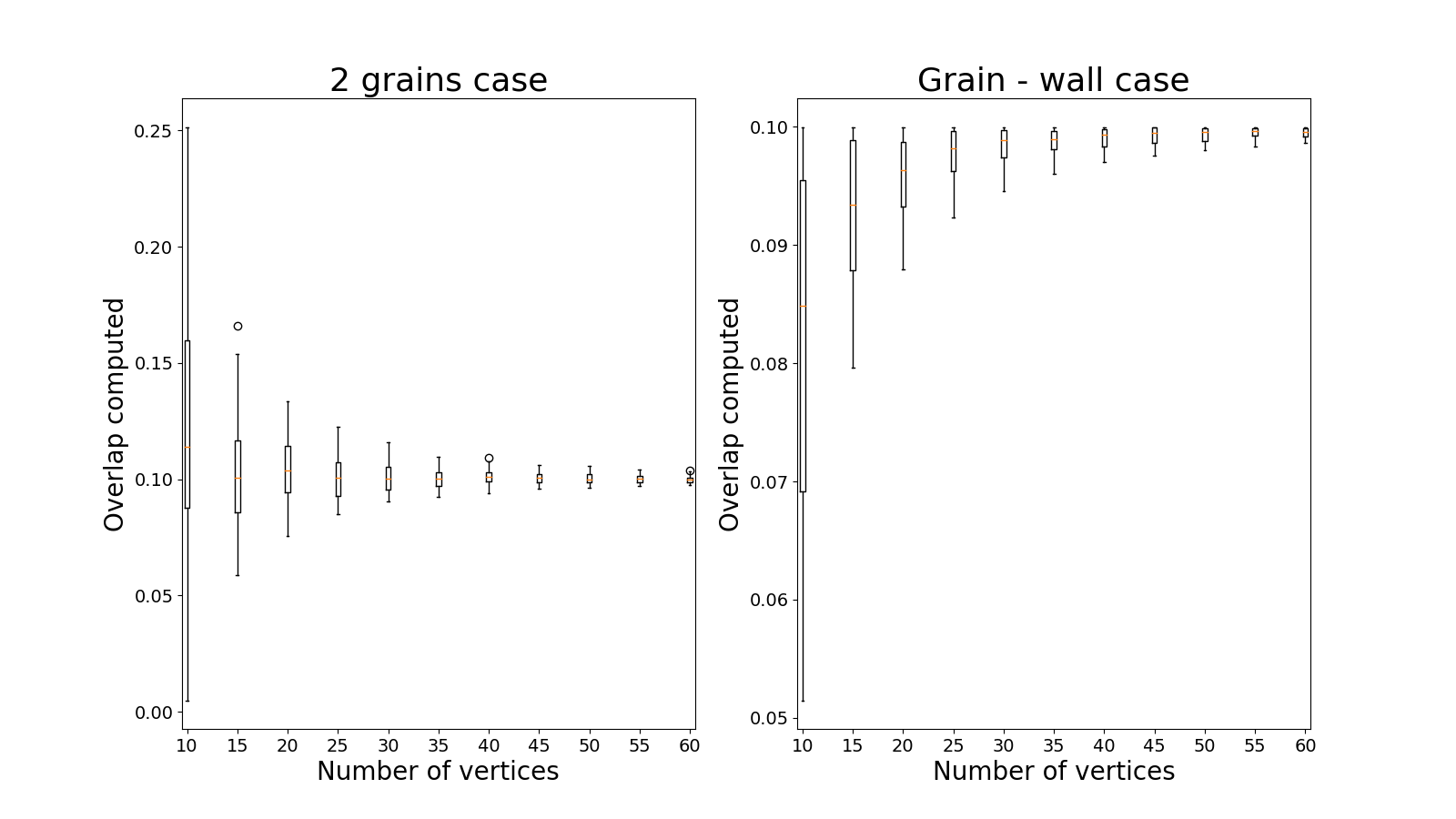}
    \caption{Influence of the grain discretization on the overlap estimation.}
    \label{Influence n border overlap 10}
\end{figure}

Different grains with 10 to 60 (with a step of 5) vertices are generated to see the influence of grain discretization. Then, the grains are randomly rotated and the overlap is estimated with the common-plane algorithm described in \ref{CP Algorithm Appendix}. The statistical result obtained with 100 repetitions is shown in figure \ref{Influence n border overlap 10}. It appears the variation of estimated overlap decreases with grain discretization. 

\begin{figure}[ht]
    \centering
    \includegraphics[width = 0.8\linewidth]{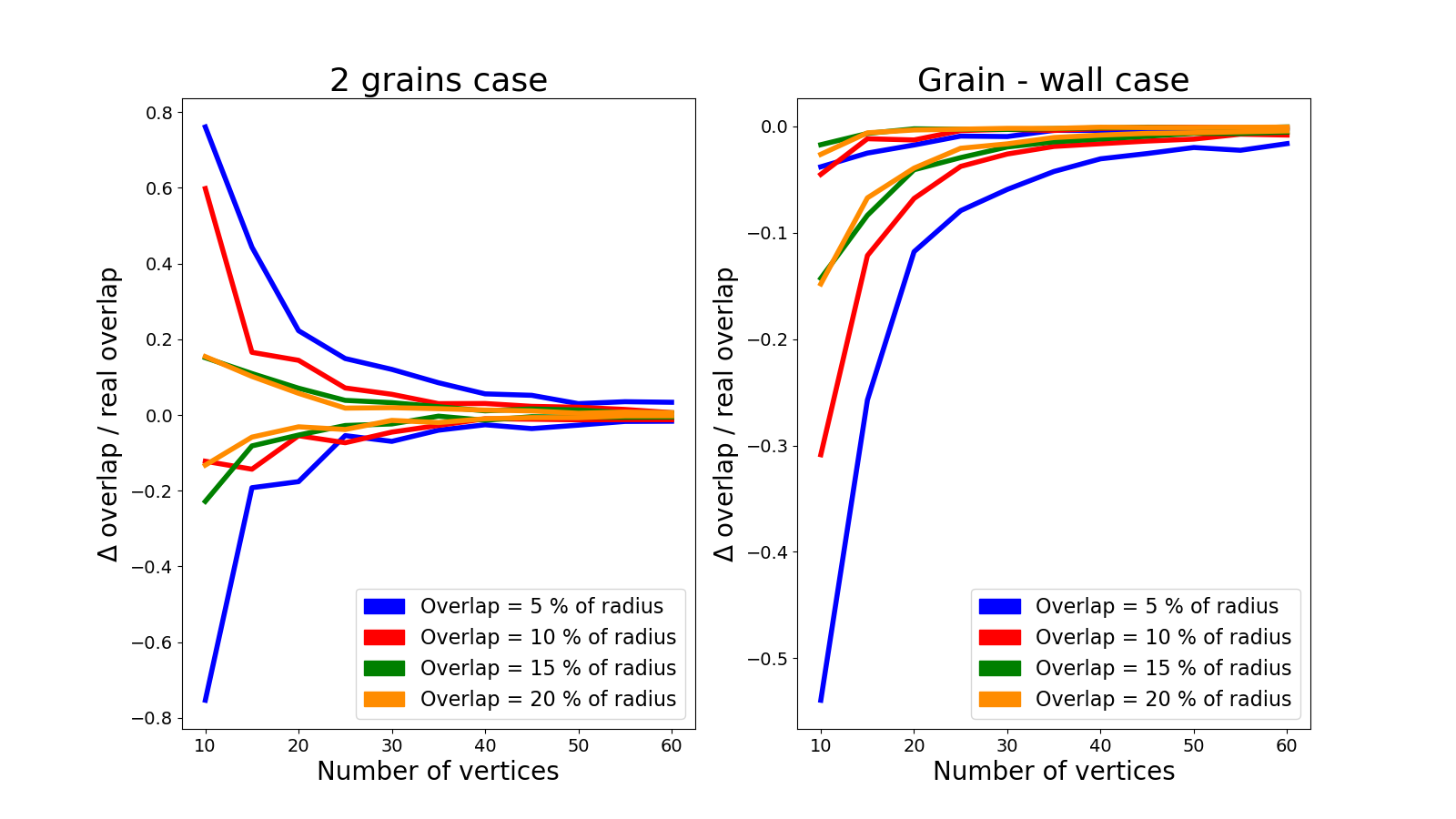}
    \caption{Influence of the grain discretization on the overlap estimation for different solicitations. Only the first and the third quartiles of each data set are plotted.}
    \label{Influence n border overlap}
\end{figure}

This study is repeated for different real overlaps. Hence, for coarse discretization and low solicitation, a theoretical contact can be not detected because of the discretization. Figure \ref{Influence n border overlap} shows that grain discretization greatly influences the results' quality for small overlap, as is the case in the granular sample.

\begin{figure}[ht]
\centering
\includegraphics[width=0.9\linewidth]{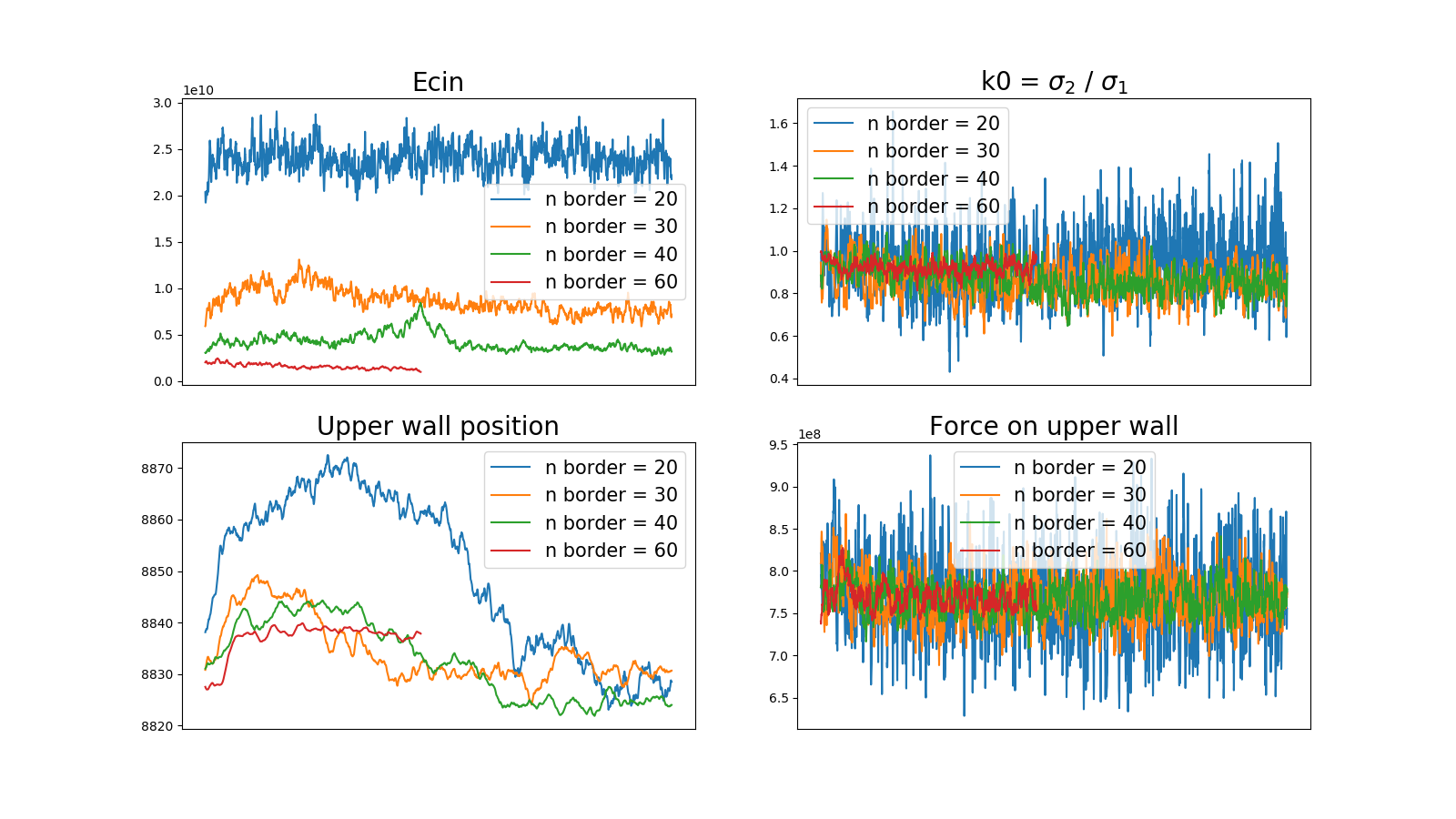}
\caption{Evolution with time of the total kinetic energy, the coefficient $k_0=\sigma_{II}/\sigma_{I}$, the upper wall position and the force applied on the upper wall for different numbers of vertices.}
\label{Trackers Sample Discretization}
\end{figure}

Secondly, a more complex case is considered: a granular media composed of 300 grains is loaded under oedometric conditions. To see the influence of grain discretization, a sample composed of perfect disks is loaded. Then the grains are divided into 20, 30, 40 or 60 vertices. For each case, 10 simulations are run. Figure \ref{Trackers Sample Discretization} shows the evolution with time of the total kinetic energy, the coefficient $k_0=\sigma_{II}/\sigma_I$, the upper wall position and the force applied on the upper wall for one run and different numbers of vertices. The number of vertices for one grain has a large influence on the global behavior of the sample. Hence, with few vertices, a lot of noise is introduced in the computing of the contacts. This noise is reflected in the estimation of the coefficient $k_0$. Moreover, the total kinetic energy stays at a large value with few vertices by grains. The equilibrium can not be reached.

\begin{figure}[ht]
\centering
\includegraphics[width=0.8\linewidth]{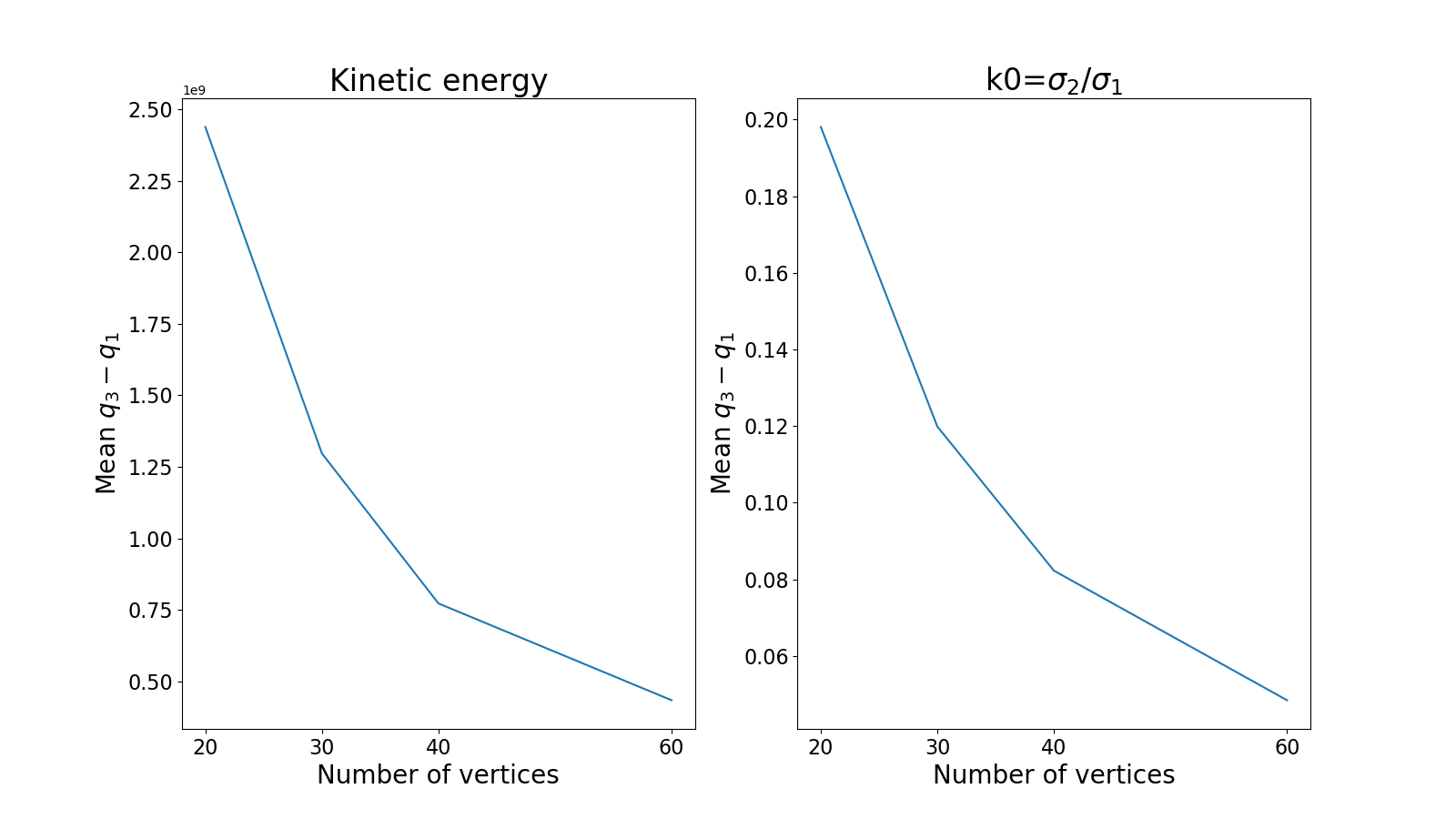}
\caption{Statistical analysis about the noise of the coefficient $k_0=\sigma_{II}/\sigma_{I}$ for different numbers of vertices.}
\label{Trackers Sample Discretization Stat}
\end{figure}

A statistical analysis of the total kinetic energy and the estimation of the coefficient $k_0$ of all the runs is presented in figure \ref{Trackers Sample Discretization Stat}. For all runs, the difference between the third quartile and the first one is estimated. Then, a mean value is computed for the different numbers of vertices. It appears the noise is decreasing with the quality of the discretization.

%%=======================================================%%

\section{A Monte Carlo method algorithm to characterize a polygonal particle}
\label{Monte Carlo method Appendix}

A polygonal particle is defined by the coordinates of the vertices and some essential information as center, surface and inertia must be known for the DEM algorithm. A Monte Carlo is used to compute them \cite{AlonsoMarroquin2009}. this method is based on the creation of a box around the grain. Then, a point is generated randomly inside the box. A point-inside-polygon test is used to determine if the point is inside or not the polygonal particle (this test is explained in the following). This step is repeated a large number of time. The mass, the center and the inertia of the particle can be determined with the equations \ref{Mass Monte Carlo Equation}, \ref{Center Monte Carlo Equation} and \ref{Inertia Monte Carlo Equation}.

\begin{equation}
    mass = \frac{A_{box}}{N_p}\displaystyle\sum_{i=1}^{N_p} \chi(\overrightarrow{X}_i)\sigma
    \label{Mass Monte Carlo Equation}
\end{equation}

\begin{equation}
    mass \times \overrightarrow{r} = \frac{A_{box}}{N_p}\displaystyle\sum_{i=1}^{N_p} \chi(\overrightarrow{X}_i)\sigma\overrightarrow{X}_i
    \label{Center Monte Carlo Equation}
\end{equation}

\begin{equation}
    I + mass \lVert\overrightarrow{r}\rVert^2 = \frac{A_{box}}{N_p}\displaystyle\sum\_{i=1}^{N_p} \chi(\overrightarrow{X}_i)\sigma\lVert\overrightarrow{X}_i\rVert^2
    \label{Inertia Monte Carlo Equation}
\end{equation}

with $A_{box}$ is the area of the box, $\overrightarrow{X}_i$ the coordinate of the random point, $N_p$ the total number of random point, $\sigma$ the surface mass, $\overrightarrow{r}$ the center of mass and $I$ the moment of inertia. The function $\chi(\overrightarrow{X}_i)$ is the characteristic function, which returns 1 if the point is inside the grain and 0 otherwise. 

This characteristic function follows the point-inside-polygon test developed by Franklin \cite{Franklin1994}. A semi-infinite ray is generated from the point in the direction of the increasing x. Every time this ray crosses an edge the point switches between inside and outside.

%%=======================================================%%

\section{Common-plane algorithm to detect contact between two polygonal particles}
\label{CP Algorithm Appendix}
This summary of the common-plane algorithm, introduced by Cundall \cite{Cundall1988} to detect a contact between two polygonal particles. In fact, the common-plane is defined as a plane that bisects the space between two particles. Then, this algorithm has been improved with a faster identification \cite{Nezami2004, Podlozhnyuk2018}.
Hence, only five candidates are possible for the common-plane.

The different geometric elements used are presented at figure 5 of \textit{Nezami et al.} \cite{Nezami2004}. It is important to notice the point A (resp. B) is the nearest vertex of the particle A (resp. B) to the particle B (resp. A). Moreover, the point M is the midpoint of the segment [AB]. The five candidates are the following :
\begin{itemize}
\item the normal to the line (AB)
\item $m_1$ defined from the vector $\overrightarrow{AA_1}$ and the point M
\item $m_2$ defined from the vector $\overrightarrow{AA_2}$ and the point M
\item $m_3$ defined from the vector $\overrightarrow{AA_3}$ and the point M
\item $m_4$ defined from the vector $\overrightarrow{AA_4}$ and the point M
\end{itemize}

The basis of the fastest algorithm is the facts that the common-plane passes through the midpoint M. Moreover, the common-plane must be located inside the space S delimited by $m_1$, $m_2$, $m_3$ and $m_4$. Finally, the common-plane should produce the smallest angle with the normal to the line (AB). As figure 6 of \textit{Nezami et al.} \cite{Nezami2004} highlights, only two cases are possible: (a) the normal to the line (AB) is inside of the surface S, the common-plane is this line; (b) the normal to the line (AB) is outside of the surface S. In the latter case, the angles between lines $m_i$ and the normal to the line (AB) are computed. The common-plane is the line with the smallest angle. 

%%=======================================================%%

\section{Find the Representative Equivalent Volume and the Representative Time Step}
\label{Look for something representative}

As for every numerical simulation, it is essential to work with a representative configuration. 

\subsection{Find the Representative Equivalent Volume}
\label{Look for REV}

In the case of the simulation campaign presented in section \ref{Section Irregular Shape}, a granular material is considered. The number of grains must be selected to obtain a representative equivalent volume. This number must stay reasonable because the PFDEM algorithm is asking for a lot of computational power. A first preliminary study is done.

To have faster results, perfect circle particles are used. Parameters from the table \ref{Parameters Acid Oedometer} are used to define the material, the discrete element model time step, and the particle's geometry. 
As described by the table \ref{Directory number of run for number of grains}, the number of grains is varying between 50 and 700. The table describes also the number of simulations run for each number of grains (between 5 and 10 times).
$15\%$ of the grains selected randomly are assumed as dissolvable. After a equilibrium is found, the radius of the dissolvable particles decreases by $1\%$ of the initial mean radius ($350\,\mu m$).

\begin{table}[ht]
    \centering
    \begin{tabular}{|l|c|c|c|c|c|c|c|}
        \hline
        Number of grains & 50 & 100 & 200 & 300 & 400 & 500 & 700\\
        \hline
        Number of runs & 10 & 10 & 10 & 9 & 10 & 5 & 5\\
        \hline
    \end{tabular}
    \caption{Different samples are tried with different numbers of grains. Each configuration is run several times to make a statistical study of the results.}
    \label{Directory number of run for number of grains}
\end{table}

\begin{figure}[ht]
    \centering
    \includegraphics[width=0.8\linewidth]{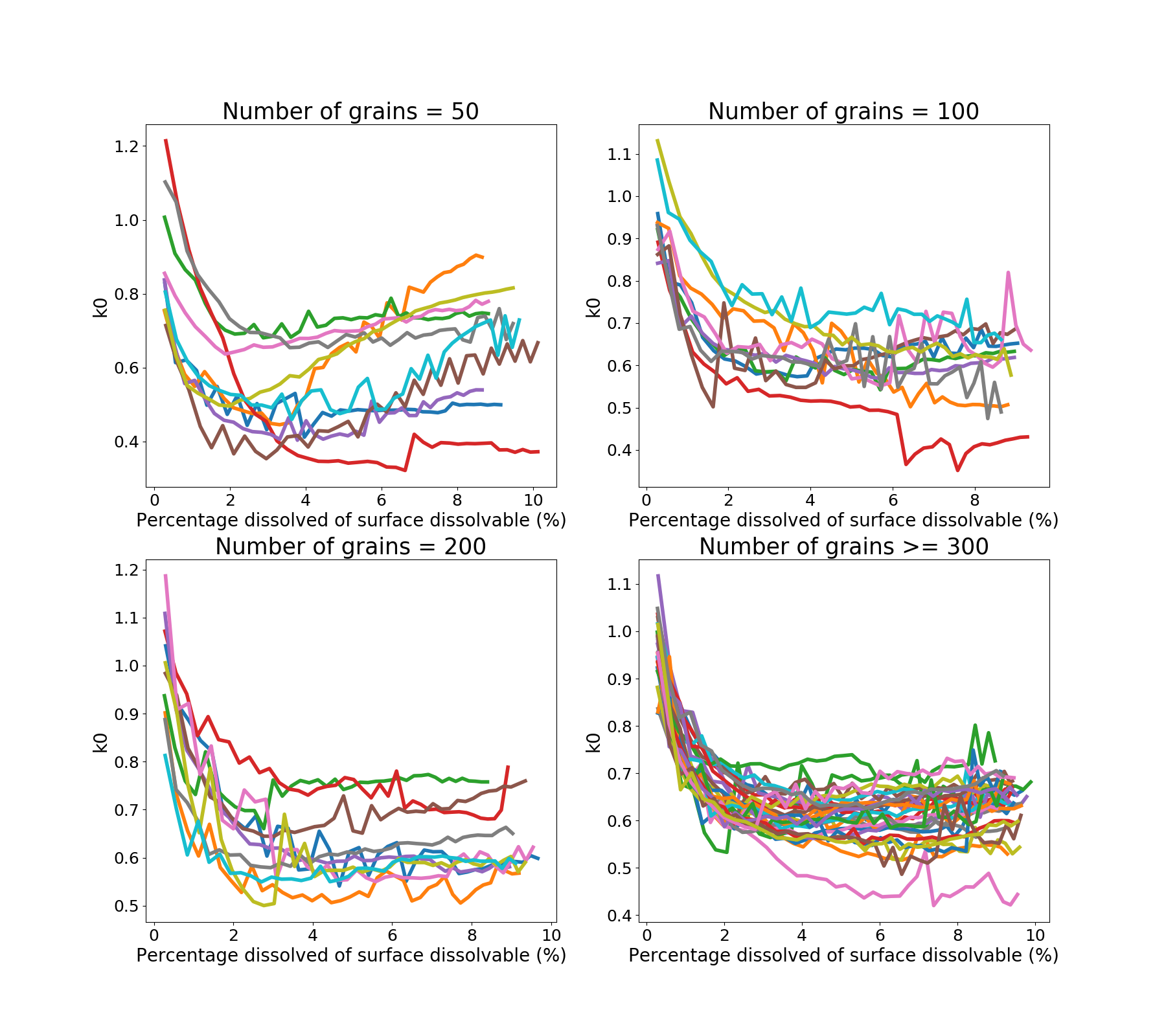}\\
    \caption{All curves of the preliminary campaign sorted following the number of grains in the sample.}
    \label{Result All Curves REV Oedo}
\end{figure}

\begin{figure}[ht]
    \centering
    \includegraphics[width=0.7\linewidth]{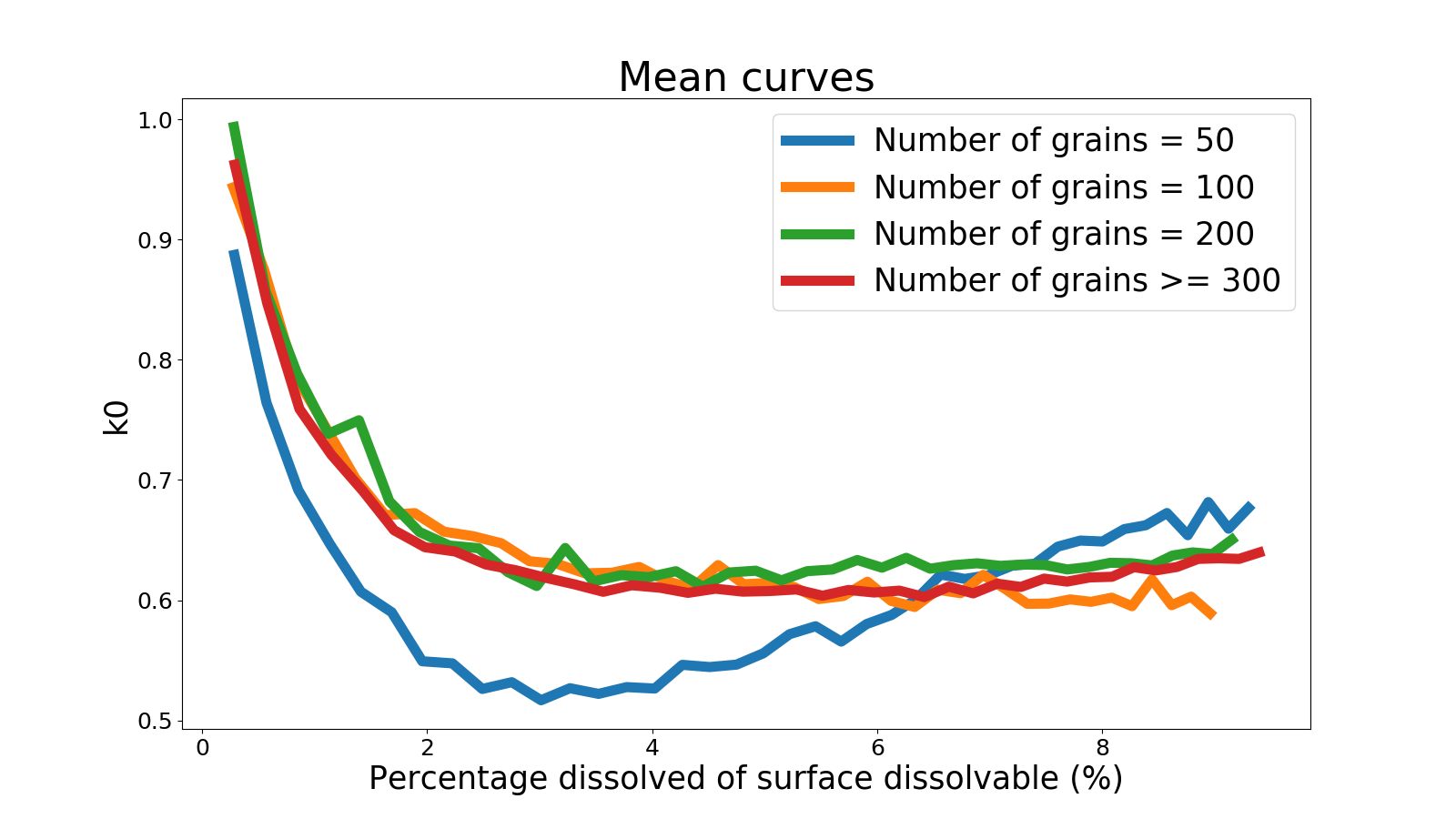}\\
    \caption{Mean curves for different numbers of grains in the sample.}
    \label{Result Mean REV Oedo}
\end{figure}

\begin{figure}[ht]
    \centering
    \includegraphics[width=\linewidth]{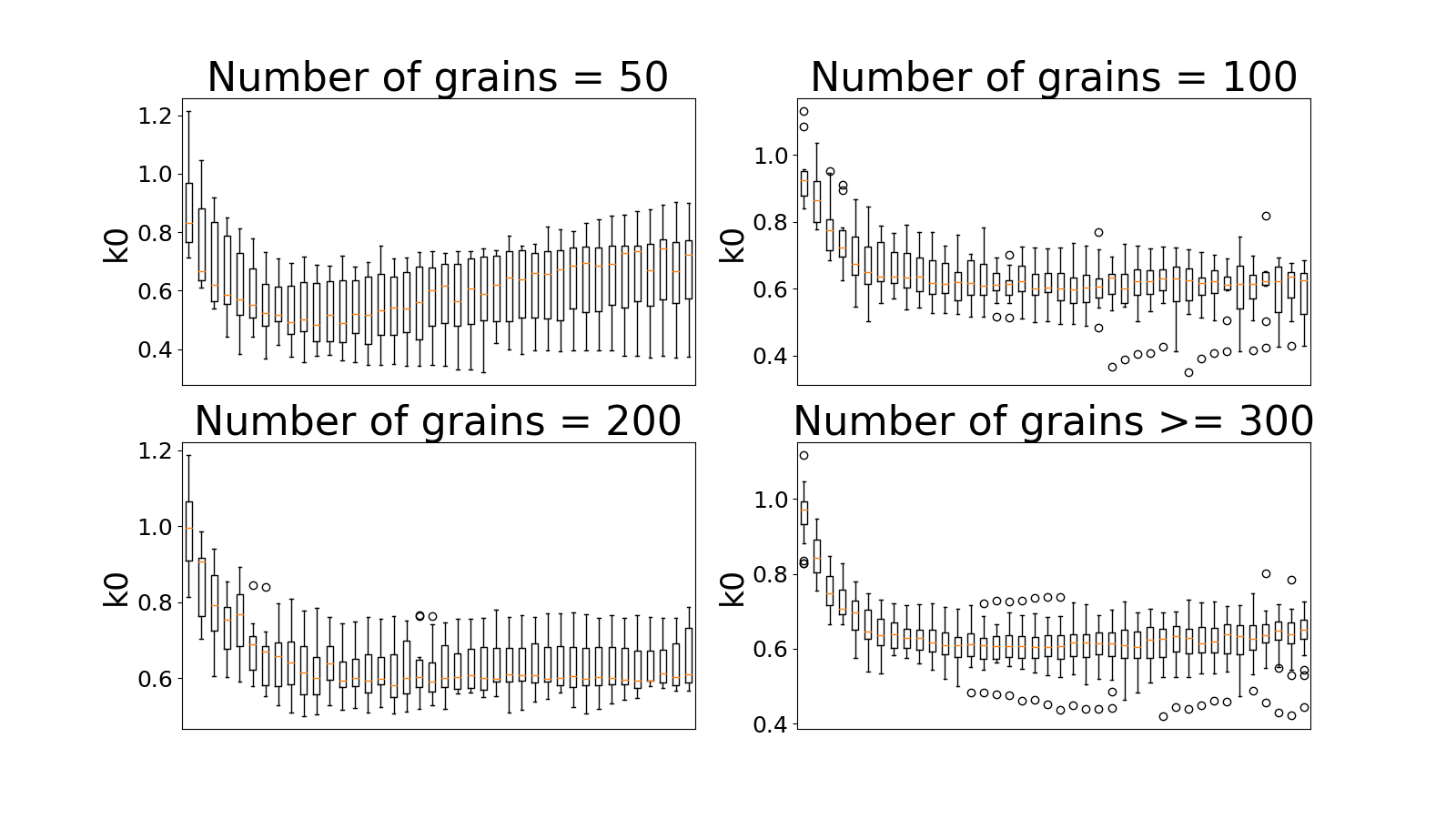}\\
    \caption{Statistical study to justify that a sample with 200 grains is a representative equivalent volume.}
    \label{Result Stat REV Oedo}
\end{figure}

The results are presented in figures \ref{Result All Curves REV Oedo}, \ref{Result Mean REV Oedo} and \ref{Result Stat REV Oedo}. A statistical study is done on the evolution of the $k0=\sigma_{II}/\sigma_I$ with dissolved material quantity to determine the representative equivalent volume. 

It appears the results obtained with only 50 grains are really random and not at all representative. Even if figure \ref{Result Mean REV Oedo} shows the fact that mean curves are the same for samples with 100, 200 or more than 300 grains, figure \ref{Result Stat REV Oedo} highlights the large variance for the case of 100 grains. The representative equivalent volume can be assumed with 200 grains as the variance aims to reduce and become reasonable. As the sample is a granular material, there is always a noise in the results. Even if the sample is with a representative element volume, several simulations at the same configuration should be done to work with a mean evolution.

\subsection{Find the Representative Time Step}
\label{Look for RTS}

Moreover, there is a dissolved material quantity between two discrete element model simulations. This quantity must stay small to obtain a representative behavior. Hence, it is assumed that the discrete element model simulation is used to find a mechanical equilibrium (short term). Whereas, the phase-field simulation represents the dissolution of the material (long term). The quantity of dissolved material must stay reasonable because the PFDEM algorithm is asking for a lot of computational power. A second preliminary study is done.

To have faster results, perfect circle particles are used. Parameters from the table \ref{Parameters Acid Oedometer} are used to define the material, the discrete element model time step, and the particle's geometry. 
According to the preliminary work about the representative equivalent volume presented in section \ref{Look for REV}, the number of grains is 300. $15\%$ of the grains selected randomly are assumed as dissolvable. 
As described by the table \ref{Directory number of run for radius reduction}, the radius of the dissolvable particles is decreasing by $0.2\,\%$, $0.5\,\%$, $1.0\,\%$ or $2.0\,\%$ of the initial mean radius ($350\,\mu m$). The table describes also the number of simulations run for each radius reduction (between 8 and 15 times).
The radius reduction between iterations defines the representative time step.

\begin{table}[ht]
    \centering
    \begin{tabular}{|p{0.3\linewidth}|c|c|c|c|c|c|c|c|}
        \hline
        Radius reduction (\% of the initial mean radius) & 0.2 & 0.3 & 0.5 & 0.6 & 0.7 & 1.0 & 2.0 & 5.0\\
        \hline
        Number of runs & 21 & 9 & 13 & 9 & 9 & 13 & 9 & 5\\
        \hline
    \end{tabular}
    \caption{Different radius reductions between iterations are tried. Each configuration is run several times to make a statistical study of the results.}
    \label{Directory number of run for radius reduction}
\end{table}

\begin{figure}[ht]
    \centering
    \includegraphics[width=0.98\linewidth]{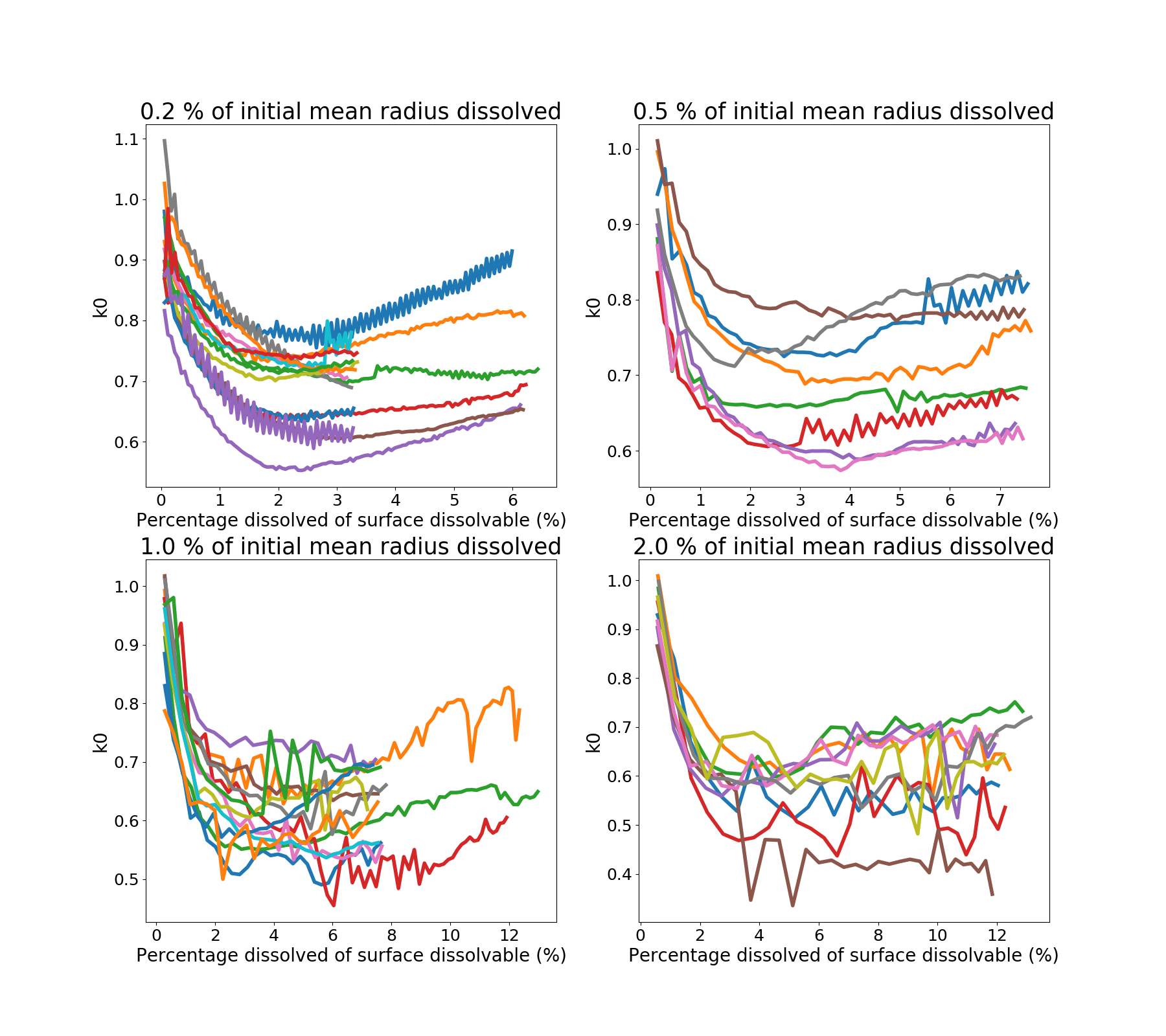}\\
    \caption{All curves of the preliminary campaign sorted following radius reduction.}
    \label{Result All Curves RTS Oedo}
\end{figure}

\begin{figure}[ht]
    \centering
    \includegraphics[width=0.9\linewidth]{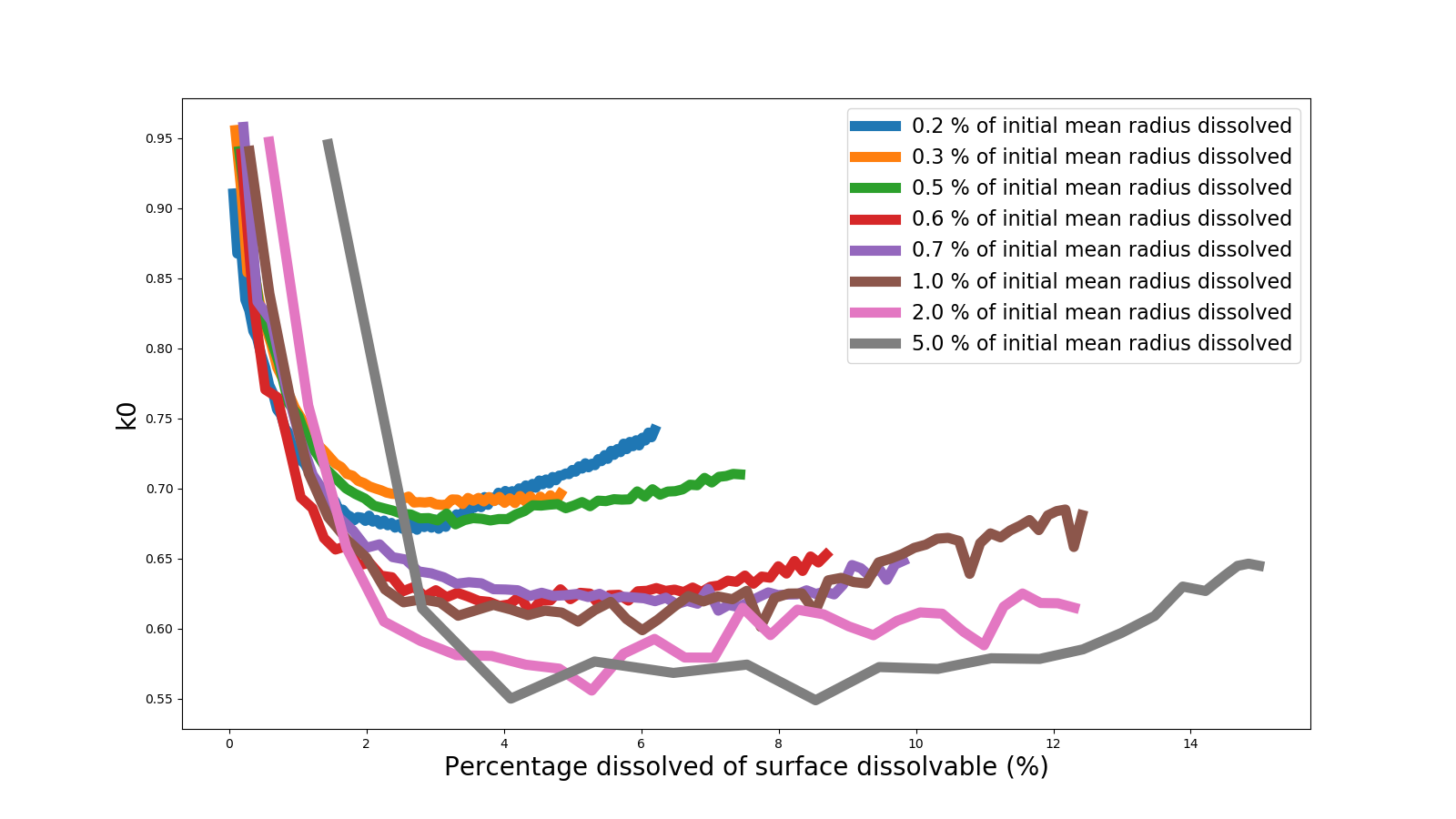}\\
    \caption{Mean curves for different radius reduction.}
    \label{Result Mean RTS Oedo}
\end{figure}

\begin{figure}[ht]
    \centering
    \includegraphics[width=\linewidth]{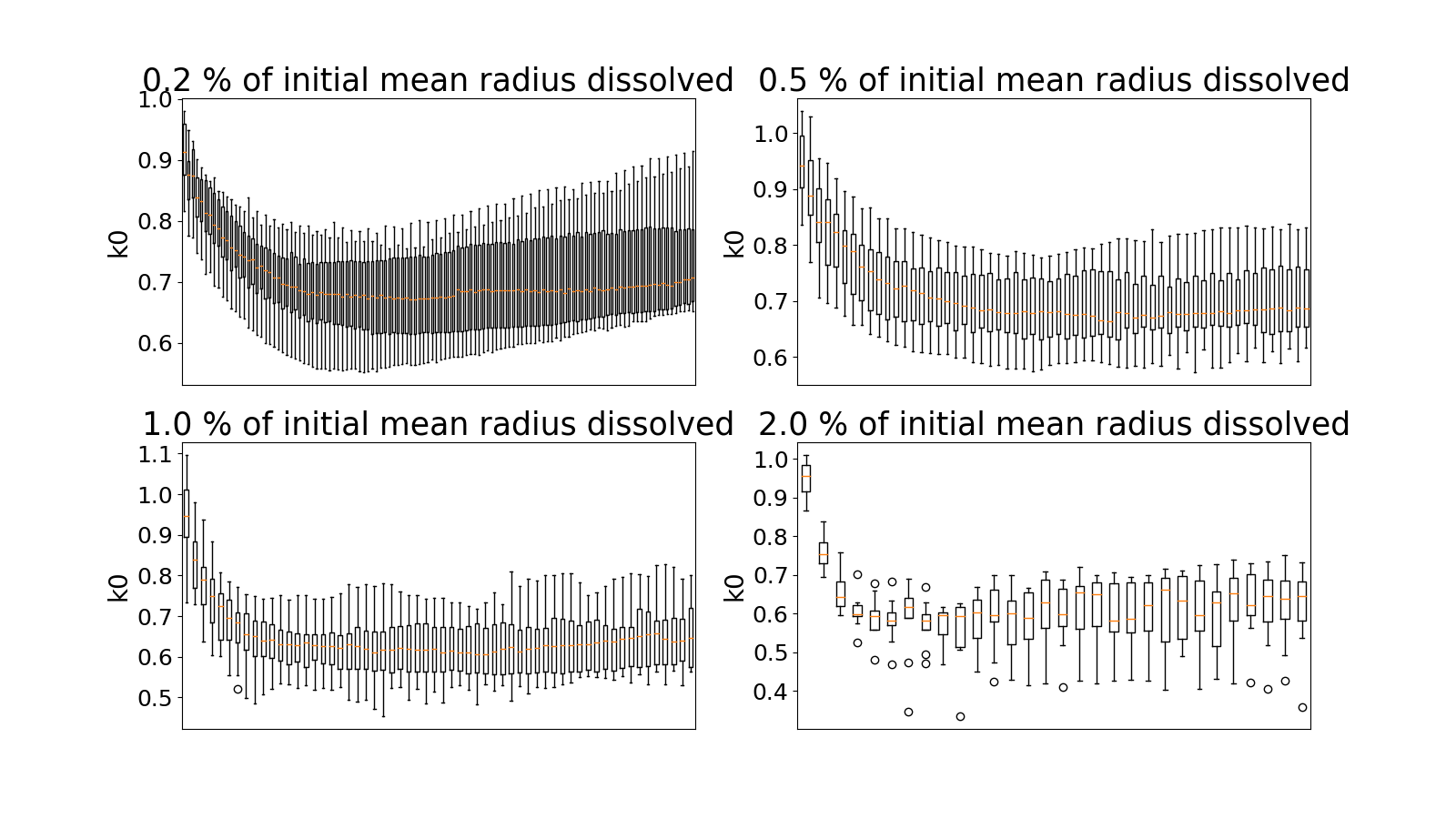}\\
    \caption{Statistical study to justify that 0.5 \% of the initial mean radius dissolved at each iteration is a representative time step.}
    \label{Result Stat RTS Oedo}
\end{figure}

The results are presented in figures \ref{Result All Curves RTS Oedo}, \ref{Result Mean RTS Oedo} and \ref{Result Stat RTS Oedo}. A statistical study is done on the evolution of the $k0=\sigma_{II}/\sigma_I$ with dissolved material quantity to determine the representative time step. 

It appears the results obtained with a radius reduction equivalent to $2.0\,\%$ of the initial mean radius are really random and not at all representative. Moreover, the steps are to large to well describe the initial fast decrease of $k0$.
Figures \ref{Result Mean RTS Oedo} and \ref{Result Stat RTS Oedo} show the curves obtained with a radius reduction lower than $0.5\,\%$ of the initial mean radius are reproducible and the initial fast decrease is well captured by those configurations.  The representative time step can be assumed with a radius reduction equivalent to $0.5\,\%$ of the initial mean radius. As the sample is a granular material, there is always a noise in the results. Even if the sample is with a representative time step, several simulations at the same configuration should be done to work with a mean evolution.

%%=======================================================%%

\section{Method to build the field of the solute diffusion coefficient}
\label{Build kappa c}

In the case of pressure solution, the problem asks for a heterogeneous solute diffusion coefficient field. The goal is to have diffusion at the contact area (to evacuate the solute generated) and in the pore area (as some fluid is assumed). However, it should not have diffusion in the grain material. The map of this coefficient is following the different steps :

\begin{enumerate}
    \item for each node of the mesh, a Boolean value is attributed, depending on the different phase variable (if $\eta_i>0.5$ and $\eta_j>0.5$ returns True; if $\eta_i<0.5$ and $\eta_j<0.5$ returns True; else return False). A Boolean map is generated.
    \item a dilation method \textit{scipy.ndimage.binary\_dilation} is applied to the Boolean map to generate a dilated map.
    \item The diffusion coefficient map is generated. For each node, if the value of the dilated map is True, $\kappa_c$ is attributed to the diffusion map; else, $0$ is attributed to the diffusion map.
\end{enumerate}

\begin{figure}[ht]
    \centering
    \includegraphics[width=0.7\linewidth]{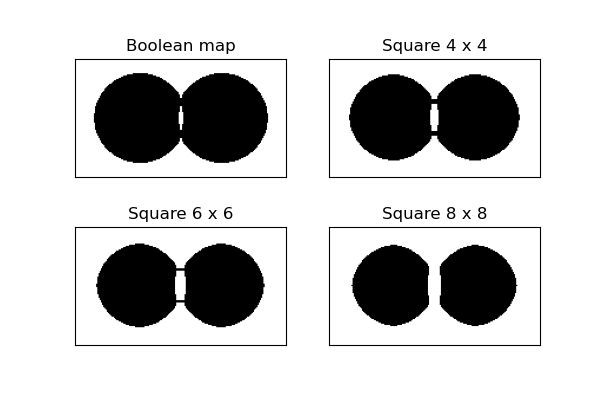}\\
    \caption{Diffusion coefficient maps for different morphological operators.}
    \label{Diffusion map by operator}
\end{figure}

As figure \ref{Diffusion map by operator} highlights, the dilation step is needed to be sure the contact zone is connected to the pore space. Depending on the configuration, the selection of the morphological operator is the main choice. The larger the size of the operator is, the more diffusion inside the grains occurs. 

%%=======================================================%%

\section{Method to move solute out of a grain}
\label{Move out solute grain}

\begin{figure}[ht]
    \centering
    \includegraphics[height = 4.5cm]{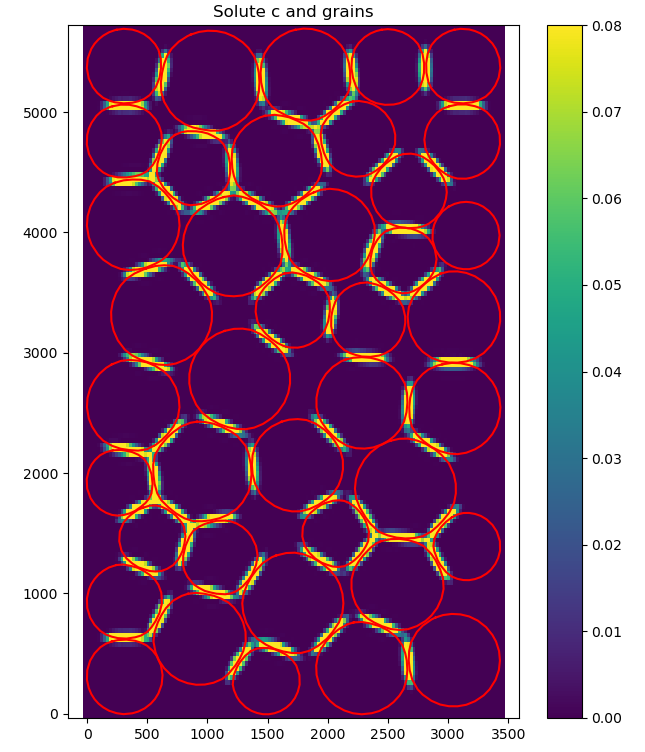}~
    \includegraphics[height = 4.5cm]{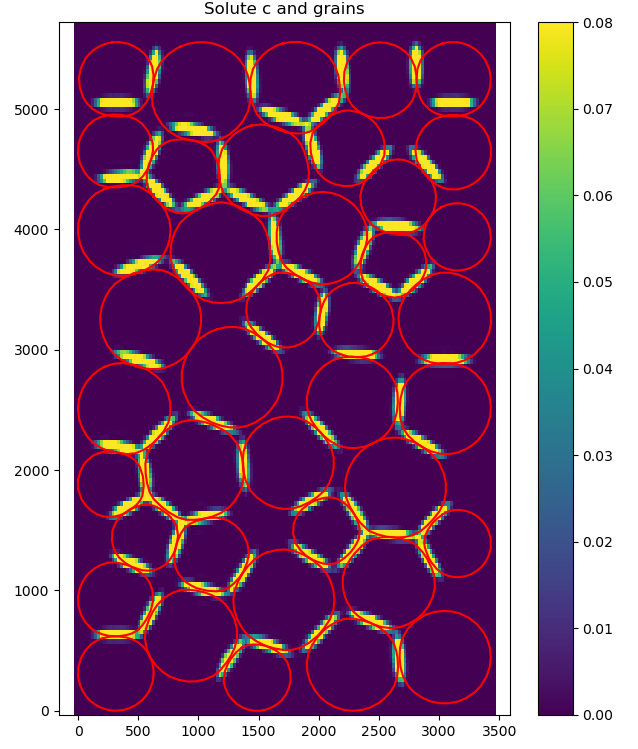}~
    \includegraphics[height = 4.5cm]{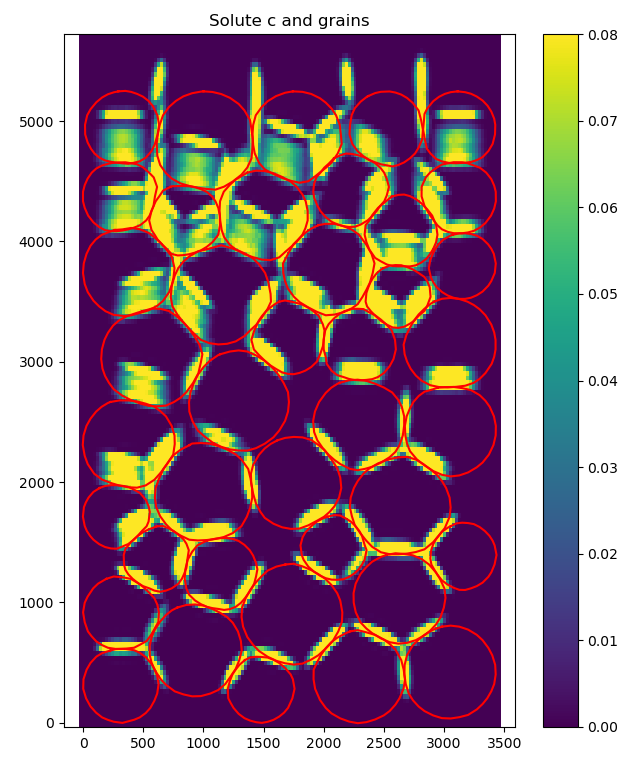}\\   
    \caption{Solute is generated initially (left). Then, the grains move after the DEM step to find a new equilibrium whereas the solute stays at the same place (center). Finally, it appears after multiple iterations  that the solute needs to follow the grains (left).}
    \label{Reason Algorithm Move Solute}
\end{figure}

Compare to the two grains pressure solution simulation, the multi-grain one faces trouble concerning the solute and the grain moves. As illustrated by figure \ref{Reason Algorithm Move Solute}, the solute generated does not follow the grain moves and it produces nonphysical results. It can be noticed the troubles are located more at the top of the sample. Hence, the simulation is made as the upper wall moves to verify the confinement pressure. Grains at the top move more than grains at the bottom. A new algorithm must be designed to move the solute with the grains.

\begin{figure}[ht]
    \centering
    \includegraphics[width=0.58\linewidth]{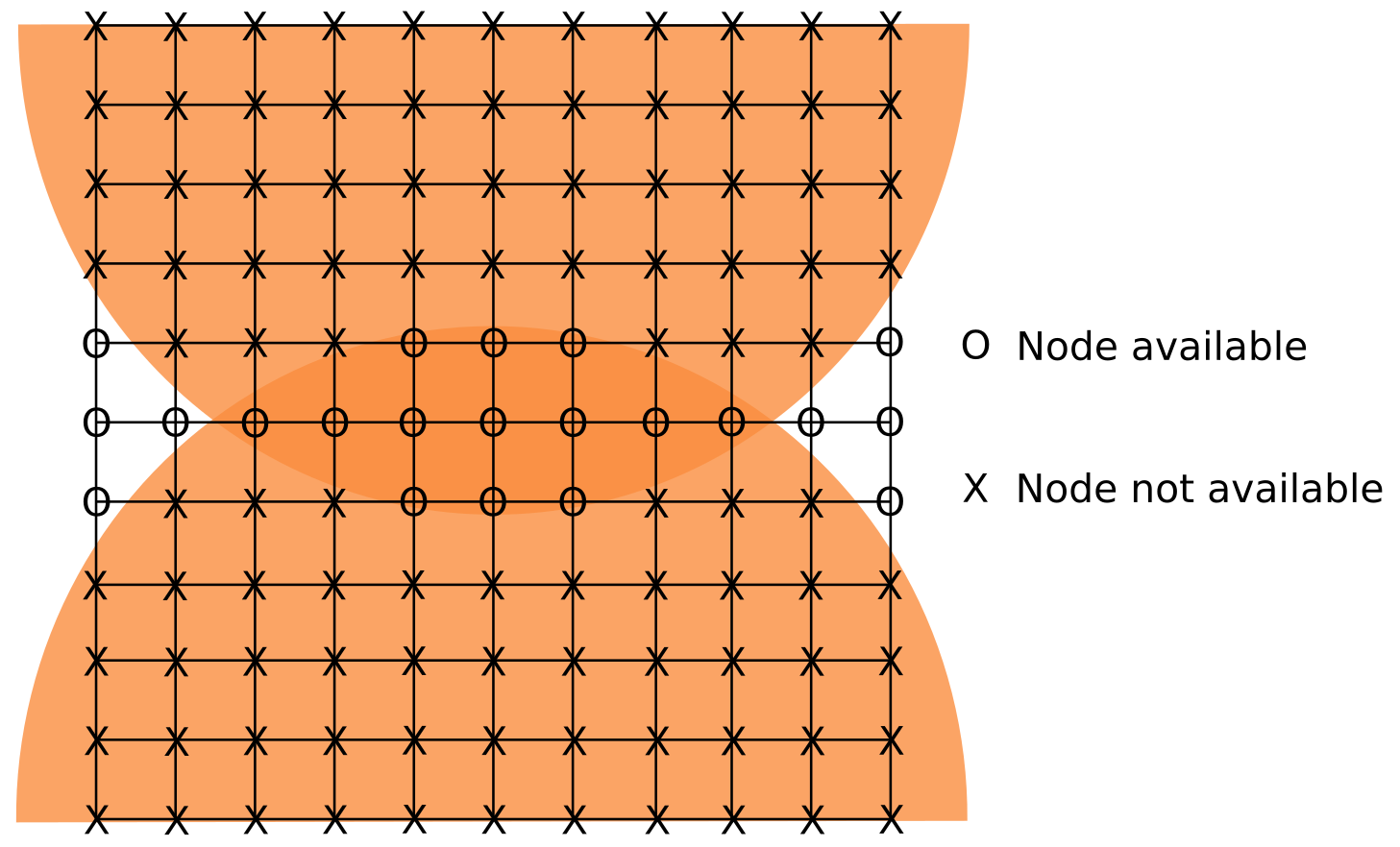}~
    \includegraphics[width=0.38\linewidth]{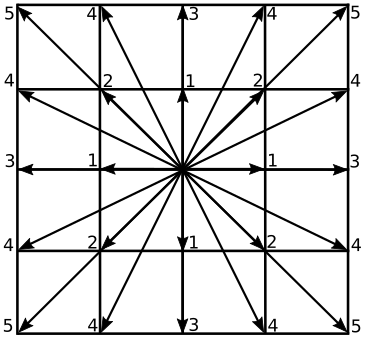}
    \caption{A node is considered available to host solute if it is a pore space or a contact area (left) and an illustration of the priorities to move the solute to the nearest node (right).}
    \label{Node Available Move Solute Priorities}
\end{figure}

The fundamental assumption is a small displacement of the grains. As illustrated by figure \ref{Node Available Move Solute Priorities}, all nodes are said available (in the pore space or in the contact area) or not (in only one grain) to host solute. Then, for nodes not available the solute is distributed to the nearest node available as illustrated in figure \ref{Node Available Move Solute Priorities}. If there are multiple nodes available at the same distance, the concentration to move is divided into equivalent parts.

The algorithm can be resumed by the following steps :
\begin{enumerate}
    \item a Boolean map is generated depending if the node can host or not solute.
    \item a dilation method \textit{scipy.ndimage.binary\_dilation} is applied to the Boolean map to generate a dilated map.
    \item For all nodes of the dilated map not available to host the solute, the nearest nodes are searched:
    \begin{enumerate}
        \item It is checked if at least one node of the priority $i$ can host solute. 
        \item If no node is available: the previous step is repeated with priority $i+1$.
        \item If at least one node is available: the solute is moved with the relation at every available node $c_{available} = c_{available} + c_{not\; available}/n_{available}$, where $c_{available}$ is the solute concentration at the available node, $c_{not\; available}$ is the solute concentration at the not available node and $n_{available}$ the number of available nodes at the priority $i$. The solute concentration at the not available node is set to $0$.
    \end{enumerate}
\end{enumerate}

%%=======================================================%%


\begin{thebibliography}{100}

\bibitem{McCartney2016} McCartney JS, S\'anchez M, Tomac I (2016) Energy geotechnics: Advances in subsurface energy recovery, storage, exchange, and waste management. Comput. and Geotech. 75: 244-256. https://doi.org/10.1016/j.compgeo.2016.01.002

\bibitem{Rattez2020} Rattez H, Veveakis M (2020) Weak phases production and heat generation control fault friction during seismic slip. Nat. Commun. 11: 350. https://doi.org/10.1038/s41467-019-14252-5

\bibitem{Lesueur2020} Lesueur M, Poulet T, Veveakis M (2020) Three-scale multiphysics finite element framework (FE3) modeling fault reactivation. Comput. Methods in Appl. Mech. and Eng. 365:112988. https://doi.org/10.1016/j.cma.2020.112988 

\bibitem{KUMAR2023} Ramesh Kumar K, Honorio H, Chandra D, Lesueur M, Hajibeygi H (2023) Comprehensive review of geomechanics of underground hydrogen storage in depleted reservoirs and salt caverns. J. Of Energy Storage. 73:108912. https://doi.org/10.1016/j.est.2023.108912

\bibitem{Brzesowsky2014} Brzesowsky RH, Spiers CJ, Peach J, Hangx SJT (2014) Time-independent compaction behavior of quartz sands. J. Geophys. Res. Solid Earth 119: 936-956. https://doi.org/10.1002/2013JB010444

\bibitem{Rohmer2016} Rohmer J, Pluymakers A, Renard F (2016) Mechano-chemical interactions in sedimentary rocks in the context of CO2 storage: Weak acid, weak effects?. Earth-Sci. Rev. 157. https://doi.org/10.1016/j.earscirev.2016.03.009.

\bibitem{Rohmer2016b} Manceau JC, Rohmer J (2016) Post-injection trapping of mobile CO 2 in deep aquifers: Assessing the importance of model and parameter uncertainties. Comput. Geosci. 20: 1251-1267. https://doi.org/10.1007/s10596-016-9588-x

\bibitem{RATTEZ2021} Rattez H, Disidoro F, Sulem J, Veveakis M (2021) Influence of dissolution on long-term frictional properties of carbonate fault gouge. Geomech. For Energy And The Environ. 26:100234. https://doi.org/10.1016/j.gete.2021.100234

\bibitem{Bjorlykke1989} Bjørlykke K, Ramm M, Saigal GC (1989) Sandstone diagenesis and porosity modification during basin evolution. Geol. Rundsch. 78:243–268. https://doi.org/10.1007/BF01988363

\bibitem{Bos2000} Bos B, Peach CJ, Spiers CJ (2000) Slip behavior of simulated gouge-bearing faults under conditions favoring pressure solution. J. of Geophys. Res. 105(B7):16,699–16,717. https://doi.org/10.1029/2000JB900089

\bibitem{Rutter1976} Rutter EH (1976) Discussion on natural strain and geological structure - The kinetics of rock deformation by pressure solution. Philos. Transact. of the R. Soc. of London. Series A, Math. and Phys. Sci. 283203–219.
http://doi.org/10.1098/rsta.1976.0079

\bibitem{Sleep1992} Sleep NH, Blanpied ML (1992) Creep, compaction and the weak rheology of major faults. Nat. 359: 687-692. https://doi.org/10.1038/359687a0

\bibitem{Tang2023} Tang X, Hu M (2023) A Reactive-Chemo-Mechanical Model for Weak Acid-Assisted Cavity Expansion in Carbonate Rocks. Rock Mech. and Rock Eng. 56:515-533. https://doi.org/10.1007/s00603-022-03077-2

\bibitem{Cundall1980} Burman B, Cundall P, Strack O (1980) A discrete numerical model for granular assemblies. Geotech. 30:331-336. https://doi.org/10.1680/geot.1980.30.3.331

\bibitem{OSullivan2011} O'Sullivan C (2011) Particulate Discrete Element modeling. https://doi.org/10.1201/9781482266498

\bibitem{Cha2019} Cha M, Santamarina J (2019) Pressure-dependent grain dissolution using discrete element simulations. Granul. Matter 21:1-10. https://doi.org/10.1007/s10035-019-0960-0

\bibitem{Alam2022} Alam M, Parol V, Das A (2022) A DEM study on microstructural behaviour of soluble granular materials subjected to chemo-mechanical loading. Geomech. for Energy and the Environ. 32: 100390. https://doi.org/10.1016/j.gete.2022.100390

\bibitem{vanDenEnde2018} van den Ende MPA, Marketos G, Niemeijer AR, Spiers CJ (2018) Investigating Compaction by Intergranular Pressure Solution Using the Discrete Element Method. J. of Geophys. Res.: Solid Earth 123: 107-124. https://doi.org/10.1002/2017JB014440

\bibitem{Guevel2022} Guével A, Rattez H, Veveakis M (2022) Morphometric description of strength and degradation in porous media. Int. J. of Solids and Struct., 241, 111454. https://doi.org/10.1016/j.ijsolstr.2022.111454

\bibitem{Binaree2019} Binaree T, Preechawuttipong I, Azéma E (2019) Effects of particle shape mixture on strength and structure of sheared granular materials. Phys. Rev. E 100:012904. https://doi.org/10.1103/PhysRevE.100.012904

\bibitem{Mollon2020} Mollon G, Quacquarelli A, Andò E, Viggiani G (2020) Can friction replace roughness in the numerical simulation of granular materials?. Granul. Matter 22:42 https://doi.org/10.1007/s10035-020-1004-5

\bibitem{Rorato2021} Rorato R, Arroyo M, Gens A, Andò E, Viggiani G (2021) Image-based calibration of rolling resistance in discrete element models of sand. Comput. and Geotech. 131:103929. https://doi.org/10.1016/j.compgeo.2020.103929

\bibitem{Garcia2009} Garcia X, Latham JP, Xiang J, Harrison J (2009) A clustered overlapping sphere algorithm to represent real particles in discrete element modeling. Geotech. 59: 779-784. https://doi.org/10.1680/geot.8.T.037

\bibitem{Rothenburg1991} Rothenburg, L, Bathurst, RJ (1991) Numerical simulation of idealized granular assemblies with plane elliptical particles. Comput. Geotech. 11:315–329 https://doi.org/10.1016/j.jmps.2017.10.003

\bibitem{Podlozhnyuk2018} Podlozhnyuk, A. (2018) Modeling superquadric particles in DEM and CFD-DEM: implementation, validation and application in an open-source framework.

\bibitem{Cundall1988} Cundall P. (1988) Formulation of a three-dimensional distinct element model-Part I. A scheme to detect and represent contacts in a system composed of many polyhedral blocks. Int. J. Rock Mech. Min. Sci. \& Geomech 25:107-116. https://doi.org/10.1016/0148-9062(88)92293-0

\bibitem{Nezami2004} Nezami EG, Hashash YMA, Zhao D, Ghaboussi J (2004) A fast contact detection algorithm for 3-D discrete element method. Comput. and Geotech. 31:575-587. https://doi.org/10.1016/j.compgeo.2004.08.002

\bibitem{AlonsoMarroquin2009} Alonso-Marroquin F, Wang Y (2009) An efficient algorithm for granular dynamics simulations with complex-shaped objects. Granul. Matter 11:317-329. https://doi.org/10.1007/s10035-009-0139-1

\bibitem{VanDerMeer2015} van der Meer F, Sluys L (2015) The Thick Level Set method: Sliding deformations and damage initiation. Comput. Methods in Appl. Mech. and Eng. 285:64-82. https://doi.org/10.1016/j.cma.2014.10.020

\bibitem{Kawamoto2018} Kawamoto R, Andò E, Viggiani G, Andrade JE (2018) All you need is shape: Predicting shear banding in sand with LS-DEM. J. of the Mech. and Phys. of Solids 111:375-392. https://doi.org/10.1016/j.jmps.2017.10.003

\bibitem{Landau1936} Landau L. (1936) The Theory of Phase Transitions. Nat. 138:840–841. https://doi.org/10.1038/138840a0

\bibitem{Cahn1958} Cahn J, Hilliard J (1958) Free energy of a nonuniform system. I. Interfacial free energy. The J. of Chem. Phys. 28:258-267. https://doi.org/10.1063/1.1744102

\bibitem{Allen1979} Allen S, Cahn J (1979) A microscopic theory for antiphase boundary motion and its application to antiphase domain coarsening. Acta Metall. 27:1085-1095. https://doi.org/10.1016/0001-6160(79)90196-2

\bibitem{Wheeler1992} Wheeler AA, Ltoettinger WJ,Mc Fadden GB (1992) A Phase-field model for isothermal phase transitions in binary alloys. Phys. Rev. A 45:7424-7439. https://doi.org/10.1103/PhysRevA.45.7424

\bibitem{Munjiza2004} Munjiza A (2004) The Combined Finite‐Discrete Element Method. https://doi.org/10.1002/0470020180

\bibitem{Shinagawa2014} Shinagawa K (2014) Simulation of grain growth and sintering process by combined phase-field/discrete-element method. Acta Mater. 66:360-369. https://doi.org/10.1016/J.ACTAMAT.2013.11.023

\bibitem{Samiei2013} Samiei K, Peters B, Bolten M, Frommer A (2013) Assessment of the potentials of implicit integration method in discrete element modeling of granular matter. Comput. and Chem. Eng. 49:183-193. https://doi.org/10.1016/j.compchemeng.2012.10.009

\bibitem{Hertz1882} Hertz H (1882) Über die Berührung fester elastischer Körper (On the contact of elastic solids). J. fur die Reine und Angew. Math. 92:156-171. https://doi.org/10.1515/crll.1882.92.156

\bibitem{Johnson1985} Johnson KL (1985) Contact Mechanics. Cambridge University Press, London. https://doi.org/10.1017/CBO9781139171731

\bibitem{DiRenzo2004} Di Renzo A, Di Maio F (2004) Comparison of contact-force models for the simulation of collisions in DEM-based granular flow codes. Chem. Eng. Sci. 59:525-541. https://doi.org/10.1016/j.ces.2003.09.037

\bibitem{Mindlin1953} Mindlin RD, Deresiewicz H (1953) Elastic spheres in contact under varying oblique forces. J. Appl. Mech. 20:327-344. https://doi.org/10.1115/1.4010702

\bibitem{Feng2023} Feng Y (2023) Thirty years of developments in contact modelling of non-spherical particles in DEM: a selective review. Acta Mech. Sin. 39:722343. https://doi.org/10.1007/s10409-022-22343-x

\bibitem{Guevel2020} Guével A, Rattez H, Veveakis M (2020) Viscous phase-field modeling for chemo-mechanical microstructural evolution: application to geomaterials and pressure solution. Int. J. of Solids and Struct. 207:230-249. https://doi.org/10.1016/j.ijsolstr.2020.09.026

\bibitem{Santamarina2009} Shin H, Santamarina J (2009) Mineral Dissolution and the Evolution of k0. J. of Geotech. and Geoenviron. Eng. 135:1141-1147. https://doi.org/10.1061/(asce)gt.1943-5606.0000053

\bibitem{Santamarina2014} Cha M, Santamarina J (2014) Dissolution of randomly distributed soluble grains: Post-dissolution k0-loading and shear. Geotech. 64:828-836. https://doi.org/10.1680/geot.14.P.115

\bibitem{Sheng2004} Sheng Y, Lawrence C, Briscoe B, Thornton C (2004) Numerical studies of uniaxial powder compaction process by 3D DEM. Eng. Comput. 21:304-317. https://doi.org/10.1108/02644400410519802

\bibitem{Cundall2004} Potyondy D, Cundall P (2004) A bonded-particle model for rock. Int. J. Rock Mech. Min. Sci. 41:1329-1364. https://doi.org/10.1016/j.ijrmms.2004.09.011

\bibitem{SacMorane2023} Sac-Morane A, Rattez H, Veveakis M (2023) Discrete Element Modeling of a fault reveals that viscous rolling relaxation controls friction weakening, Submitted to Granul. Matter. https//doi.org/10.1002/essoar.10508662.1

\bibitem{DeMeer1997} De Meer S, Spiers CJ, Peach CJ (1997) Pressure solution creep in gypsum: Evidence for precipitation reaction control. Phys. and Chem. of the Earth 22:33-37. https://doi.org/10.1016/s0079-1946(97)00074-8

\bibitem{Spiers1990} Spiers CJ, Schutjens PMTM, Brzesowky RH, Peach CJ, Liezenberg JL, Zwart HJ (1990) Experimental determination of constitutive parameters governing creep of rocksalt by pressure solution. Geol. Soc. Spec. Publ. 54:215-227. https://doi.org/10.1144/GSL.SP.1990.054.01.21

\bibitem{Gratier1993} Gratier JP (1993) Experimental pressure solution of halite by an indenter technique. Geophys. Res. Lett. 20:1647-1650. https://doi.org/10.1029/93GL01398

\bibitem{Urai2008} Urai J, Schléder Z, Spiers CJ, Kukla P (2008) Flow and Transport Properties of Salt Rocks. 

\bibitem{Aitken1938} Aitken A (1938) XX.—Studies in Practical Mathematics. II. The Evaluation of the Latent Roots and Latent Vectors of a Matrix. Proc. of the Royal Soc. of Edinburgh 57:269-304. https://doi.org/10.1017/s0370164600013808

\bibitem{Dysthe2002} Dysthe DK, Podladchikov Y, Renard F, Feder J, Jamtveit B (2002) Universal Scaling in Transient Creep. Phys. Rev. Lett. 89:1-4. https://doi.org/10.1103/PhysRevLett.89.246102

\bibitem{Zheng2015} Zheng J, Hryciw RD (2015) Traditional soil particle sphericity, roundness and surface roughness by computational geometry. Geotech. 65:494-506. https://doi.org/10.1680/geot.14.P.192

\bibitem{Lu2021} Lu R, Cheng C, Nagel T, Milsch H, Yasuhara H, Kolditz O, Shao H (2021) What process causes the slowdown of pressure solution creep. Geomech. and Geophys. for Geo-Energy and Geo-Resour. 7:1-11. https://doi.org/10.1007/s40948-021-00247-4

\bibitem{Moose2020} Permann CJ, Gaston DR, Andr{\v{s}} D, Carlsen RW, Kong F, Lindsay AD, Miller JM, Peterson JW, Slaughter AE, Stogner RH, Martineau RC (2020) {MOOSE}: Enabling massively parallel multiphysics simulation. SoftwareX 11:100430. https://doi.org/10.1016/j.softx.2020.100430

\bibitem{Franklin1994} Franklin R (1994) How do I find if a point lies within a polygon. In: comp.graphics.algorithms FAQ, www.faqs.org/faqs/graphics/ algorithms-faq/, p. Subject 2.03

\end{thebibliography}
\end{document}